\documentclass[12pt,twoside]{article}
\usepackage{amssymb,amsmath,graphicx,amsmath}
\usepackage[colorlinks=true,linkcolor=blue,urlcolor=blue,citecolor=blue]{hyperref}

\usepackage{times}
\usepackage{color}

\graphicspath{{./}{}}
\newcommand{\uu}{{u}}

\usepackage{pifont}

\usepackage{bbm}
\skip\footins 39pt plus 4pt minus 2pt
\def\footnoterule{\kern-19pt\hrule width.5in\kern18.6pt}\floatsep 12pt plus 2pt minus 2pt
\renewcommand{\baselinestretch}{1}
\renewcommand{\epsilon}{\varepsilon}

\graphicspath{{}}

\begin{document}

\newcommand{\red}{\color{red}}

\newcommand{\blue}{\color{blue}}
\newcommand{\fehler}[1]{{\color{red}#1}}
\newcommand{\qed}{\hfill\raisebox{-0.5mm}[0mm][0mm]{$\Box$}}
\newcommand{\st}{\!\!\!/}
\newcommand{\ksl}{k\!\!\!/}
\newcommand{\dsl}{\partial\!\!\!/}
\newcommand{\Asl}{A\!\!\!/}
\newcommand{\Dsl}{D\!\!\!\!/}
\newcommand{\half}{\frac{1}{2}}
\newcommand{\tbt}{{\bar \theta}\theta}
\newcommand{\A}{\alpha}
\newcommand{\B}{\beta}
\newcommand{\G}{\gamma}
\newcommand{\D}{\delta}
\newcommand{\E}{\varepsilon}
\newcommand{\T}{\theta}
\newcommand{\ts}{
   {\raisebox{-0.5ex}{\parbox{0.5ex}{
      \setlength{\unitlength}{0.5ex}
      \begin{picture}(1,6)
         \thinlines
         \put(0.5,-1){\line(0,1){7}}
      \end{picture}
   }}}
}
\newcommand{\lts}{
   {\raisebox{0ex}{\parbox{0.5ex}{
      \setlength{\unitlength}{0.5ex}
      \begin{picture}(1,12)
         \thinlines
         \put(0.5,-0.25){\line(0,1){12}}
      \end{picture}
   }}}
}
\newcommand{\N}{\mathbb{N}}
\newcommand{\R}{\mathbb{R}}
\newcommand{\Rscript}{\scriptstyle\mbox{\scriptsize \rm I}\!\mbox{\scriptsize\rm R}}
\newcommand{\rme}{{\mathrm{e}}}
\newcommand{\rmd}{{\mathrm{d}}}
\newcommand{\tr}{\mbox{tr}}
\newcommand{\nn}{\nonumber}
\newcommand{\weiter}{\nonumber \\ & & }
\newcommand{\nicht}[1]{}
\newlength{\lang}
\newcommand {\eq}[1]{(\ref{#1})}
\newcommand {\Eq}[1]{Eq.\hspace{0.55ex}(\ref{#1})}
\newcommand {\Eqs}[1]{Eqs.\hspace{0.55ex}(\ref{#1})}
\newcommand {\Sec}[1]{section~\ref{#1}}
\newcommand {\ds}{\displaystyle}
\newcommand {\tx}{\textstyle}
\newcommand {\scr}{\scriptstyle}
\newcommand {\scrscr}{\scripscriptstyle}
\newcommand {\ind}[1]{\mathrm{#1}}
\newcommand {\p}{\partial}
\newcommand{\trint}{\int\!\!\!\int\!\!\!\int}
\newcommand{\bn}{:\hspace*{-0.5ex}}
\newcommand{\en}{\hspace*{-0.5ex}:}
\newcommand{\1}{\,{\bf 1}\,}
\newcommand{\llongrightarrow}{-\!\!\!-\!\!\!\!\rightarrow}
\newcommand{\lllongrightarrow}{-\!\!\!-\!\!\!-\!\!\!\!\rightarrow}
\newcommand{\llllongrightarrow}{-\!\!\!-\!\!\!-\!\!\!-\!\!\!\!\rightarrow}
\newcommand{\lllllongrightarrow}{-\!\!\!-\!\!\!-\!\!\!-\!\!\!-\!\!\!\!\rightarrow}
\newcommand{\llllllongrightarrow}{-\!\!\!-\!\!\!-\!\!\!-\!\!\!-\!\!\!-\!\!\!\!\rightarrow}
\newcommand{\lllllllongrightarrow}{-\!\!\!-\!\!\!-\!\!\!-\!\!\!-\!\!\!-\!\!\!-\!\!\!\!\rightarrow}
\newcommand{\llllllllongrightarrow}{-\!\!\!-\!\!\!-\!\!\!-\!\!\!-\!\!\!-\!\!\!-\!\!\!-\!\!\!\!\rightarrow}
\newcommand{\lllllllllongrightarrow}{-\!\!\!-\!\!\!-\!\!\!-\!\!\!-\!\!\!-\!\!\!-\!\!\!-\!\!\!-\!\!\!\!\rightarrow}
\newcommand{\llllllllllongrightarrow}{-\!\!\!-\!\!\!-\!\!\!-\!\!\!-\!\!\!-\!\!\!-\!\!\!-\!\!\!-\!\!\!-\!\!\!\!\rightarrow}
\newcommand{\lllllllllllongrightarrow}{-\!\!\!-\!\!\!-\!\!\!-\!\!\!-\!\!\!-\!\!\!-\!\!\!-\!\!\!-\!\!\!-\!\!\!-\!\!\!\!\rightarrow}
\newcommand{\llllllllllllongrightarrow}{-\!\!\!-\!\!\!-\!\!\!-\!\!\!-\!\!\!-\!\!\!-\!\!\!-\!\!\!-\!\!\!-\!\!\!-\!\!\!-\!\!\!\!\rightarrow}
\newcommand{\setcurrentlabel}[1]{\def\@currentlabel{#1}}
\newbox{\expbox}
\newlength{\explength}
\newcommand{\EXPhelp}[6]{\sbox{\expbox}{\ensuremath{#4#1}}\settowidth{\explength}{\rotatebox{90}{\ensuremath{#4\left#5\usebox{\expbox}\right#6}}}\ensuremath{#4\left#5\usebox{\expbox}\right#6_{{#4\hspace*{0.4ex}}\hspace*{-0.22\explength}#2}^{{#4\hspace*{0.4ex}}\hspace*{-0.22\explength}#3}}}
\newcommand{\EXPdu}[3]{\mathchoice{\EXPhelp{#1}{#2}{#3}{\displaystyle}{<}{>}}{\EXPhelp{#1}{#2}{#3}{\textstyle}{<}{>}}{\EXPhelp{#1}{#2}{#3}{\scriptstyle}{<}{>}}{\EXPhelp{#1}{#2}{#3}{\scriptscriptstyle}{<}{>}}}
\newcommand{\EXP}[2]{\mathchoice{\EXPhelp{#1}{#2}{}{\displaystyle}{<}{>}}{\EXPhelp{#1}{#2}{}{\textstyle}{<}{>}}{\EXPhelp{#1}{#2}{}{\scriptstyle}{<}{>}}{\EXPhelp{#1}{#2}{}{\scriptscriptstyle}{<}{>}}}
\newcommand{\BRAhelp}[6]{\sbox{\expbox}{\ensuremath{#4#1}}\settowidth{\explength}{\rotatebox{90}{\ensuremath{#4\left#5\usebox{\expbox}\right#6}}}\ensuremath{#4\left#5\usebox{\expbox}\right#6_{{#4\hspace*{0.4ex}}\hspace*{-0.2\explength}#2}^{{#4\hspace*{0.4ex}}\hspace*{-0.2\explength}#3}}}
\newcommand{\BRA}[2]{\mathchoice{\BRAhelp{#1}{}{#2}{\displaystyle}{(}{)}}{\BRAhelp{#1}{}{#2}{\textstyle}{(}{)}}{\BRAhelp{#1}{}{#2}{\scriptstyle}{(}{)}}{\BRAhelp{#1}{}{#2}{\scriptscriptstyle}{(}{)}}}
\newbox{\atbox}
\newlength{\atlengtha}
\newlength{\atlengthb}
\newcommand{\AThelp}[4]{\sbox{\atbox}{\ensuremath{#3#1}}\settoheight{\atlengtha}{\ensuremath{#3\usebox{\atbox}}}\settodepth{\atlengthb}{\ensuremath{#3\usebox{\atbox}}}#3\addtolength{\atlengtha}{0.1ex}
#3\addtolength{\atlengthb}{0.75ex}\addtolength{\atlengtha}{\atlengthb}#1\rule[-\atlengthb]{0.1ex}{\atlengtha}_{\raisebox{0.12ex}{\ensuremath{\,#4#2}}}}\newcommand{\AT}[2]{\mathchoice{\AThelp{#1}{#2}{\displaystyle}{\scriptstyle}}{\AThelp{#1}{#2}{\textstyle}{\scriptstyle}}{\AThelp{#1}{#2}{\scriptstyle}{\scriptscriptstyle}}{\AThelp{#1}{#2}{\scriptscriptstyle}{\scriptscriptstyle}}}
\newlength{\picheight}\newcommand{\AxesPicture}[4]{\settowidth{\picheight}{\rotatebox{90}{$\tx#4$}}{\arraycolsep0.5ex
\renewcommand{\arraystretch}{1.3}
$\begin{array}{cc}
\mbox{\parbox{\picheight}{\rotatebox{90}{$~~\tx#4$}}}
 & \mbox{\epsfxsize=#2\textwidth\parbox{#2\textwidth}{\epsfbox{#1}}} \\
~ & \mbox{$~~\tx#3$}
\end{array}$}}
\newbox{\picbox}
\newlength{\picwidth}
\newcommand{\AxesPictureRight}[4]{\sbox{\picbox}{\epsfxsize=#2\textwidth\parbox{#2\textwidth}{\epsfbox{#1}}}
\settowidth{\picheight}{\rotatebox{90}{$\tx#4$}}\settowidth{\picwidth}{\rotatebox{90}{\usebox{\picbox}}}
\centerline{\arraycolsep0.5ex
\renewcommand{\arraystretch}{1.3}
$\begin{array}{cr}
\mbox{\parbox{\picheight}{\rotatebox{90}{\parbox{\picwidth}{\hfill{$\tx#4$}}}}}
 & \mbox{\usebox{\picbox}} \\
~ & \mbox{$\tx#3$}
\end{array}$}}
\newcommand{\AxesPictureSpecial}[6]{\centerline{$\raisebox{#5\textwidth}{\rotatebox{90}{$\tx#4$}}~
{{\epsfxsize=#2\textwidth\parbox{#2\textwidth}{\epsfbox{#1}}}}$$\hspace*{-2ex}\raisebox{#6\textwidth}{$\tx#3$}$}}
\newcommand{\NOhere}{\parbox{1cm}{\epsfxsize=1cm\epsfbox{./eps/isno-ani.eps}}}
\newcommand{\NO}{\marginpar{$\!\!\parbox{1.8cm}{\epsfxsize=1.8cm\epsfbox{./eps/isno-ani.eps}}$}}

\newcommand{\beq}{\begin{equation}}
\newcommand{\eeq}{\end{equation}}
\newcommand{\bea}{\begin{eqnarray}}
\newcommand{\eea}{\end{eqnarray}}
\newcommand{\bal}{\begin{align}}
\newcommand{\eal}{\end{align}}
\def\beginincorrect{

\noindent{\unitlength1mm
\begin{picture}(100,5)
\put(0,0){\line(0,1){5}}
\put(0,5){\line(1,0){179}}
\put(179,0){\line(0,1){5}}
\put(78,2){\mbox{\normalsize*****incorrect*****}}
\end{picture}}\newline
\scriptsize}
\def\endincorrect{

\noindent{\unitlength1mm
\begin{picture}(100,5)
\put(0,0){\line(0,1){5}}
\put(0,0){\line(1,0){179}}
\put(179,0){\line(0,1){5}}
\put(74,1){\mbox{\normalsize*****end incorrect*****}}
\end{picture}}

\normalsize}

\newcommand{\bilderscale}{0.35}
\newcommand{\textbilderscale}{0.25}

\newcommand{\fig}[2]{\includegraphics[width=#1]{#2}}
\newcommand{\pfig}[2]{\parbox{#1}{\includegraphics[width=#1]{#2}}}
\newcommand{\Fig}[1]{\includegraphics[width=\columnwidth]{#1}}
\newcommand{\rfig}[2]{\includegraphics[width=#1\columnwidth]{#2}}
\newlength{\bilderlength}
\newcommand{\usebilderscale}{\bilderscale}
\newcommand{\bilderskip}{\hspace*{0.8ex}}
\newcommand{\textdiagram}[1]{\renewcommand{\usebilderscale}{\textbilderscale}\diagram{#1}\renewcommand{\usebilderscale}{\bilderscale}}
\newcommand{\diagram}[1]{\settowidth{\bilderlength}{\bilderskip\includegraphics[scale=\usebilderscale]{#1}\bilderskip}\parbox{\bilderlength}{\bilderskip\includegraphics[scale=\usebilderscale]{#1}\bilderskip}}
\newcommand{\Diagram}[1]{\settowidth{\bilderlength}{\includegraphics[scale=\usebilderscale]{#1}}\parbox{\bilderlength}{\includegraphics[scale=\usebilderscale]{#1}}}

\newcommand{\LWC}{\cite{ChauveLeDoussalWiese2000a,LeDoussalWieseChauve2003}}

\newcommand{\ovl}[1]{\overline{#1}}
\newcommand{\mtin}[1]{\mbox{\tiny {#1}}}
\newcommand{\ca}[1]{{\cal #1}}
\newcommand{\sfrac}[2]{{\textstyle\frac{#1}{#2}}}

\title{\vspace{-1.5cm}\sffamily\Large\bfseries
Field Theory of Disordered Elastic Interfaces at 3-Loop Order: The $\beta$-Function}

\author{{\sffamily\bfseries  Kay J\"org
Wiese$^{1}$, Christoph Husemann\(^2\) and Pierre Le Doussal$^1$}
\\ \small $^{1}$CNRS-Laboratoire de Physique Théorique de l'Ecole Normale Sup\'erieure, PSL Research University,\\ \small Sorbonne Universit\'es, UPMC, 24 rue Lhomond, 75005 Paris, France.\\
\small \(^{2}\)Carl Zeiss AG, Carl Zeiss Promenade 10, D-07745 Jena, Germany}

\date{}\maketitle
\begin{abstract}
We calculate the effective action for disordered elastic manifolds in the ground state (equilibrium) up to 3-loop order. This yields the  renormalization-group $\beta$-function   to third order in $\epsilon=4-d$, in an expansion in the dimension $d$ around the upper critical dimension $d=4$. The calculations are  performed using exact RG, and several other techniques, which allow us to resolve consistently the problems associated with the cusp of the renormalized disorder.
\end{abstract}

\section{Introduction}\label{intro}

Disordered systems are notoriously difficult to treat, since naive perturbation theory leads to absurd results, as exemplified by the phenomenon of dimensional reduction \cite{ParisiSourlas1979}. Two main paths out of this dilemma have been pursued: Replica symmetry breaking \cite{MezardParisiVirasoro}, and the functional renormalization group. The latter goes back to the work by Wilson \cite{WilsonKogut1974}
and Wegner and Houghton \cite{WegnerHoughton1973}. For disordered systems these methods were first used by Daniel Fisher \cite{Fisher1985b}. However it took  until 1992 that Narayan and Fisher \cite{NarayanDSFisher1992a,NarayanDSFisher1992b}, shortly thereafter followed by Natterman, Stepanow, Tang and Leschhorn \cite{NattermannStepanowTangLeschhorn1992}, recognized that the disorder correlator, which plays the role of the coupling constant in the functional renormalization group (FRG) treatment, has to assume a cuspy form. The physical origin of this cusp lies in the metastability of the zero-temperature states which dominate the partition function, as recognized by Balents, Bouchaud and M\'ezard \cite{BalentsBouchaudMezard1996} in 1996. Only in 2006   this property was given a precise meaning as an observable, which can be measured in a numerical simulation both for the statics \cite{LeDoussal2006b,MiddletonLeDoussalWiese2006}, the driven dynamics \cite{LeDoussalWiese2006a,RossoLeDoussalWiese2006a}, and in an  experiment \cite{LeDoussalWieseMoulinetRolley2009}. This has nicely been reviewed in \cite{Bouchaud2006}. It was important conceptually, since the very existence of the cusp had in the early days questioned the validity of the method. Once this question of principle   solved, it remained the problems of feasibility {\em and }practicality: First, whether there is a controlled loop or \(\epsilon \)-expansion, and second how to implement a method which makes sense of the cusp in this loop expansion, and more particularly of the derivatives at the cusp. The latter change sign, depending on whether the limit is taken for positive or negative argument, not to mention the additional problems arising for a higher-dimensional field \cite{LeDoussalWiese2005a}. While these problems were conceptually simpler to solve for depinning \cite{LeDoussalWieseChauve2002}, due to the Middleton-theorem \cite{Middleton1992}, for the statics the question is more delicate. A consistent solution has been given at 2-loop order, based on renormalizability, recursive construction, or consistency schemes (the ``sloop-algorithm'' to be discussed below) \cite{ChauveLeDoussalWiese2000a,LeDoussalWieseChauve2003}, or exact RG \cite{ChauveLeDoussal2001,ScheidlDincer2000,LeDoussal2008}.
At 3-loop order, we performed several independent calculations.
Here we give their result for the $\beta$-function.

The analysis of the fixed point will be published separately \cite{HusemannWiese2017}. There we will extract the roughness exponent $\zeta$, obtain the fixed-point functions $R$ to 3-loop order,   give the corrrection-to-scaling exponent $\omega$, as well as the momentum dependent 2-point function.

\section{Model and basic definitions}

The equilibrium problem is
defined by the partition function ${\cal Z} := \int {\cal D}[u]\,
\exp(-{\cal H}[u]/T)$ associated to the Hamiltonian (energy)
\begin{equation} \label{ham}
 {\cal H}[u]= \int \rmd^d x\, \frac{1}{2} \left[\nabla u(x)\right]^2 +\frac{m^2}{2} \left[u(x)-w\right]^2 +V\big(u(x),x\big)
\ .
\end{equation}
In order to simplify notations, we will often note \begin{equation}\label{2.2}
\int_x f(x):=\int \rmd^d x\, f(x)\ ,
\end{equation}
and in momentum space
\begin{equation}\label{2.3}
\int_q \tilde f(q):=\int  \frac{\rmd^d q}{(2\pi)^d}\tilde f(q)\ .
\end{equation}
The Hamiltonian (\ref{ham}) is the sum of the elastic energy \(\int_x \frac{1}{2} \left[\nabla u(x)\right]^2 \) plus the confining potential  \(\frac{m^2}2 \int_x \left[u(x)-w\right]^2 \) which tends to suppress
fluctuations away from the  ordered state $u(x)={w}$, and
a random potential \(V(u,x)\) which enhances them. \(w\) is, up to a  factor of $m^2$, an applied external force, which is useful to measure the renormalized disorder  \cite{LeDoussal2006b,MiddletonLeDoussalWiese2006,LeDoussalWiese2006a,RossoLeDoussalWiese2006a,LeDoussalWieseMoulinetRolley2009,LeDoussalWiese2008a,LeDoussalMuellerWiese2007}, or properly define avalanches \cite{LeDoussalWiese2006a,RossoLeDoussalWiese2006a,LeDoussalWiese2009a,LeDoussalWiese2011b,LeDoussalMiddletonWiese2008,LeDoussalWiese2008c,RossoLeDoussalWiese2009a}.
The resulting roughness
exponent $\zeta$
\begin{equation}\label{lf5}
\overline{\left<[u(x) - u(x')]^2\right>} \sim |x-x'|^{2 \zeta}
\end{equation}
can be  measured in experiments.

Here and below $\left<\dots \right>$
denote thermal averages and $\overline {(\dots) }$
disorder ones. In the zero-temperature limit, the partition function is dominated by  the ground state, and  we may  drop the explicit thermal averages.
 The random potential can without
loss of generality \LWC\ be chosen Gaussian with second cumulant \begin{equation}\label{corrstat}
\overline{V (u, x) V(u',x')} =: R_0(u-u') \delta^d(x-x') \ .
\end{equation}
\(R_0(u)\)  takes various forms: Periodic systems
are described by a periodic function $R_0(u)$, random-bond disorder by a short-ranged function, and random-field disorder of
variance $\sigma$ by $R(u) \simeq - \sigma |u|$ at large $u$.

To average over disorder, we replicate the partition function \(n\)
times,  \(\overline{{\cal Z}^n} =:\rme^{-\cal S}\), which defines the effective action \(\cal S\),
\begin{equation}
{\cal S}[u]= \sum_{a=1}^n \int_x\frac{1}{2T} \left[\nabla u_\alpha (x)\right]^2
+ \frac{m^2}{2T}u_a(x)^2 -\frac1{2T^2}\sum_{a,b=1}^n\int_x R_0\big(u_{a}(x)-u_{b}(x)\big)\ .
\end{equation}
 We used the notations introduced in Eqs.~(\ref{2.2}) and (\ref{2.3}).
In presence of external sources \(j_a \), the \(n\)-times replicated action becomes
\begin{equation}\label{lf6}
{\cal Z}[j] := \int \prod_{a=1}^n {\cal D}[u_a]\, \exp\!\left(- {\cal S}[u] +
\int_x \sum_a j_a(x) u_a(x)\right) \ ,
\end{equation}
from which all static observables can be obtained.
$a$ runs from 1 to $n$, and the limit of zero  replicas $n=0$
is implicit everywhere.

\section{3-loop $\beta$-function}

In generalization of Eq.~(3.43) of  \LWC, we obtain the following functional renormalization group equation for the renormalized, dimensionless disorder correlator \(\tilde R(u)\),
\begin{eqnarray}
-m\partial_{m} \tilde R(u) &=& (\epsilon-4\zeta) \tilde R(u) + \zeta u \tilde R'(u)+  \textstyle \frac{1}{2}
{\tilde R''(u)}^{2}-\tilde R''(u)\tilde R''(0)\nn \\
 && +\left({\textstyle \frac{1}{2}} + \epsilon\, {\cal C}_{1} \right)
\Big[\tilde R''(u) {\tilde R'''(u)}^{2}-\tilde R''(0) {\tilde R'''({u})}^{2} - \tilde R''(u) {
\tilde R'''({0^+})}^{2}  \Big] \nn \\
&& + {\cal C}_{2} \Big[ {\tilde R'''(u)}^4-2 {\tilde R'''(u)}^{2} {\tilde R'''(0^+)}^{2}
\Big]+ {\cal C}_{3}\,\big[\tilde R''(u)-\tilde R''(0)\big]^2 {\tilde R''''(u)}^2 \rule{0mm}{3ex} \nn \\
&&+ {\cal C}_{4}\, \Big[  \tilde R''(u) {\tilde R'''(u)}^{2}{\tilde R''''(u)}-\tilde R''(0)
{\tilde R'''(u)}^{2}{\tilde R''''(u)}
-\tilde R''(u) { \tilde R'''(0^+)}^{2} {\tilde R''''(0)}  \Big] \ .\qquad \ \ \
\label{betafinal}
\end{eqnarray}
The coefficients are \begin{eqnarray}
{\cal C}_{1} & =&\frac{1}{4} +\frac{\pi ^{2}}{9}-  \frac{\psi'
(\frac{1}{3})}{6}  = -0.3359768096723647... \label{C1}\\ \label{C2}
{\cal C}_{2}
&=& \frac{3}{4}\zeta (3)+\frac{\pi ^{2}}{18}-\frac{\psi'
(\frac{1}{3})}{12} = 0.6085542725335131...\\ \label{C3}
{\cal C}_{3}&=&  \frac{\psi'
(\frac{1}{3})}{6} -\frac{\pi ^{2}}{9}= 0.5859768096723648... \\
{\cal C}_{4} \label{C4}
&=& 2+\frac{\pi ^{2}}{9}-\frac{\psi' (\frac{1}{3})}{6}= 1.4140231903276352... \ .
\end{eqnarray}
The first line contains the rescaling and 1-loop terms, the second line the 2-loop terms, and the last two lines the three 3-loop terms. Note that \(\mathcal{C}_1=\frac14 -\mathcal{C}_3\), and \({\cal C}_4=2-{\cal C}_3 = \sqrt 2-0.000190372...  \)

\section{Lifting ambiguities in a non-analytic theory, summary} \label{sec:list}
Ambiguities arise in a perturbative computation of the effective
action if one uses a non-analytic
action.   To resolve this issue, several methods have been designed, of which
we  list the most important ones below. Some failed attempts at 2-loop order are described
in Ref.~\cite{LeDoussalWieseChauve2003}.
In addition, the physics
of the problem
requires the theory to
be renormalizable, potential and without super-cusp, which gives  valuable
checks on the values of the ``anomalous'' graphs.
\smallskip

\smallskip

{\bf 1) Exact RG}. The starting point of exact RG (ERG) methods are
exact relations between functionals, for reviews
see \cite{BergesTetradisWetterich2002,LeDoussal2008}. A systematic but
straightforward
expansion in $\epsilon$
combines the anomalous terms from naive perturbation theory
in a way that makes them automatically non-ambiguous. This method and  the corresponding
derivation of the $\beta$-function is discussed in Section
\ref{sec:ERG}.

\smallskip

{\bf 2) Elimination of sloops.}
The idea, which will be explained in detail in Section \ref{sec:sloops} below,
is as follows: Since the propagator \(\left<\tilde u_a(k)\tilde u_b(-k)\right>=T
\delta_{ab}/(k^2+m^2)\) is diagonal in replica space, each contraction in a
diagram reduces the number of free replica sums by at most one. Doing a
contraction which does not constrain the number of replicas further   counts as
a factor of \(T=0\), and can thus
be set to zero. Further contracting such diagrams generates a set of
identities which, remarkably, is sufficient to obtain unambiguously the
2-replica projection without any further assumption. In some sense, it uses in a
non-trivial way the constraint that we
are working with a true $T=0$ theory. \smallskip

\smallskip

{\bf 3) Recursive construction:} An efficient method is to
construct diagrams recursively. The idea is to identify in a first
step parts of the diagram, which can be computed without
ambiguity. This is e.g.\ the 1-loop chain-diagram  discussed in Section
\ref{sec:sloops}. In a
second step, one treats the already calculated
sub-diagrams as effective vertices. In general, these vertices
have the same analyticity properties, namely are derivable twice,
and then have a cusp. (Compare\ $R (u)$ with $[ R'' (u)-R''
(0)]R''' (u)^{2}-R'' (u)R''' (0^{+})^{2}$, which is a contribution to the $\beta$-function at 2-loop order). By construction, this
method ensures renormalizability, at least as long as there is
only one possible path. However it is not more general than the
demand of renormalizability diagram by diagram, discussed below.

\smallskip

{\bf 4) Renormalizability diagram by diagram:}  Renormalizability  diagram by
diagram is the key to all proofs of perturbative renormalizability in
field-theory, see e.g.\
\cite{BogoliubovParasiuk1957,Hepp1966,Zimmermann1969,BergereLam1975,DDG2,DDG4,WieseHabil,Collins}.
These methods define a  subtraction operator $\mathrm{\bf R}$.
Graphically it can be constructed by drawing a box around each
sub-divergence, which leads to a {\em forest} or {\em nest} of
sub-diagrams (the counter-terms in the usual language), which have
to be subtracted, rendering the diagram {\em finite}. The advantage
of this procedure is that it explicitly assigns all counter-terms
to a given diagram, which finally yields a proof of perturbative
renormalizability. If we demand that this proof goes through for
the {\em functional} renormalization group, the counter-terms must
necessarily have the same functional dependence on $R(u)$ as the
diagram itself. In general, the counter-terms are less ambiguous,
and this procedure can thus be used to lift ambiguities in the
calculation of the diagram itself. By construction this procedure
is very similar to the recursive construction discussed under
point 3, and it is build in to the ERG approach.

It has some limitations though. Indeed, if one applies this
procedure to the 3-loop calculation, one finds that it renders
unique all but one ambiguous diagram, namely
\begin{equation}\label{lf161}
\diagram{3loopi}\ ,
\end{equation}
which has no subdivergence. Thus there are no counter-terms which
could lift the ambiguities. This diagram must therefore be computed
directly and we found that it can be obtained unambiguously by the
sloop elimination method.
We   give an explanation of this
method in section \ref{sec:sloops};  it is also well documented, see
section VD of \cite{LeDoussalWieseChauve2003}.

We will not give a detailed explanation of this method here, since we will not need it, and it is well documented, see section VD of \cite{LeDoussalWieseChauve2003}.
\smallskip

{\bf 5) Reparametrization invariance:} From standard field theory,
one knows that RG functions are not
unique, but
depend on the renormalization scheme. Only critical exponents are
unique. This is reflected in the freedom to reparametrize the
coupling constant $g$ according to $g  \longrightarrow  \tilde g
(g) $ where $\tilde g (g)$ is a smooth function, which has to be
invertible in the domain of validity of the RG $\beta$-function.

Here we have chosen a scheme, namely defining $R(u)$ from the
exact zero momentum effective action, using dimensional
regularization, and a mass. One can explore the freedom in
performing reparametrizations. In the functional RG framework,
reparameterizations are also functional, of the form
\begin{equation}\label{lf163}
R (u) \ \longrightarrow \ \hat R (u) = \hat R[R] (u)\ .
\end{equation}
Of course the new function $\hat R (u)$ does not have the same
meaning as $R(u)$. Perturbatively this reads
\begin{equation}\label{lf164}
R (u) \ \longrightarrow \ \hat R (u) = R (u) + B (R,R) (u) +O
(R^{3}) \ ,
\end{equation}
where $B (R, R)$ is a functional of $R$. For consistency, one has
to demand that $B (R,R)$ has the same analyticity properties as
$R$, at least at the fixed point $\tilde R=\tilde R^{*}$, i.e.\ $B
( R,R)$ should as $R$ be twice differentiable and then have a
cusp.
Details can be found in Section \ref{newrepara}.

\smallskip

As far as applicable, all methods, who are are genuinely different,
give consistent results. This is strong evidence that the problem has a
unique field theory, which we identify in this paper to 3-loop order.
In particular, the ambiguities which arise in perturbation theory due to the
cusp turn out to be superficial and are absent in our treatment.
Let us now turn to actual calculations using these methods.
We start with the ERG approach. We will  then use renormalized field theory and a
combination of the above-mentioned methods. Let us stress that each of these
two calculations was done independently by one of the authors, which
serves as a non-trivial check of the RG $\beta $-function such obtained.

\section{Calculation via the Exact Renormalization Group}
\label{sec:ERG}
In this section we derive the 3-loop flow equations by means of the exact renormalization group (ERG). In condensed matter this RG is sometimes called ``functional RG'' because it is based on exact flow equations formulated for thermodynamic functionals. To avoid possible confusions, we will use the term ``functional RG'' only in the sense of perturbative field theory, i.e.\ as a loop expansion.

\subsection{Set-up of ERG equation}
\label{sec:ERGset-up}
For each realization of the random potential \(V\), let $Z_V$ be the  partition function.
By the standard replica trick we average the logarithm of $Z_V$ over disorder
\begin{align}
 \ovl{\ln Z_V} = \lim_{n\to 0} \frac 1n \left(\ovl{Z_V^n}-1 \right)
\end{align}
by introducing replicas of the field. The replicated partition function is written as a functional integral
\begin{align}
 e^{W[j]} ={\cal N}_0 \int \prod_x \prod_{a=1}^n \mathrm{d} u_a(x)\, e^{-S[u] + (j,u) }\ .
\end{align}
It depends on an external replica-dependent field $j_a(x)$ with $a=1\ldots n$. We choose $\ca{N}_0 =(\ovl{Z_V^n})^{-1}$ such that $e^{W[0]}=1$.  We denote $(j,u)= \sum_a\int\mathrm{d}^dx \, j_a(x) u_a(x)$ and the replicated action is given by
\begin{align}
 S[u] = \frac{1}{2T} \sum_{a}\int\mathrm{d}^dx \big[(\nabla_x u_a)^2 + m^2 u_a(x)^2\big] - \frac{1}{2T^2} \sum_{a,b} \int\mathrm{d}^dx \; R_0\big(u_a(x)-u_b(x)\big) \, .
\end{align}
Correlation functions and other observables averaged over disorder can be calculated from replica averages obtained from a polynomial expansion of $W[j]$, see Ref.~\cite{LeDoussal2008} for details. For example, the connected 2-point correlation function is given by
\begin{align}
  \ovl{\langle u(x) u(y) {\rangle}}_{V} - \ovl{\langle u(x) {\rangle}}_{V}  \ovl{\langle u(y) {\rangle}}_{V}  &= \lim_{n\to 0} \Big[ \langle u_1(x)u_1(y) {\rangle}_{\mtin{rep}} -  \langle u_1(x)u_2(y) {\rangle}_{\mtin{rep}}\Big]\ ,
\end{align}
where $\langle u_a(x)u_b(y) {\rangle}_{\mtin{rep}} = \frac{\delta^2}{\delta j_a(x)\delta j_b(y)}_{| j=0} W[j]$. Note that $\langle u_a(x) {\rangle}_{\mtin{rep}}= \frac{\delta}{\delta j_a(x)}_{| j=0} W[j]=0$ since $S[u]=S[-u]$.

The mass $m^2>0$ provides an infrared regularization, and we are interested in the limit of $m^2\to 0$. The ERG is set up by successively lowering the parameter $m^2$, which is our RG scale. Since the action $S[u]$ depends on $m$ only via its quadratic part in the fields, the scale derivative of $W[j]$ can be expressed by a Polchinski-type equation
\begin{align}
 \dot{W}[j] = \frac{\mathrm{d}}{\mathrm{d} m} W[j]= -\frac{1}{2} \left( \frac{\delta W}{\delta j}, \dot{q} \frac{\delta W}{\delta j}\right) -\frac{1}{2} \mbox{Tr} \left[ \dot{q} \frac{\delta^2 W}{\delta j^2}\right] \, .
\end{align}
Here $q_{ab}(p)= T^{-1}(p^2 +m^2)\delta_{ab}$ denotes the bare inverse   propagator   and the derivative with respect to the scale $m$ is denoted by a dot. Due to momentum conservation the inverse   propagator in Fourier space has only diagonal elements. The inner product between two fields $u$ and $v$ sums over the replica index and the spatial dependence
\begin{align}
(u,v) := \sum_{a}\int\mathrm{d}^dx \; u_a(x)v_a(x)\, .
\end{align}
We use matrix notation such that, for example,
\begin{align}
\left( \frac{\delta W}{\delta j}, \dot{q} \frac{\delta W}{\delta j}\right) = \sum_{a,b}\int\mathrm{d}^dx  \int\mathrm{d}^dy \;\frac{\delta W}{\delta j_a(x)}  \dot{q}_{ab}(x,y) \frac{\delta W}{\delta j_b(y)} \, .
\end{align}
The second term in $S[u]$ is invariant under a shift with a replica-independent field. This is expressed by the so-called statistical tilt symmetry (STS)
\begin{align}
 W[j+\tilde{j}] =W[j] + \frac 12 (\tilde{j},g\tilde{j}) +(j,g\tilde{j})\ ,
\end{align}
 where $\tilde{j}$ is a replica-independent field and $g_{ab}(q)={q_{ab}(q)}^{-1}$. It follows that the thermal propagator
\begin{align}
\lim_{n\to 0} \sum_a \frac{\delta^2 W}{\delta j_a(x)\delta j_b(y)} \bigg|_{ j=0}= \lim_{n\to 0}\sum_a g_{ab}(x,y)
\end{align}
is not renormalized.

A Legendre transform of $W[j]$ allows   us to write a more convenient expansion in loops. For
this we define a functional map $u_a \mapsto J_a[u]$ such that
$\frac{\delta}{\delta j_a(x)} W[j] \Big|_{ j_a=J_a[u]} =u_a(x)$. This map
exists since the second functional derivative of $W$ is positive for $m>0$ at
$j=0$. The Legendre transform is defined as
\begin{align}
 \Gamma[u] = - W[J[u]] + (J[u],u)
\end{align}
and is called the effective action.
Therefore $\frac{\delta}{\delta u_a(x)} \Gamma[u] = J_a[u](x)$ and
$
 \frac{\delta^2 \Gamma}{\delta u^2} = \left(  \frac{\delta^2 W}{\delta j^2}\Big|_{ j=J[u]} \right)^{-1}
$
is the inverse full propagator.

The Legendre transformed version  of the statistical tilt symmetry reads
\begin{align}
 \Gamma[u+\tilde{u}] =\Gamma[u] + \frac 12 (\tilde{u},q \tilde{u}) +     (u,q\tilde{u})
\end{align}
with the field $\tilde{u}$ again being replica-independent. Because there is no thermal self-energy we write $\Gamma[u]=\frac{1}{2} (u,g^{-1}u) -\hat{\Gamma}[u]$. The flow equation for $\hat{\Gamma}$ follows from $\dot{\Gamma}=-\dot{W}$ and reads
\begin{align}\label{eq:GammaDot}
 \dot{\hat{\Gamma}}[u] = \frac 12 \mbox{Tr} \left[ g\dot{q} \left(1- g \frac{\delta^2 \hat{\Gamma}}{\delta u^2}\right)^{-1}\right] \ .
\end{align}
In the limit of $m\to \infty$ the effective action becomes the bare action without regularization
\begin{align}
 \lim_{m\to \infty} \Gamma[u] = S[u]\big|_{m=\infty} \, .
\end{align}

\subsection{Replica expansion hierarchy}
We expand the effective action in the number of replica sums
\begin{align}
\Gamma[u] = \frac{1}{2} (u,g^{-1}u) -\frac{1}{2 T^2} \sum_{a,b} R[u_{ab}]
- \sum_{n\ge 3} \frac{1}{n!T^n} \sum_{a_1,\ldots ,a_n}
S^{(n)}[u_{a_1},\ldots ,u_{a_n}]\ ,
\end{align}
where we use the shorthand notation $u_{ab}(x)=u_a(x)-u_b(x)$. Due to STS the one-replica term is given by the bare inverse thermal propagator. The two-replica term is a scale-dependent functional that depends on $u_{ab}(x)$ only. The initial condition for $R$ is local and given by the bare disorder distribution function
\begin{align}
 \lim_{m\to \infty} R[u] = \int \mathrm{d}^dx\; R_0(u(x))\ .
\end{align}
Higher replica terms are not present in the bare action but they are generated by the RG flow. STS implies that
\begin{align}\label{eq:STScumulants}
  S^{(n)}[u_{a_1},\ldots, u_{a_n}]= S^{(n)}[u_{a_1}+v,\ldots, u_{a_n}+v]
\end{align}
for any field $v(x)$. It follows that the two-replica term $S^{(2)}[u_a,u_b]=R[u_{ab}]$ is a functional of only one field. Because of the sum over all replica indices,  we assume all $n$-replica terms or $\Gamma$-cumulants to be symmetric under permutation of the replica fields.

We use the following compressed notation for functional derivatives of $n$-replica terms to denote $p_1$ derivatives with respect to field $u_{a_1}$ and similarly $p_i$ derivatives with respect to field $u_{a_i}$ for $i=1,...n$
\begin{align}
 S^{(n)}_{p_1\ldots p_n}[u_{a_1\ldots a_n}](x_1,\ldots ,x_{p_{\mtin{max}}}) = \sfrac{\delta}{\delta u_{a_1}(x_1)}\ldots \sfrac{\delta}{\delta u_{a_1}(x_{p_1})}  \sfrac{\delta}{\delta u_{a_2}(x_{p_1+1})} \ldots \sfrac{\delta}{\delta u_{a_n}(x_{p_{\mtin{max}}})} S^{(n)}[u_{a_1\ldots a_n}]
\end{align}
with the total number of derivatives $p_{\mtin{max}}=\sum_{i=1}^n p_i$ and the short-hand notation \\$ S^{(n)}[u_{a_1\ldots a_n}]=
S^{(n)}[u_{a_1},\ldots ,u_{a_n}]$. For example, using permutation symmetry, the
second functional derivative of $\Gamma$ is given by
\begin{align}\nonumber
 \frac{\delta^2 \hat{\Gamma}}{\delta u_a(x) \delta u_b(x)} &=\sum_{n=2}^{\infty} \frac{1}{(n-1)! T^n} \sum_{a_2\ldots a_n}\Big[ S_{20\ldots 0}^{(n)}[u_a,u_{a_2},\ldots, u_{a_n}](x,y) \delta_{ab}\\
&\hspace{2cm}  + \frac{1}{T} S_{110\ldots 0}^{(n+1)}[u_a,u_b,u_{a_2}, \ldots,
u_{a_n}](x,y)\Big]+ \frac{1}{T^2} S_{11}^{(2)}[u_a,u_b](x,y) \ .
\label{eq:DerivativeGamma}
\end{align}
Symmetrization over fields is denoted by curly brackets, that is, $\{ \ldots \}$ is the symmetrization of $\ldots$ over all its variables. Differentiating Eq.~(\ref{eq:STScumulants}) and using permutation symmetry implies that
\begin{align}
0&= \frac{\delta^2}{\delta v(x) \delta v(y)}\bigg|_{v=0} S^{(n)}[u_{a_1}+v,\ldots, u_{a_n}+v] \\ \nonumber
&= n \{S_{20\ldots 0}[u_{a_1}\ldots u_{a_n}](x,y)\} +2n(n-1) \{S_{110\ldots 0}[u_{a_1},\ldots, u_{a_n}](x,y)\} \ .
\end{align}
Because we are interested in the limit of the number of replica indices $n\to 0$ , we are free to add any function that depends on $k<n$ replicas to a $n$-th cumulant. This ``gauge invariance'' will be used later to get rid of constant terms in the cumulants.

Via Legendre transformation there is a one-to-one correspondence of $\Gamma$-cumulants to cumulants obtained from a replica expansion of $W[j]$, see Ref.~\cite{LeDoussal2008}. Therefore, the $\Gamma$-cumulants can be used to calculate observables. In particular, the exact 2-point correlation function averaged over disorder is given by
\begin{align}\label{eq:TwoPointCorrelation}
\ovl{\langle u(p) u(-p) \rangle}_V = \lim_{n\to 0} \left(\frac{\delta^2
\Gamma}{\delta u_a(p) \delta u_b(-p)}\bigg|_{u=0}  \right)^{-1}_{a=b} =
\frac{T}{m^2+p^2} - \frac{1}{\big[m^2+p^2\big]^2} \frac{\delta^2 R[u]}{\delta
u(p) \delta u(-p)}\bigg|_{u=0} \ .
\end{align}
Here, compared to leading-order perturbation theory, the second derivative of the bare function $R_0(u)$ is replaced by the second derivative of the renormalized functional $R[u]$.

In order to obtain RG equations for each $\Gamma$-cumulant, we expand the
inverse in Eq.~(\ref{eq:GammaDot}) in a geometric series
\begin{align}\label{eq:GammaGeometric}
 \dot{\hat{\Gamma}}[u] =\frac12 \sum_{l\ge 0}  \mbox{Tr} \left[ g\,\dot{q} \!  \left(g \frac{\delta^2 \hat{\Gamma}}{\delta u^2}\right)^l\right]\ ,
\end{align}
insert Eq.~(\ref{eq:DerivativeGamma}), and count the number of replica sums. The
propagators $g$ and $g\dot{q}g=-\dot{g}$ are diagonal in replica space. Replica
sums arise from second derivatives of $\hat{\Gamma}$, their matrix products, and
one additional sum from the trace. Therefore, in order to calculate the flow
equation of the $n$-th cumulant, the geometric series in
Eq.~(\ref{eq:GammaGeometric}) does not contribute for $l>n$. On the other hand,
a term in the geometric series of $l$-th order contributes to cumulants to all
orders $n\ge l$. That is, for any initial action the RG
flow generates cumulants to all orders.

The term $l=0$ and the one-replica term in $l=1$ in
Eq.~(\ref{eq:GammaGeometric}) are  constants and can therefore  be neglected due to
gauge invariance. Evaluating the two-replica contributions in the $l=1$
and $l=2$ terms give the flow equation
\begin{align}\nonumber \label{eq:dotR}
\dot{R}[u ] &= \int_{x_1,x_2}   \dot{g}(x_1,x_2) \left[ T  \ca{R}''[u](x_2,x_1)  + S_{110}^{(3)}[0,0,-u](x_2,x_1)  \right] \\ & \;\;+\frac 12 \int_{x_1,...,x_4}  \left[ \frac{\mathrm{d}}{\mathrm{d}m} g(x_1,x_2)g(x_3,x_4) \right]  \ca{R}''[u](x_2,x_3) \ca{R}''[u](x_4,x_1)  \ ,
\end{align}
where $\ca{R}''[u]=R''[u]-R''[0]$ with
$
R''[u](x,y) = S^{(2)}_{20}[u_1,u_2](x,y)_{|u_1 - u_2 = u}
$.
The evaluation at zero field arises in terms of coinciding replica indices. We note that Eq.~(\ref{eq:dotR}) is a non-linear integro-differential equation for a functional. Similar equations can be obtained for higher cumulants, $\dot{S}^{(3)}$ and $\dot{S}^{(4)}$; they are given in Appendix~\ref{app:ReplicaExpansionRGE}. Due to the $l=1$ term in Eq.~(\ref{eq:GammaGeometric}) there is a contribution from $S^{(m+1)}$ to $\dot{S}^{(m)}$. Therefore, in order to obtain exact solutions for the $\Gamma$-cumulants, one has to consider the full, infinitly large hierarchy. Note that, formally, up to now no approximations were made; in particular, we do not encounter ambiguities when a cusp in the second derivative of the local disorder distribution function develops.

\subsection{$\epsilon$-expansion for $T=0$}
Since we cannot treat an infinite hierarchy, we perform an  additional  expansion in $\epsilon=4-d$. To this aim we split the disorder distribution functional \(R[u]\) into a local  and a non-local part
\begin{align}
 R[u] = \int \mathrm{d}^d x \; R(u(x)) + \hat{R}[u]  \, .
\end{align}
If $u(x)=u_0$ is a constant field, then $\hat{R}[u_0]=0$, so that only the local part contributes, $R[u_0]=L^d R(u_0)$, where $L^d=\int\mathrm{d}^d x$ is the  volume of the system. Note that $R[u]$ and $\hat R[u]$  are functionals, whereas $R(u)$ is a function. For $m\to \infty$ the disorder-distribution function has only a local part, which we  assume to be {\em small}. That is, $R_0(u)$ and all its derivatives are uniformly bound by a small constant\footnote{Due to the formation of a cusp, this consideration does not apply to derivatives at the cusp, which  become infinite. We will discuss this later.}. We also assume that the local part  $R(u)$ of the renormalized  disorder distribution function remains small. Then the $\epsilon$-expansion can be obtained by expanding the replica expansion hierarchy in $R(u)$. From now on we set the temperature  to $T=0$. Because the rescaled temperature $T$ becomes small for small \(m\), the $\epsilon$-expansion can also be obtained for $T>0$ by a composite expansion in $T$ and $R(u)$.

Suppose that $R(u) \sim \ca{O}(\epsilon)$. Since for $T=0$, Eq.~(\ref{eq:dotR}) is quadratic in $\ca{R}''$, the non-local part of the renormalized disorder distribution function will be $\hat{R}[u] \sim \ca{O}(\epsilon^2)$. A similar argument for the higher $\Gamma$-cumulants gives $S^{(n)} \sim \ca{O}(\epsilon^n)$ for $n\ge 3$. The assumption that also all derivatives of the cumulants remain of the same order has to be checked; we will do so up to order $\epsilon^4$, that is, 3-loop order. With this method the 2-loop order was already obtained in Ref.~\cite{LeDoussal2008}.

The 1-loop equation can be obtained by an expansion to second order in $\epsilon$ and can directly be  read off from Eq.~(\ref{eq:dotR}). We use that
\begin{align}
 R''[u](x_1,x_2) = R''\big(u(x_1)\big) \delta(x_1-x_2) + \hat {R}''[u](x_1,x_2) \, ,
\end{align}
where the second term is already $\ca{O}(\epsilon^2)$ and does not contribute to Eq.~(\ref{eq:dotR}) at 1-loop order. We therefore find
\begin{align}\label{eq:oneloopbeta}
 \dot{R}[u] = \frac 12 \int_{x_1,x_2}  \left[ \frac{\mathrm{d}}{\mathrm{d}m} g(x_1,x_2)^2 \right] \ca{R}''\big(u(x_1)\big) \ca{R}''\big(u(x_2)\big) +\ca{O}(\epsilon^3)\ ,
\end{align}
where $\ca{R}''(u) := R''(u) - R''(0)$. The local part is obtained by inserting the constant field $u(x)=u_0$ and dividing by $L^d$
\begin{align}\label{R-1loop-ERG}
 \dot{R}(u_0) = \frac 12  \dot{I}_1 \ca{R}''(u_0)^2 +\ca{O}(\epsilon^3)
\end{align}
where after Fourier transformation $I_1 = \int _p \;g(p)^2 \sim \ca{O}(\frac{m^{-\epsilon}}{\epsilon})$, and so $\dot{I}_1 \sim \ca{O}(1)$.
The diagram $I_{1}$ is evaluated in Eq.~(\ref{I1}). In order to have the simplist possible formulas, we will absorb a factor of $\epsilon I_{1}|_{{m=1}}$ into the renormalized disorder, see Eq.~(\ref{6.43}). This  effectively   sets $I_{1}$ to $m^{{-\epsilon}}/\epsilon$. For an $n$-loop integral $I_{n}$ we will have to evaluate the  ratio $I_{n}/I_{1}^{n}$. We believe this to be the most convenient convention for obtaining standardized expressions.

Up to rescaling Eq.~(\ref{R-1loop-ERG}) is the standard 1-loop FRG equation. The solution of this flow equation corrects $R_0(u)\sim \ca{O}(\epsilon)$ to  the renormalized $R(u)$ to order $\epsilon^2$. The non-local part in terms of this renormalized disorder-distribution function is given by
\begin{align}\label{eq:OneLoopNonLocal}
 \hat{R}[u] = \frac 12 \int_{x_1,x_2} g(x_1,x_2)^2 \ca{R}''(u(x_1)) \ca{R}''(u(x_2))  - \frac 12 I_1 \int_{x} \ca{R}''(u(x))^2 +\ca{O}(\epsilon^3)
\end{align}
Superficially, the $\epsilon$-expansion seems to work. However, we assumed that $\hat{R}[u]$ is of order $\epsilon^2$ and likewise all derivatives of $\hat{R}[u]$. In fact, the existence of the cusp in $R''(u)$ of the 1-loop solution appearing at a finite scale
destroys our assumptions. Due to this cusp, $R''(u)$ is not differentiable at $u=0$. The left- and right-sided limits of $R'''$ exist but do not coincide $R'''(0^+)=-R'''(0^-)$. The fourth derivative $R''''(u)$ is uniformly bound for $u\neq 0$ but it is infinity  at $u=0$.

If we would only need up to two derivatives of $R$, the $\epsilon$-expansion would work without caveats. However, even the computation of the 2-point correlation function in Eq.~(\ref{eq:TwoPointCorrelation}) requires a second derivative of the non-local part of $R[u]$, that is, a second derivative of Eq.~(\ref{eq:OneLoopNonLocal}) at zero field. There enters a third and fourth derivative of $R(u(x))$ that have to be evaluated at $u(x)=0$. Furthermore, in the derivation of higher orders in $\epsilon$, that is, higher orders in the expansion of the replica hierarchy in $R(u)$, one encounters higher derivatives as well.

For the calculation of observables via analytic continuation it suffices to evaluate derivatives of $R(u)$ at $u=0^\pm$, if the left- and right-sided limits give the same result for the observable. For example, the ambiguity is avoided if odd derivatives of $R(u)$ enter the equations only squared. However, we work with fluctuating fields $u(x)$ and, for example, the second derivative of Eq.~(\ref{eq:OneLoopNonLocal}) contains a term $R'''(u(x_1))R'''(u(x_2))$. While this is a square term we have to ensure that either $u(x)\to 0^+$ or $u(x)\to 0^-$ uniformly for all $x$. That is, {\em ambiguities can be avoided if we restrict to non-crossing configurations.}

Note that none of our methods can handle observables involving crossing configurations. However, handling those is not necessary for the present purpose. The calculation of correlation functions requires only configurations which are infinitesimally close to a uniform one (see e.g. the discussion in section V.E of [23]). Hence the two methods presented here are consistent, and consistent with each other.

From now on, the limit of two fields $u_a(x)$ and $u_b(x)$ being equal in a $\Gamma$-cumulant is understood as
\begin{align}\label{eq:weakcont}
 \lim_{u_b\to u_a} S^{(n)}[u_a,u_b, \ldots] := \lim_{u_0\to 0^+} S^{(n)}[u_a,u_a+u_0, \ldots]
\end{align}
where $u_0$ is a constant field. That is, all fields are assumed to be close to a uniform configuration. We demonstrate in the next two subsections that in this weak limit the 3-loop $\beta$-function can be derived consistently. That is, it does not matter if the right limit $u_0\to 0^+$ or left limit $u_0\to 0^-$ are taken in Eq.~(\ref{eq:weakcont}).

\subsection{$\epsilon$-expansion to 2-loop order}
As an instructive example we review the 2-loop ERG  calculation done in Refs.~\cite{ChauveLeDoussal2001,LeDoussal2008}. In order to obtain Eq.~(\ref{eq:dotR}) to order $\epsilon^3$ we have to compute $S_{110}^{(3)}[0,0,-u]$ and $\ca{R}''[u](x_2,x_3) \ca{R}''[u](x_4,x_1)$ to this order. We first concentrate on the second term. Note that we expand in the renormalized disorder distribution function $R(u)$. This gives
\begin{align}
 \ca{R}''[u](x,y)= \ca{R}''(u(x)) \delta(x-y) + \hat{R}''[u](x,y) -\hat{R}''[0^+](x,y)\ .
\end{align}
$\ca{R}''(u(x))$ is of order \( \epsilon\) and we have to expand the non-local part $\hat{R}''$ to second order in $R(u)$, that is, we have to insert the second derivative of Eq.~(\ref{eq:OneLoopNonLocal})
\begin{align}\label{eq:TwoLoopNonLocal}
 \hat{R}''[u](z_1,z_2)&= \delta^d(z_1-z_2) \bigg[ \int_x  g(z_1,x)^2 R''''(u(z_1)) \ca{R}''(u(x)) -I_1 R'''(u(z_1))^2 \\
 &\quad\quad\quad \quad\quad\quad- I_1 R''''(u(z_1)) \ca{R}''(u(z_1)) \bigg] +g(z_1,z_2)^2 R'''(u(z_1)) R'''(u(z_2)) +\ca{O}(\epsilon^3)\nonumber.
\end{align}
As described above, in order to calculate $\hat{R}''[0^+](z_1,z_2)$ we first insert a constant field $u_0$ and then take the limit $u_0\to 0^+$ to obtain
\begin{align}\label{eq:TwoLoopNonLocalZero}
 \hat{R}''[0^+](z_1,z_2)= \big[ g(z_1,z_2)^2 -\delta^d(z_1-z_2) I_1 \big] R'''(0^+)^2 +\ca{O}(\epsilon^3)\,.
\end{align}
It is important to note that in this ``weak limit'' we obtain no ambiguities since $R'''(0^+)^2=R'''(0^-)^2$ has a straightforward analytic continuation.

The feedback from $S^{(3)}$ is calculated by retaining only terms of order $\epsilon^3$. The calculation is relegated to appendix \ref{app:ReplicaExpansionRGE}. The result from Eq.~(\ref{eq:dotS3}) is
\begin{align} \label{eq:dotS3twoloop}
\dot{S}^{(3)}[u_{abc}] &=
  \int_{x_1,...,x_6} \left[ \frac{\mathrm{d}}{\mathrm{d}m} g(x_1,x_2)g(x_3,x_4)g(x_5,x_6) \right] \times \\ \nonumber & \hspace{3cm}\times \Big( 3 \ca{R}''[u_{ab}](x_2,x_3) \ca{R}''[u_{ac}](x_4,x_5) \ca{R}''[u_{ac}](x_6,x_1)  \\ \nonumber & \hspace{3cm} ~~~~~\;  - \ca{R}''[u_{ab}](x_2,x_3) \ca{R}''[u_{bc}](x_4,x_5) \ca{R}''[u_{ac}](x_6,x_1)\Big) +\ca{O}(\epsilon^4) \, .
\end{align}
To this order it integrates to
\begin{align}
 S^{(3)}[u_{abc}] &= \left\{ 3\,\mbox{tr} \big[ g\ca{R}''_{ab}g\ca{R}''_{ab}g\ca{R}''_{ac}\big] -\mbox{tr} \big[ g\ca{R}''_{ab}g\ca{R}''_{bc}g\ca{R}''_{ac}\big]\right\} +\ca{O}(\epsilon^4)\, ,
\end{align}
where $\ca{R}_{ab}=\ca{R}[u_{ab}]$. Here the trace is over real space. The symmetrization over fields $\{\ldots \}$ can be written as $S^{(3)}[u_{abc}]= \frac 12 (A_1+A_2+A_3)+\ca{O}(\epsilon^4) $ with
\begin{align}\label{eq:A123}
A_1&=  \mbox{tr} \big[ g\ca{R}''_{ab}g\ca{R}''_{ab}g(\ca{R}''_{ac}+\ca{R}''_{bc})\big] \\ \nonumber
A_2&=  \mbox{tr} \big[ g\ca{R}''_{ab}g(\ca{R}''_{ac}-\ca{R}''_{bc})g(\ca{R}''_{ac}-\ca{R}''_{bc})\big]\\ \nonumber
A_3&= \sfrac 13 \mbox{tr} \big[ g(\ca{R}''_{ac}+\ca{R}''_{bc})g(\ca{R}''_{ac}+\ca{R}''_{bc})g(\ca{R}''_{ac}+\ca{R}''_{bc})\big]\ .
\end{align}
In a next step we take the functional derivatives of these terms with respect to $u_a(x)$ and $u_b(y)$. Then the limit $b\to a$ is performed by first replacing $u_b(y)$ by $u_a(x)+u_0$ and then sending $u_0\to 0^+$. The remaining fields $u_a(x)$ only occur in the combination $u_a(x)-u_c(x)$ and can directly be set  to zero.
When taking the functional derivatives of Eqs.~(\ref{eq:A123}), it is helpful to remember that we set $b=a$ afterwards. Therefore $A_2$ does not contribute to $S_{110}^{(3)}[0,0,-v]$ and $A_1$ contributes only if the derivatives act on the first two $\ca{R}_{ab}$ in the trace. Finally, the term $A_3$ is a symmetric functional in $u_{ac}$ and $u_{bc}$ and can be symmetrically expanded in $u_{ab}$ as outlined in the appendices of Refs.~\cite{LeDoussal2008,BalentsLeDoussal2004}. Setting $u_c(x)=u(x)$, we obtain
\begin{align}\label{eq:S3DiffTwoLoop}
 S^{(3)}_{110}[0,0,-u](x_1,x_2) =&\; 2 \big[R'''(x_1) R'''(x_2)- R'''(0^+)^2\big]  g(x_1,x_2) \int_y  \; g(x_1,y)g(y,x_2) \ca{R}''(y) \nn\\&+\ca{O}(\epsilon^4)\ .
\end{align}
 In the above equation and from now on we use the  shorthand notation ${R}''(x):={R}''(u(x))$ (and similar for higher derivatives expect for $x=0^+$ and $x=u$).
Inserting Eqs.~(\ref{eq:TwoLoopNonLocal}), (\ref{eq:TwoLoopNonLocalZero}), and (\ref{eq:S3DiffTwoLoop}) into Eq.~(\ref{eq:dotR}) we arrive at the 2-loop result
\begin{align}\label{eq:TwoLoopBeta}
\dot{R}[u] &= \int_{x_1,...,x_3} \left[\sfrac{\mathrm{d}}{\mathrm{d}m} g(x_1,x_2) g(x_1,x_3)\right] \big[ g(x_2,x_3)^2 -\delta(x_2-x_3) \big] \ca{R}''(x_1) \\ \nonumber & \hspace{6cm} \times \big[ R'''(x_2) R'''(x_3) -R'''(0^+)^2\big] \\ \nonumber & \;\;
+ \int _{x_1,...,x_3} \left[\sfrac{\mathrm{d}}{\mathrm{d}m} g(x_1,x_2)^2 \right] \big[ g(x_2,x_3)^2 -\delta(x_2-x_3) \big] \ca{R}''(x_1) R''''(x_2) \ca{R}''(x_3)\\ \nonumber & \;\;
+ \int _{x_1,...,x_3} \left[\sfrac{\mathrm{d}}{\mathrm{d}m} g(x_1,x_2)^2 \right] g(x_1,x_2)g(x_1,x_3) \ca{R}''(x_1) \big[ R'''(x_2) R'''(x_3) -R'''(0^+)^2\big]
\end{align}
The 2-loop $\beta$-function known from FRG calculations \cite{ChauveLeDoussalWiese2000a,LeDoussalWieseChauve2003} is the local contribution and is obtained by inserting a constant field and dividing by $L^d$
\begin{align}
 \dot{R}(u) &= \frac 12  \dot{I}_1 \ca{R}''(u)^2 + (\dot{I}_A -I_1\dot{I}_1) \ca{R}''(u) \Big[ R'''(u)^2-R'''(0^+)^2\Big] + {\ca O}(\epsilon^4)\ ,
\label{5.35}
\end{align}
where
\begin{align}
I_A= \int _{p_1,p_2} \; g(p_1) g(p_2) g(p_1+p_2)^2  \sim \frac{m^{-2\epsilon}}{\epsilon^2}\ .
\end{align}
The $\ca{O}(\frac{1}{\epsilon}) $  term in $\dot{I}_A$ is cancelled by $I_1 \dot{I}_1$, ensuring a finite $\beta$-function.
The non-local part integrates to
\begin{align}\label{eq:NonLocalTwoLoop}
 \hat{R}[u]=A[u] + \int_x B(u(x)) + {\ca O}(\epsilon^4)
\end{align}
with contributions
\begin{align} \nn
A[u] &= \frac 12 \int _{x_1,x_2}g(x_1,x_2)^2\; \ca{R}''(x_1) \ca{R}''(x_2) + \frac 12 \int _{y_1,y_2,z} g(y_1,z)^2 g(y_2,z)^2\; \ca{R}''(y_1)\ca{R}''(y_2)R''''(z) \\
\nn
& \quad +\int  _{x_1,x_2,y}   g(x_1,x_2)^2 g(x_1,y)g(x_2,y)  \Big[ R'''(x_1)R'''(x_2) -R'''(0^+)^2\Big] \ca{R}''(y)\\
& \quad - I_1 \int  _{x_1,x_2}  g(x_1,x_2) \; \ca{R}''(x_1) \Big[R'''(x_2)^2 -R'''(0^+)^2 + \ca{R}''(x_2) R''''(x_2) \Big]
\end{align}
and
\begin{align}
 B(u) &= -\frac 12 I_1 \ca{R}''(u)^2 + (I_1^2- I_A) \ca{R}''(u) \Big[R'''(u)^2 -R'''(0^+)^2\Big] +\frac 12 I_1^2 \ca{R}''(u)^2 R''''(u)\ .
\end{align}

\subsection{$\epsilon$-expansion to 3-loop order}
In 3-loop order we have to compute $S_{110}^{(3)}[0,0,-u]$ and $\ca{R}''[u](x_2,x_3) \ca{R}''[u](x_4,x_1)$ in Eq.~(\ref{eq:dotR}) to order $\epsilon^4$. The flow equation for the three-replica cumulant at $T=0$, see Eq.~(\ref{eq:dotS3}), is given by
\begin{align}\label{eq:dotS3T0}
\dot{S}^{(3)}[u_{abc}] &=\int _{x_1,x_2} \;\dot{g}(x_1,x_2)  \left\{ \sfrac 32 S_{1100}^{(4)}[u_{aabc}](x_1,x_2) \right\} \\ \nonumber
&\;\; +\int _{x_1,...,x_4} \left[ \sfrac{\mathrm{d}}{\mathrm{d}m} g(x_1,x_2)g(x_3,x_4) \right] \\ \nonumber & \hspace{3cm}\times \left\{ 3 \ca{R}''[u_{ab}](x_2,x_3) \left[S_{110}^{(3)}[u_{aac}](x_4,x_1) - S_{110}^{(3)}[u_{abc}](x_4,x_1)\right] \right\} \\ \nonumber
&\;\; + \int _{x_1,...,x_6}\left[ \sfrac{\mathrm{d}}{\mathrm{d}m} g(x_1,x_2)g(x_3,x_4)g(x_5,x_6) \right] \\ \nonumber & \hspace{3cm} \times \big\{ 3 \ca{R}''[u_{ab}](x_2,x_3) \ca{R}''[u_{ac}](x_4,x_5) \ca{R}''[u_{ac}](x_6,x_1) \\ \nonumber & \hspace{3cm}\; ~~~~~ - \ca{R}''[u_{ab}](x_2,x_3) \ca{R}''[u_{bc}](x_4,x_5) \ca{R}''[u_{ac}](x_6,x_1)\big\} \ .
\end{align}
Because there is a feeding term from the fourth $\Gamma$-cumulant $S^{(4)}$ we have to calculate $S^{(4)}$ to order $\epsilon^4$. The only contribution in Eq.~(\ref{eq:dotS4}) is $S^{(4)}_4$ that integrates in this order to
\begin{align}\label{eq:S4threeloop}
 S^{(4)}[u_{abcd}] &= 3 \Big\{ 4\, \mbox{tr}\Big[ g\mathrm{R}''_{ab} g\ca{R}''_{ac} g\ca{R}''_{ad} g\ca{R}''_{ad} \Big]  \
+ 2\,  \mbox{tr}\Big[ g\ca{R}''_{ab} g\ca{R}''_{ac} g\ca{R}''_{cd} g\ca{R}''_{ac} \Big] \\ &\hspace{2cm} \nonumber
-4\,  \mbox{tr}\Big[ g\ca{R}''_{ab} g\ca{R}''_{ac} g\ca{R}''_{cd} g\ca{R}''_{ad} \Big]
+ \mbox{tr}\Big[ g\ca{R}''_{ab} g\ca{R}''_{bc} g\ca{R}''_{cd} g\ca{R}''_{ad} \Big] \Big\} + \ca{O}(\epsilon^5)\ ,
\end{align}
where again $\ca{R}''_{ab}(x,y)=\ca{R}''[u_{ab}](x,y)$.
In order to obtain $S_{1100}^{(4)}[u_{aabc}]$ the equation has to be symmetrized over replica fields and two functional derivatives have to be taken. This lengthy but straightforward calculation is not reproduced here. The limit of identical replica fields in the first and second entry again has to be taken in the weak limit.

For brevity we introduce the symbol $\frac{\tilde{\mathrm{d}}}{\mathrm{d}m_g} $ that formally denotes a scale derivative that acts only on the propagators $g$ that were differentiated in the initial 1PI flow equations, see Eqs.~(\ref{eq:dotR}), (\ref{eq:dotS3}), and (\ref{eq:dotS4}). These formal ``derivatives'' do not act on cumulants nor on $g$'s that arise otherwise. In this sense the 2-loop contribution to $\dot{S}^{(3)}$, see Eq.~(\ref{eq:dotS3twoloop}), can be written as
\begin{align}
 \frac{\tilde{\mathrm{d}}}{\mathrm{d}m_g} \frac 12 \big[A_1+A_2+A_3\big]\ ,
\end{align}
where $A_1$, $A_2$, and $A_3$ are given in Eq.~(\ref{eq:A123}). This term was easily integrated in 2-loop order since a scale derivative acting on $R$ gives an additional order of $\epsilon$, that is, $\frac{\tilde{\mathrm{d}}}{\mathrm{d}m_g} $ could be replaced by $\frac{\mathrm{d}}{\mathrm{d}m}$. Here we also need the next order, so we have to calculate
\begin{align}
   \left(\frac{\mathrm{d}}{\mathrm{d}m} -\frac{\tilde{\mathrm{d}}}{\mathrm{d}m_g}\right) \frac 12 \big[A_1+A_2+A_3\big]\ ,
\end{align}
and reinsert the 1-loop result for $\dot{R}[u]$  from Eq.~(\ref{eq:oneloopbeta}) to obtain this expression to order $\epsilon^4$. It  is sufficient to insert the local part of $R$ into $A_1$, $A_2$, and $A_3$.

Apart from the feeding from $S_{1100}^{(4)}[u_{aabc}]$, there are two more 3-loop contributions to the flow of $S^{(3)}$. One arises by inserting also non-local contributions to 1-loop order from Eq.~(\ref{eq:OneLoopNonLocal}) into $A_1$, $A_2$, and $A_3$. And, finally, there is a cross term $R\times S^{(3)}$ from the third line of Eq.~(\ref{eq:dotS3T0}). Here we can insert the 2-loop solution $S^{(3)}=\frac 12 (A_1+A_2+A_3)$ with local $R$'s into the right-hand-side of the flow equation to obtain the complete result at 3-loop order. These 3-loop contributions are easily integrated since scale derivatives acting on cumulants would introduce additional loops. The details of this calculation and the resulting functional $S^{(3)}[u_{abc}]$ to 3-loop order are given in Appendix~\ref{app:S33loop}. In order to obtain $S_{110}^{(3)}[0,0,-u]$ it is again convenient to use a symmetric replica expansion. Setting $u_a=u_b$ again requires the weak limit; in addition to potentially problematic terms $\sim R'''(0^+)^2$ we also encounter $R'''(0^+)R^{(5)}(0^+)$.

Now we turn to the term $\ca{R}''[u](x_2,x_3) \ca{R}''[u](x_4,x_1)$  in Eq.~(\ref{eq:dotR}). In 3-loop order we have to insert
\begin{align}
 \ca{R}''[u](x,y)= \ca{R}''(u(x)) \delta(x-y) + \hat{R}''[u](x,y) -\hat{R}''[0^+](x,y)\ ,
\end{align}
where $\hat{R}$ is the non-local contribution of the 2-loop solution Eq.~(\ref{eq:NonLocalTwoLoop}). Taking two functional derivatives and taking the weak limit with non-crossing configurations produces anomalous terms $R'''(0^+)^2$, $R'''(0^+)R^{(5)}(0^+)$, $R''(0^+)R''''(0^+)$, and $R''(0^+)R^{(6)}(0^+)$.
Inserting the obtained expressions for \newline $\ca{R}''[u](x_2,x_3) \ca{R}''[u](x_4,x_1)$ and $S_{110}^{(3)}[0,0,-u]$ in Eq.~(\ref{eq:dotR}) allows us to rearrange terms such that they are total derivatives acting on the propagators $g$ only. In summary we obtain to 3-loop order
\newcommand{\splitt}{\end{align} \begin{align}}
\begin{align}
 \dot{R}[u] =\beta_{\mtin{1loop}}[u] + \beta_{\mtin{2loop}}[u] + \beta{\mtin{3loop}}[u]
\end{align}
with 1- and 2-loop contributions $\beta_{\mtin{1loop}}[u] + \beta_{\mtin{2loop}}[u]$ given by Eq.~(\ref{eq:TwoLoopBeta}) and the 3-loop contribution
\begin{align}\nn
\beta_{\mtin{3loop}}[u] &= \int _{x_1,x_2,y,z} \left[ \sfrac{\mathrm{d}}{\mathrm{d}m} g_{x_1x_2}^2 g_{x_1z}g_{x_2z} g_{yz}^2\right] \Big[R'''_{x_1}R'''_{x_2} -R'''(0^+)^2 \Big] \ca{R}''_y R''''_z \\  \nonumber
& -2 I_1 \int_{x_1,x_2,y} \left[ \sfrac{\mathrm{d}}{\mathrm{d}m} g_{x_1x_2}^2 g_{x_1y}g_{x_2y}\right] \ca{R}''_y  \Big( 3\Big[ R'''_{x_1} R'''_{x_2} R''''_{x_2} -R'''(0^+)^2R''''(0^+) \Big] +R'''_{x_1}\ca{R}''_{x_2} R^{(5)}_{x_2} \Big)\\ \nonumber
& - I_1 \int_{x,y,z} \left[ \sfrac{\mathrm{d}}{\mathrm{d}m} g_{xz}^2 g_{yz}^2\right] \Big[ {R'''_x}^2 -R'''(0^+)^2 +\ca{R}''_xR''''_x\Big] \ca{R}''_x R''''_z \\ \nonumber
& +\sfrac 12 \int _{x_1,x_2,y_1,y_2} \left[ \sfrac{\mathrm{d}}{\mathrm{d}m} g_{x_1x_2}^2g_{x_1y_1}^2 g_{x_2y_2}^2\right] R''''_{x_1}R''''_{x_2} \ca{R}''(y_1)\ca{R}''(y_2) \\ \nonumber
& + \sfrac 12  I_1^2 \int _{x_1,x_2} \left[ \sfrac{\mathrm{d}}{\mathrm{d}m} g_{x_1x_2}^2\right] \Big[{R'''_{x_1}}^2-R'''(0^+)^2   +\ca{R}''_{x_1}R''''_{x_1}\Big] \Big[{R'''_{x_2}}^2-R'''(0^+)^2 +\ca{R}''_{x_2}R''''_{x_2}\Big] \\ \nonumber
& -I_1 \int  _{x_1,x_2,y} \left[ \sfrac{\mathrm{d}}{\mathrm{d}m} g_{x_1x_2}^2 g_{x_1y} g_{x_2y} \right] \Big[ R'''_{x_1}R'''_{x_2} -R'''(0^+)^2 \Big]     \Big[ {R'''_y}^2 -R'''(0^+)^2 + \ca{R}''_x R''''_x\Big] \\ \nonumber
& +\sfrac 12 \int  _{x_1,x_2,y_1,y_2} \left[ \sfrac{\mathrm{d}}{\mathrm{d}m} g_{x_1x_2}^2 g_{y_1y_2}^2 g_{x_1y_1} g_{x_2y_2} \right]  \Big[ R'''_{x_1}R'''_{x_2} -R'''(0^+)^2 \Big]\Big[ R'''_{y_1}R'''_{y_2} -R'''(0^+)^2 \Big] \\ \nonumber
& +4 \int  _{x_1,x_2,y,z} \left[ \sfrac{\mathrm{d}}{\mathrm{d}m} g_{x_1x_2}g_{yz}g_{x_1z}g_{x_2z}^2 g(x_1y) \right]  \Big[ R'''_{x_1}R'''_{x_2}R''''_z -R'''(0^+)^2R''''(0^+) \Big] \ca{R}''_x\\ \nonumber
& +\int  _{x_1,x_2,y,z} \left[ \sfrac{\mathrm{d}}{\mathrm{d}m} g_{x_1z}^2g_{x_2z}^2g_{x_1y}g_{x_2y} \right]   \Big[ R'''_{x_1}R'''_{x_2} R''''_z -R'''(0^+)^2R''''(0^+) \Big]\\ \nonumber
&+2\int  _{x,y_1,y_2,z} \left[ \sfrac{\mathrm{d}}{\mathrm{d}m} g_{xz}^2g_{y_1z}^2g_{y_2z}g_{y_2x} \right] R'''_x\ca{R}''_{y_1} \ca{R}''_{y_2} R^{(5)}_z \\ \nonumber
&-\sfrac 12  I_1 \int  _{y_1,y_2,z} \left[ \sfrac{\mathrm{d}}{\mathrm{d}m} g_{y_1z}^2g_{y_2z}^2 \right] \ca{R}''_{y_1}\ca{R}''_{y_2} \Big[ 3{R''''_z}^2 + 4 R'''_zR^{(5)}_z +\ca{R}''_z R^{(6)}_z\Big] \\ \nonumber
&+ \sfrac 16 \int  _{y_1,y_2,y_3,z} \left[ \sfrac{\mathrm{d}}{\mathrm{d}m} g_{y_1z}^2g_{y_2z}^2 g_{y_3z}^2\right] \ca{R}''_{y_1}\ca{R}''_{y_2}\ca{R}''_{y_3} R^{(6)}_z \\ \nonumber
& + \int  _{x_1,x_2,y_1,y_2} \left[ \sfrac{\mathrm{d}}{\mathrm{d}m} g_{x_1x_2}^2 g_{x_1y_1} g_{x_1y_2} g_{x_2y_1}g_{x_2y_2} \right] R''''_{x_1}R''''_{x_2} \ca{R}''_{y_1} \ca{R}''_{y_2}\\ \nonumber
& + \sfrac 12 \int  _{x_1,x_2,x_3,x_4}  \left[ \sfrac{\mathrm{d}}{\mathrm{d}m} g_{x_1x_2} g_{x_3x_4} g_{x_1x_3}g_{x_1x_4} g_{x_2x_3}g_{x_2x_4} \right] \\ \nonumber
   &\hspace{3cm} \times \Big\{ \Big[R'''_{x_1}R'''_{x_2} -R'''(0^+)^2\Big] \Big[ R'''_{x_3}R'''_{x_4} -R'''(0^+)^2\Big] -R'''(0^+)^4 \Big\} \\ \nonumber
& +\int  _{x,y} \left[ \sfrac{\mathrm{d}}{\mathrm{d}m} g_{xy}^2\right] \ca{R}''_x \Big\{ (I_1^2-I_A) R''''_x\Big[ {R'''_{x}}^2-R'''(0^+)^2\Big] \\ \nonumber& \quad  +(5I_1^2-4I_A) \Big[ {R'''_x}^2R''''_x -R'''(0^+)^2R'''(0^+) \Big] +(3I_1^2 -2I_A) \ca{R}''_x {R''''_x}^2 \\  &\quad  +(4I_1^2-2I_A) \ca{R}''_x R'''_x R^{(5)}_x + \sfrac 12 I_1^2 {\ca{R}''_x}^2 R^{(6)}_x \Big\}
\label{eq:3loopbeta}
\end{align}
Here we once again introduced shorthand notations $g_{xy}:=g(x,y)$ and $R_x= R(u(x))$ except for $x=0^+$ and likewise for derivatives of $R$.

Inserting a constant field and dividing by $L^d$ gives the 3-loop contribution to the $\beta$-function
\begin{align}\nonumber
 \beta_{\mtin{3loop}}(u) &=   \big(4\dot{I}_l +\dot{I}_m -6 I_1 \dot{I}_A +(5I_1^2-4I_A)  \dot{I}_1\big)\Big[R'''(u)^2 R''''(u) - R'''(0^+)^2R''''(0^+) \Big] \ca{R}''(u) \\      \label{eq:localbeta3loop}
&\quad +\big[ \dot{I}_j -2I_A  \dot{I}_1\big)\big]  R''''(u)^2 \ca{R}''(u)^2\\ \nonumber
&\quad +\frac 12\left(  I_1^2 \dot{I}_1 -2I_1\dot{I}_A  +\dot{I}_m +\dot{I}_i\right) \Big[ R'''(u)^2 -R'''(0^+)^2\Big]^2  -\frac 12 \dot{I}_i R'''(0^+)^4 \quad +\ca{O}(\epsilon^5)
\end{align}
with the following integrals\begin{align}
 I_1&=\int _p \; g(p)^2  \\ \nonumber
I_A&= \int _{p_1,p_2} \; g(p_1)g(p_2)g(p_1+p_2)^2 \nonumber \\ \nonumber
I_m &= \int_{p_1,p_2,p_3} \; g(p_1)g(p_2)g(p_1+p_2+p_3)g(p_3)g(p_1+p_2)^2 \\ \nonumber
I_l &= \int_{p_1,p_2,p_3}\;  g(p_1)g(p_2)g(p_1+p_2)g(p_3)g(p_1+p_2+p_3)^2 \\ \nonumber
I_j&=   \int_{p_1,p_2,p_3} \; g(p_1)g(p_2)g(p_3)^2g(p_1+p_2+p_3)^2 \\ \nonumber
I_i &=  \int_{p_1,p_2,p_3} \; g(p_1)g(p_2)g(p_3)g(p_1+p_3)g(p_2+p_3)g(p_1-p_2)
\end{align}
which are calculated in  appendix \ref{a:Diagrams}. It turns out that the combinations occurring in Eq.~(\ref{eq:localbeta3loop}) are finite for $\epsilon\to 0$, so our counting of orders in $\epsilon$ is consistent, and the theory is 3-loop renormalizable.
Due to gauge invariance we can add any scale-dependent function to $R(u)$ that does {\em not} depend on the fields. In this way we can drop all constants from the $\beta$-function. The constants in Eq.~(\ref{eq:localbeta3loop}) arise directly from Eq.~(\ref{eq:3loopbeta}). In the derivation of the latter we neglected gauge terms in $S^{(3)}$ and $S^{(4)}$, so these constants are arbitrary.

Assuming non-crossing configurations, that is, using Eq.~(\ref{eq:weakcont}) for taking limits $u_a-u_b\to 0$, allows us to derive all anomalous terms in the $\beta$-function without ambiguities. With this assumption, the $\epsilon$-expansion is a straightforward expansion of the exact hierarchy of flow equations for the $\Gamma$-cumulants in powers of the effective local disorder distribution function $R(u)$. Presumably, this will  work to all orders in $\epsilon$.

Crossing configurations could not be treated and are an open problem. It is doubtful that the standard $\epsilon$-expansion can be applied. This is because $R(u)$ and all its derivatives are not a small parameter suitable for an expansion if $u=0$ cannot be avoided.

In order to make contact with the result obtained by an alternative method later in Section \ref{sec:field-theory}, we rescale
\begin{align}\label{R-rescale}
 R(u)=\frac{1}{\epsilon I_1} m^{-4\zeta} \tilde{R}(um^\zeta)
\end{align}
where $\zeta$ is the roughness exponent. The rescaled function $\tilde{R}$ still depends on the RG scale $m$ and satisfies the RG equation to  3-loop order given in Eq.~(\ref{betafinal}).

\subsection{2-point correlation function}

The  expressions  obtained above allow us to extract the non-local part, relevant to evaluate the full 2-point correlation function \cite{HusemannWiese2017}. The latter is obtained as follows:  The Fourier transform of Eq.~(\ref{eq:TwoPointCorrelation}) reads
\begin{align}\label{eq:TwoPointCorrelationRealSpace}
 \ovl{\langle u(x)u(y)\rangle}_V = T g(x,y) -{ \int_{z,z'} g(x,z)g(y,z')} R''[0^+](z,z').
\end{align}
Because of the limit $n\to 0$ in Eq.~(\ref{eq:TwoPointCorrelation}), where $n$ is the number of replica fields, this is an exact expression; in particular, there are no contributions of three- or higher-replica terms to the 2-point function. As in Sec.~\ref{sec:ERG}, the expression $R''[0^+](z,z')$ denotes the second functional derivative of $R[u]$ with respect to $u(z)$ and $u(z')$ that is evaluated in a weak limit $u(x)\approx \mbox{const.} \to 0$. A precise definition of the weak limit is given in Eq.~(\ref{eq:weakcont}).

For the expansion in the renormalized local disorder function $R(\uu)$ with a constant field $\uu$ we   use that
\begin{align}
R''[0^+](z,z')= R''(0^+)\delta(z-z') + \tilde{R}''[0^+](z,z') ,
\end{align}
where the non-local part $\tilde{R}''[0^+](z,z')$ has to be expanded to 2-loop order, like in the derivation of the 3-loop $\beta$-function. Taking two derivatives of Eq.~(\ref{eq:NonLocalTwoLoop}), evaluated at a constant field $\uu$, and sending $\uu\to 0^+$ or  $\uu\to 0^-$ we find
\begin{align} \nn
\tilde{R}''[0^+](z_1,z_2) &= \delta_{z_1z_2} \Big[-I_1 R'''(0^+)^2 +(5I_1^2-4I_A) R'''(0^+)^2 R''''(0^+) \Big] \\
&+ g(z_1,z_2)^2 \Big[R'''(0^+)^2 -6I_1R'''(0^+)^2 R''''(0^+) \Big] \nonumber\\
&+ 2g(z_1,z_2) R'''(0^+)^2 R''''(0^+) \int_x \; g(x,z_1)g(x,z_2)\big[g(x,z_1)+g(x,z_2)\big] \nonumber \\ &+ R'''(0^+)^2 R''''(0^+) \int _x \;g(x,z_1)^2 g(x,z_2)^2   .
\end{align}

\section{Effective action and  $\beta$-function via field theory}
\label{sec:field-theory}

\subsection{Calculation using the sloop elimination method}
\label{sec:sloops}
Here we discuss  a different way to do the contractions, using ``excluded replicas'',  which will finally lead to a rather efficient algorithm for calculating the anomalous terms.

We start by a 1-loop diagram involving two disorder vertices, after having done one Wick-contraction. For simplicity of notation we are not writing space-indices and momentum integrals, which are unimportant for the following discussion.
\begin{equation}\label{lf47}
\diagram{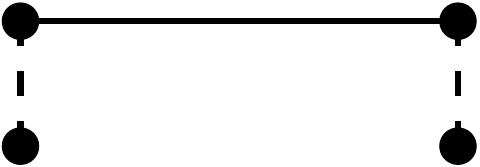} =\frac1{2 T^{3}}\sum_{abc}R'(u_{a}-u_{b}) R'(u_{a}-u_{c}) \ .
\end{equation}
At the next step, the following  contractions are possible (restoring the integral)
\begin{eqnarray} \label{e2}
&&\!\!\!\!\diagram{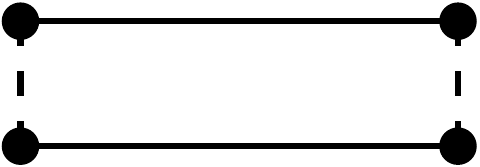} +\diagram{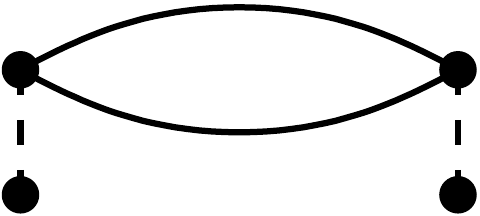}-
\diagram{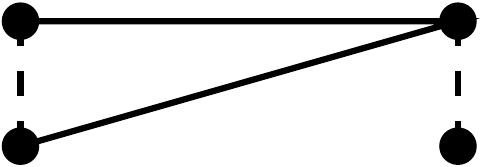}- \diagram{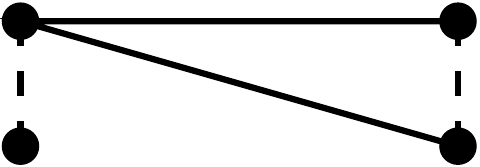} \nn\\
&&= \frac1{2 T^{2}} \left[ \sum_{ab}R''(u_{a}-u_{b})^{2}  +\sum_{abc}R''(u_{a}-u_{b}) R''(u_{a}-u_{c}) -2 \sum_{ab}R''(0) R''(u_{a}-u_{b})  \right] \ I_1.~~~~~ \qquad
\end{eqnarray}
The second term is a 3-replica contribution (contribution to the third cumulant of the disorder), thus not of interest to us. The correction to the disorder at 1-loop order therefore consists   of the first and last term, equivalent to the first and last two diagrams,   \begin{equation}\label{6.3}
\delta^{(1)} R(u) = \left[ \frac{1}2 R''(u)^2-R''(u)R''(0) \right] I_1\ .
\end{equation}
This is equivalent to the result obtained in Eq.~(\ref{R-1loop-ERG}).

An alternative approach consists in remarking that in Eq.\ (\ref{lf47}) the terms $a=b$, and $a=c$ could be dropped, since they are constants, thus will not be contracted in the next step. We thus start from
\begin{equation}\label{lf47-bis}
\diagram{1line} =\frac1{2 T^{3}}\sum_{b\neq a\neq c}R'(u_{a}-u_{b}) R'(u_{a}-u_{c})
\ ,
\end{equation}
which after one Wick-contraction leads to
\begin{eqnarray} \label{e2b}
&&\diagram{repconsloop} +\diagram{sloop} \nn\\
&&= \frac1{2 T^{2}} \left[ \sum_{a\neq b}R''(u_{a}-u_{b})^{2}  +\sum_{b\neq a\neq c}R''(u_{a}-u_{b})
R''(u_{a}-u_{c})   \right] I_1 \nn\\
&&= \frac1{2 T^{2}} \left[ \sum_{a b}R''(u_{a}-u_{b})^{2}(1-\delta_{ab})  +\sum_{abc}R''(u_{a}-u_{b})
R''(u_{a}-u_{c}) (1-\delta_{ab})(1-\delta_{ac})  \right] I_1 \qquad \nn\\
&& = \frac1{2 T^{2}} \left[ \sum_{a b}\left[R''(u_{a}-u_{b})^{2}
 -2R''(u_a-u_b)R''(0)\right]+\sum_{abc}R''(u_{a}-u_{b})
R''(u_{a}-u_{c})  \right] I_1\ .
\end{eqnarray}
The 2-replica term (the double sum) is, as expected, the same result as obtained in Eq.~(\ref{e2}). While the second line contains only excluded replica sums, there can not be any ambiguity. The latter may only appear in the ensuing projection onto non-excluded replica sums. This is indeed the case for the hat diagram \(\sim R''(u)R'''(0^+)^2\), as the reader is invited to check on his own, starting from \begin{eqnarray}
\lefteqn{ \sum_{a,b} \delta^{(2)}_{A} R(u_a-u_b)}\nn\\  &=& \Big[ \sum_{a \neq b} R''(u_a-u_b) (R'''(u_a-u_b))^2
+ \sum_{a \neq b, a \neq c} R''(u_a-u_b) R'''(u_a-u_b) R'''(u_a-u_c) \nonumber
\\
&& - \frac{1}{2} \sum_{a \neq b, a \neq c , b \neq c} R''(u_a-u_b)
R'''(u_a-u_c) R'''(u_b-u_c)
+ \frac{3}{2} \sum_{a \neq b, a \neq c} R''(u_a-u_b) R'''(u_a-u_c)^2  \nonumber
\\
&& + \frac{1}{2} \sum_{a \neq b, a \neq c , a \neq d} R''(u_a-u_b)
R'''(u_a-u_c) R'''(u_a-u_d)\Big] I_A\ .
\label{deltaA}
\end{eqnarray}
We will therefore in the following present an improved projection method,
which we have termed the ``sloop-elimination'' method. (The name may be thought
of as as ``super''-partner of a normal loop, thus sloop, which cancels part
of it.)

The idea of the method is very simple. Let us consider the second term on the  the second line of Eq.~(\ref{e2}). It is a three-replica term proportional to the temperature. In a
$T=0$ theory such a diagram should not appear, thus it can identically be
 set to zero:
\begin{equation}\label{d1}\sum_{abc}
\diagram{sloop} =\frac{1}{2T^2} \sum_{abc} R''(u_a-u_b) R''(u_a-u_c)I_1  \equiv 0\ .
\end{equation}
Projecting such
terms to zero at any stage of further contractions is very natural
in our present calculation (and also e.g.\ in the exact RG
approach, where terms are constructed recursively and such
forbidden terms must be projected out). It is valid only when (i)
the summations over replicas are free (ii) the term inside the sum
is non-ambiguous. These conditions are met for any diagram with
sloops, provided the vertices have at most two derivatives. (One
can in fact start from vertices which either have no derivative
or exactly two.)
Subtracting this term from Eq.~(\ref{e2b}) immediately yields the result (\ref{6.3}).

While this could also have been done directly,
let us illustrate the power of the  procedure on an example. We want to contract the expression (\ref{d1}) with a third vertex $R$
\begin{equation}\label{id2}
 0=\sum_{abc}
\diagram{sloop}  \sum_{de} R(u_d-u_e) \equiv \frac{1}{T^4} \sum_{a \neq b ,
a \neq c ,d\neq e} R''(u_a-u_b) R''(u_a-u_c) R(u_d-u_e)   \ .
\end{equation}
where we have already dropped constant terms which will disappear after the contractions.  Also note that implicitly here and in the following the vertices are at
points $x,y,z$ in that order.
 We will contract the third vertex
twice, once with the first and once with the second , i.e.\ look
at the term proportional to $I_A=\int_{x,y,z}g(x-y)^2 g(x-z) g(y-z)$.

Performing the first contraction between points $x$ and $z$ yields
\begin{equation}\label{id3}
\frac{1}{T^3} \Big[ \sum_{a \neq b , a \neq c , a \neq e} \!\!\!\!\!R'''(u_a-u_b)
R''(u_a-u_c)  R'(u_a-u_e)- \sum_{a \neq b , a \neq c , b \neq e}\!\!\!\!\!\!\! R'''(u_a-u_b)  R''(u_a-u_c)  R'(u_b-u_e) \Big]
\equiv 0  \ .
\end{equation}
Similarly, the second contraction  yields (with the standard combinatorial factor of $1/2$)
\begin{eqnarray}
&& \!\!\!\frac{1}{T^2} \Big[ \frac{1}{2} \sum_{a \neq b , a \neq c , a \neq e}
R'''(u_a-u_b)  R'''(u_a-u_c)  R''(u_a-u_e)
\nn\\&&\quad +  \sum_{a \neq b , a \neq c } R'''(u_a-u_b)  R'''(u_a-u_c)  R''(u_a-u_c)  \nonumber
\\
&& \quad+  \frac{1}{2} \sum_{a \neq b ,  b \neq e} R'''(u_a-u_b)  R'''(u_a-u_b)  R''(u_a-u_e)
\nn\\ &&\quad- \frac{1}{2} \sum_{a \neq b ,  a \neq c , b \neq c} R'''(u_a-u_b)
R'''(u_a-u_c)  R''(u_a-u_c) \Big]  I_A\equiv 0   \ . \label{id4}
\end{eqnarray}
This non-trivial identity tells us that the sum of all the terms (or
diagrams) thus generated upon contractions  must vanish. Stated
differently: A sloop, as (\ref{d1}) as well as the sum of all its descendents
vanishes. Note that this is {\em not} true for each single term,
but only for the sum.

A property that we request from a proper $p$-replica term is that upon
one self contraction it gives a $( p-1)$-replica term. It may also
give $T$ times a $p$-replica term (a sloop) but this is zero at $T=0$,
so we can continue to contract. Thus we have generated several
non-trivial projection identities. The starting one is that the
2-replica part of (\ref{d1}) is zero, since (\ref{d1}) is a
proper 3-replica term. This is what is meant by the
 symbol ``$\equiv$'' above and the last identity is the one we now use.

Indeed  compare (\ref{id4}) with (\ref{deltaA}). One
notices that all terms apart from the first in (\ref{deltaA}) appear
in (\ref{id4}). They also have  the same relative coefficients, apart from
the third one of (\ref{deltaA}). Thus one can use (\ref{id4}) to
simplify (\ref{deltaA}):
\begin{equation}\label{lf79}
\sum _{a,b}\delta_{A}^{(2)} R(u_a-u_b) = \Big[ \sum_{a \neq b} R''(u_a-u_b) R'''(u_a-u_b)^2
+  \sum_{a \neq b, a \neq c} R''(u_a-u_b) R'''(u_a-u_c)^2 \Big]
I_A  \ .
\end{equation}
The function $R'''(u)^2$, which appears in the last term, is
continuous at $u=0$. It is thus obvious how to rewrite this
expression using free summations and extract the 2-replica part
\begin{eqnarray}\label{delta2AR}
 \delta^{(2)}_{A} R(u) &=&  \Big [ \big(R''(u) - R''(0)\big) R'''(u)^2- R'''(0^+)^2 R''(u) \Big] I_A\ .
\end{eqnarray}
This coincides with the contribution of diagram A in the ERG approach, see the second term of Eq.~(\ref{5.35}).
We can write diagrammatically the subtraction that has been performed as
\begin{equation}\label{lf168}
\delta_{A}^{(2)} R = \diagram{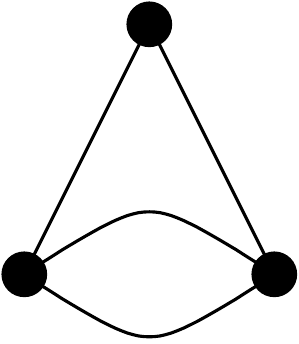}-\diagram{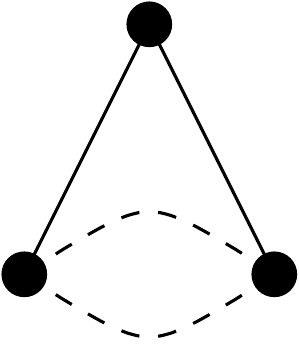} \ ,
\end{equation}
where the loop with the dashed line represents the sub-diagram with the
sloop, i.e.\ the term (\ref{id4}) (with in fact the same global
coefficient). The idea is
that subtracting sloops is allowed since they  vanish.
The advantage of the method is that all intermediate results are uniquely defined.

There are other possible identities, which are descendants of
other sloops. For instance a triangular sloop gives, by a similar
calculation:
\begin{eqnarray}\label{lf169}
\diagram{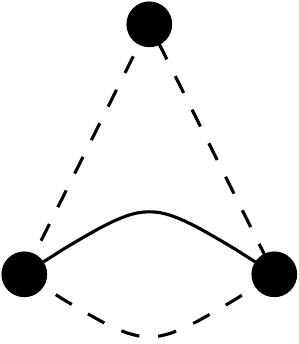} &=& R''(0) \sum_{a \neq b} R'''(u_a-u_b)^2 + \sum_{a \neq b, a \neq c}
R''(0) R'''(u_a-u_b) R'''(u_a-u_c) \nonumber \\
&& + \sum_{a \neq b, b \neq c} R''(u_b-u_c)R'''(u_a-u_b)^2
+ \sum_{a \neq c, b \neq c, c \neq d } R'''(u_a-u_c) R'''(u_b-u_c) R''(u_c-u_d)
\nonumber  \ . \\
&&
\end{eqnarray}
This however does not prove useful to simplify $\delta^{(2)}_{A} R$.

Remains to calculate the 3-loop diagrams, shown on Fig~\ref{f:all-3-loop}. This is achieved in
 appendix \ref{app:C}.
Since the above method generates a large number of identities, one
can wonder whether they are all compatible. We have checked that this is indeed so, but
we have not attempted a general proof.

\subsection{The effective action up to 3-loop order}
\begin{figure}
\fig{\textwidth}{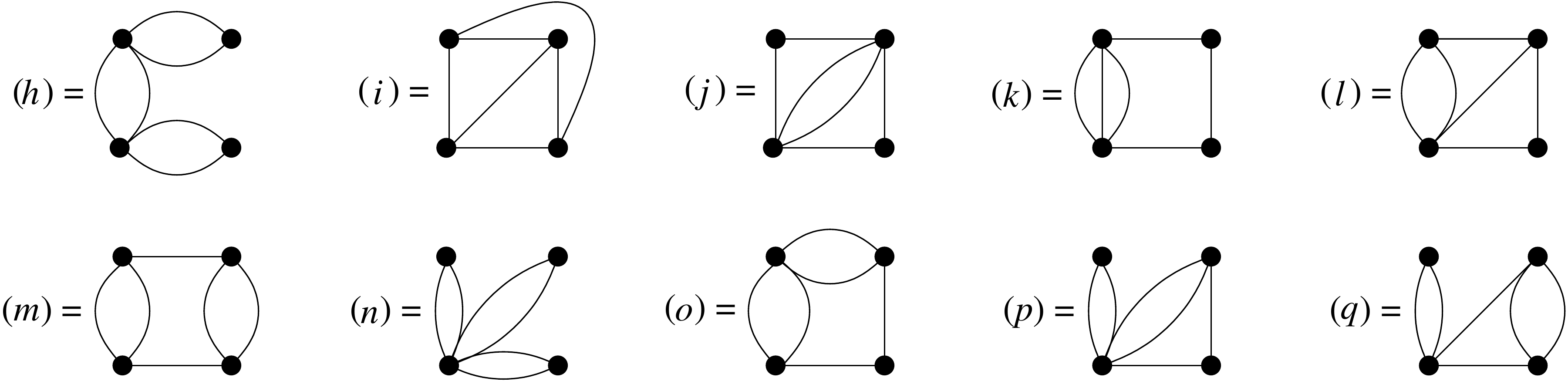}
\caption{Diagrams at 3-loop order (without insertion of lower order counter-terms)}
\label{f:all-3-loop}
\end{figure}

Using the sloop elimination method exposed in the preceding section, we have calculated all diagrams up to 3-loop order. They are presented graphically on figure \ref{f:all-3-loop}, and given below. The expressions intervening in the sloop-projection algorithm are collected in appendix \ref{app:C}. Here we give the final result for the effective action, before discussing how to obtain the $\beta$-function in the next section.

The effective dimensionfull renormalized disorder to 3-loop order reads
\begin{equation}
R_{\mathrm {eff}}(u)= R (u) + \delta^{ (1)}R (u)+ \delta ^{(2)}R (u) + \delta
^{(3)}R (u)  + \ldots
\end{equation}
The 1-loop term is, noting $R''_u:=R''(u)$, \(R''_0:=R''(0) \), \(R'''_0:=R'''(0^+)\)  etc.
\begin{equation}\label{deltaR-1loop}
\delta^{(1)} R (u) =\half  \left[R''_{u}{}^{2}-R''_{u}R''_{0} \right]I_{1}\ .
\end{equation}
The 2-loop term is \begin{eqnarray}\label{deltaR-2loop}
\delta^{(2)} R (u) &=& \left[R''_{u} R'''_{u}{}^{2}-R''_{0}
R'''_{u}{}^{2} - R''_{u}
R'''_{0}{}^{2}  \right] I_{A}+ \half \left[(R''_{u}-R''_{0})^{2}R''''_{u}\right]I_B\ .\ \ \ \
\end{eqnarray}
The 3-loop terms read
\begin{eqnarray}
\label{deltaR-3loop}
\delta^{(3)} R (u)& =& (h)+ (i)+ (j)+ (k)+ (l)+ (m)+ (n)+ (o)+ (p)+ (q)\\
 (h) &=& \frac{1}{2} (R''_{u}-R''_{0})^2 {R''''_{u}}^2 I_{h} \\
 (i) &=& \frac{1}{2} \left( {R'''_{u}}^4-2 {R'''_{u}}^{2} {R'''_{0}}^{2}
\right) I_i \\
(j) &=& (R''_{u}-R''_{0})^2 {R''''_{u}}^2   I_j  \\
(k) &=& 0 \\
 (l) &=& 4  \Big(R''_{u} {
          R'''_{u}}^{2}R''''_{u}-R''_{0} { R'''_{u}}^{2}R''''_{u}   -R''_{u} { R'''_{0}}^{2} R''''_{0}  \Big)   I_l \\
 (m) &=& \frac{1}{2}\left({R'''_{u}}^4-2 {R'''_{u}}^{2}{R'''_{0}}^{2}
\right) I_m \\
 (n) &=& \frac{1}{6} \left(R''_{u}-R''_{0}\right)^3 R^{(6)}_{u} I_n \\
 (o) &=& \Big( R''_{u}  R''''_{u} {R'''_{u}}^2- R''_{0} R''''_{u}
{R'''_{u}}^2-R''_{u}{R'''_{0}}^{2} R''''_{0}\Big) I_o \\
 (p) &=& 2(R''_{u}-R''_{0})^2 R'''_{u} R^{(5)}_{u} I_p \\
 (q) &=& (R''_{u}-R''_{0}) R''''_{u} \left( {R'''_{u}}^2- {R'''_{0}}^2
\right) I_{q}
\ .
\end{eqnarray}

\subsection{Derivation of the RG-equation to 3-loop order}
Let us now discuss in general the strategy to renormalize theories, whose
interaction  is not  a single coupling-constant, but a whole function,
here the disorder-correlator $R(u)$.
We denote by $R_0$ the bare disorder --~this is the object in
which perturbation theory is carried out~-- and by  $R$ the
renormalized disorder, i.e.\ the corresponding term in the effective
action  $\Gamma$.

We define the dimensionless bilinear 1-loop, trilinear 2-loop and
quadrilinear 3-loop functions
\begin{eqnarray}
 \delta ^{(1)}(R,R) &:=&  \delta^{(1)} R \\
 \delta ^{(2)}(R,R,R) &:=&  \delta^{(2)} R\\
 \delta ^{(3)}(R,R,R,R) &:=&  \delta^{(3)} R
\end{eqnarray}
where if all arguments are the same, we only give this one argument, e.g.\ $\delta ^{(1)}(R)=\delta ^{(1)}(R,R)$,
 $\delta ^{(2)}(R)=\delta ^{(2)}(R,R,R)$ and
 $\delta ^{(3)}(R)=\delta ^{(3)}(R,R,R,R)$.
For different arguments we use the multilinear formulas
\begin{eqnarray}\label{lf2}
f(x,y) &:=& \frac12 \Big[f(x+y)-f(x)-f(y)\Big] \ ,\\
g (x,y,z) &:=& \frac{1}{6}\Big[ g (x+y+z) - g (x+y)- g (y+z) - g (x+z)+ g (x)+ g (y)+ g (z) \Big] \ ,  \\
h (w,x,y,z) &:=&\frac{1}{24}\Big[h(w+x+y+z)-h(w+x+y)-h(w+x+z)-h(w+y+z)\nn \\
&&\qquad -h(x+y+z) +h(w+x) +h(w+y)+h(w+z)+h(x+y)+h(x+z)\nn \\
&& \qquad +h(y+z)-h(w)-h(x)-h(y)-h(z)\Big]\ .
\end{eqnarray}
Schematically, the renormalized disorder is
\begin{eqnarray}
R &=& {R}_0
+ \delta ^{(1)}R ({R}_0)
+ \delta ^{(2)}R ({R}_0) + \delta ^{(3)}R
({R}_0)
+ { O} (R_{0}^{5})\label{rr0}
\ ,
\end{eqnarray}
calculated in the preceding section (where we had not explicitly written an index 0 to indicate the bare disorder).
The inversion of relation (\ref{rr0})  is \begin{eqnarray}
{R}_0 &=& R-\delta ^{(1)} (R) -\delta ^{(2)} (R)+2\delta ^{(1)} (R,\delta ^{(1)} (R)) \nonumber \\
&&-\delta ^{(3)} (R)+3\delta ^{(2)} (R,R,\delta ^{(1)} (R))
+2\delta ^{(1)} (R,\delta ^{(2)} (R))\nonumber \\
&& -\delta ^{(1)} (\delta ^{(1)} (R))-4 \delta
^{(1)} (R,\delta ^{(1)} (R,\delta ^{(1)} (R)))+{ O} (R^{5})\label{r0r}
\ .
\end{eqnarray}
Since an $n$-loop integral scales like \(m^{-n \epsilon}\)  the $\beta$-function  is directly read off from (\ref{rr0}),
\begin{equation}
-m\partial_{m}R\ts_{{R}_0} = \epsilon \Big[  \delta ^{(1)} ({R}_0)+ 2\delta ^{(2)} ({R}_0) + 3\delta ^{(3)}
({R}_0)\Big]+ { O} ({R}_0^{5})\label{betar0}
\ .
\end{equation}
However, we need the $\beta $-function in terms of $R$, for which we
replace $R_{0}$ by $R$, using Eq.~(\ref{r0r}),\begin{eqnarray}
-m\partial_{m}R\ts_{{R}_{0}} &=&
\epsilon \Big[ \delta ^{(1)} (R)  +2\delta ^{(2)} (R)-2\delta ^{(1)} (R,\delta ^{(1)} (R)) \nonumber \\
&&\quad +3\delta ^{(3)} (R)-6\delta ^{(2)} (R,R,\delta ^{(1)} (R))
-2\delta ^{(1)} (R,\delta ^{(2)} (R))\nonumber \\
&&\quad   +\delta ^{(1)} (\delta ^{(1)} (R),\delta ^{(1)} (R))+4 \delta
^{(1)} (R,\delta ^{(1)} (R,\delta ^{(1)} (R)))\Big] +{ O} (R^{5})
\ .
\end{eqnarray}
Using the results from Eqs.~(\ref{deltaR-1loop}), (\ref{deltaR-2loop}) and (\ref{deltaR-3loop}), this is, printing one diagram and its
counter-terms (as dictated by the renormalization group \(\bf R\)-operation) per line:
\begin{eqnarray}
-m\partial_{m}R_{u} &=& \left( \textstyle \frac{1}{2}
{R''_{u}}^{2}-R''_{u}R''_{0}\right) (\epsilon I_{1})\nn \\
 && + \left(R''_{u} {R'''_{u}}^{2}-R''_{0} {R'''_{u}}^{2} - R''_{u} {
R'''_{0}}^{2}  \right) \epsilon \left (2 I_{A}- I_{1}^{2}
\right)\nn \\
&& + \left((R''_{u}-R''_{0})^{2} R''''_{u}\right)\epsilon
 \left(I_{B}-I_{1}^{2}  \right)
 \nn \\
&& + (R''_{u}-R''_{0})^2 ( R''''_{u})^2 \epsilon
\left(\textstyle \frac{3}{2} I_{h}-6I_{1}I_{B}+\frac{9}{2} I_{1}^{3}
\right)\nn  \\
&& +{\textstyle \frac{3}{2}} \Big( {R'''_{u}}^4-2 {R'''_{u}}^{2} {R'''_{0}}^{2}
\Big) (\epsilon  I_i )\nn \\
&& + (R''_{u}-R''_{0})^2 {R''''_{u}}^2 \epsilon \left(3 I_j-2I_A I_{1}
\right)\rule{0mm}{3ex} \nn \\
&&+ \Big(  R''_{u} { R'''_{u}}^{2}R''''_{u}-R''_{0} {R'''_{u}}^{2}R''''_{u}
-R''_{u} { R'''_{0}}^{2} R''''_{0}  \Big)   \epsilon \left(12 I_l-12I_{1}I_{A}+4I_{1}^{3} \right)\nn \\
&&+\left({R'''_{u}}^4-2 {R'''_{u}}^{2}{R'''_{0}}^{2}
\right) \epsilon
\left(\textstyle\frac{3}{2}I_{m}+\frac{1}{2}I_{1}^{3}-2 I_{1}I_{A}
\right)\nn \\
&&+  \left(R''_{u}-R''_{0}\right)^3 R^{(6)}_{u} \epsilon \textstyle
\half \left(I_{n}-3I_{1}I_{B}+2I_{1}^{3} \right)\nn \\
&&+ \Big( R''_{u}  R''''_{u} {R'''_{u}}^2- R''_{0} R''''_{u}
{R'''_{u}}^2 -R''_{u}{R'''_{0}}^{2} R''''_{0}\Big)  \epsilon (3I_{o}-4I_{1}I_{A}+I_{1}I_{B} )\nn \\
&&+  (R''_{u}-R''_{0})^2 R'''_{u} R^{(5)}_{u} \epsilon 6 ( I_{p}-I_{1}I_{A}-I_{1}I_{B}+I_{1}^{3})\nn \\
&&+  (R''_{u}-R''_{0}) R''''_{u} \left({R'''_{u}}^2 -{R'''_{0}}^{2}
\right)  \epsilon (3 I_{q}-3 I_{1}I_{A}-2 I_{1}I_{B}
+2I_{1}^{3})   \label{fullbeta}\ .
\end{eqnarray}
On this form, one can explicitly check renormalizability. Since we
kept the amplitudes of sub-divergences, as for instance that of the
2-loop bubble-chain diagram, one can exactly see, where these terms
come from. Actually the form given above is unique, even though several
 diagrams have the same functional dependence on $R$.

Let us now proceed to simplify the above equation. In order to do so,
we have to choose a renormalization-scheme. We calculate the 3 leading
terms in the $\epsilon $-expansion of each diagram, i.e.\ up to order
$1/\epsilon $ for the 3-loop diagrams, up to order $\epsilon ^{0}$ for
the 2-loop diagrams and up to order $\epsilon $ for the 1-loop
diagram. In order to have the final result as simple as possible, we
 absorb a factor of $\epsilon I_1$ into  $R$. This means that an $n$-loop integral has
to be normalized by $(\epsilon I_{1})^{n}$. It is with this
normalization  that
the amplitudes are given in appendix \ref{a:Diagrams}.
The advantage of this procedure is that integrals take the most simple
form, and there are no spurious terms like $\psi (1)$ or $\zeta (2)$.
By this way, the 1-loop diagram is automatically subtracted completely
and one never has to worry about its finite parts. However, we have a
choice of how to subtract diagrams at  2-loop order. The most
common choice is to subtract the divergent part only. The advantage of
this procedure is that the 2-loop $\beta $-function takes the simplest
form, with the combination of $\epsilon ( 2 I_{A}-I_{1}^{2})$ in the
second line of  (\ref{fullbeta}) replaced by $\frac{1}{2}$. The
disadvantage is that then diagrams like (q) do not vanish, but have an
amplitude proportional to (see last line of (\ref{fullbeta}))
$ I_{q} - I_{1} I_{A} $ (since $I_{B}= I_{1}^{2}$, and in our
normalizations this is exact in any subtraction scheme).
Now if at second order, we only subtract the diverging part of $I_{A}$
this combination becomes
\begin{eqnarray}\label{lf3}
\lefteqn{I_{q} - I_{1} \times \mbox{diverging part of } I_{A}}\nn \\
&& \qquad = I_{1} \times
\mbox{fintite part of } I_{A} = O\left(\frac{1}{\epsilon } \right)\ .\qquad
\end{eqnarray}
We therefore chose to always subtract the diagram exactly. At order 3
at which we are working here, this means that we have to keep the
finite part of $I_{A}$. This is sufficient, since the 1-loop integral
is normalized to have no finite part, and since from the 3-loop
integrals one only needs the diverging part anyway. Let us now use that
\begin{eqnarray}\label{lf4}
I_{B}&=& I_{I}^{2} \nn \\
I_{h}&=&I_{n}= I_{1}^{3} \nn \\
I_{p}&=& I_{q} = I_{1}I_{A}
\end{eqnarray}
to restate the $\beta $-function:
\begin{eqnarray}
-m\partial_{m}R_{u} &=& \left( \textstyle \frac{1}{2}
{R''_{u}}^{2}-R''_{u}R''_{0}\right) (\epsilon I_{1})\nn \\
 && + \left(R''_{u} {R'''_{u}}^{2}-R''_{0} {R'''_{u}}^{2} - R''_{u} {
R'''_{0}}^{2}  \right) \epsilon \left (2 I_{A}- I_{1}^{2}
\right)\nn \\
&& +{\textstyle \frac{3}{2}} \Big( {R'''_{u}}^4-2 {R'''_{u}}^{2} {R'''_{0}}^{2}
\Big) (\epsilon  I_i )\nn \\
&& + (R''_{u}-R''_{0})^2 {R''''_{u}}^2 \epsilon \left(3 I_j-2I_A I_{1}
\right)\rule{0mm}{3ex} \nn \\
&&+ \Big(  R''_{u} { R'''_{u}}^{2}R''''_{u}-R''_{0} {R'''_{u}}^{2}R''''_{u}
-R''_{u} { R'''_{0}}^{2} R''''_{0}  \Big)  \epsilon \left(12 I_l-12I_{1}I_{A}+4I_{1}^{3} \right)\nn \\
&&+\left({R'''_{u}}^4-2 {R'''_{u}}^{2}{R'''_{0}}^{2}
\right) \epsilon
\left(\textstyle\frac{3}{2}I_{m}+\frac{1}{2}I_{1}^{3}-2 I_{1}I_{A}
\right)\nn \\
&&+ \Big( R''_{u}  R''''_{u} {R'''_{u}}^2- R''_{0} R''''_{u}
{R'''_{u}}^2 -R''_{u}{R'''_{0}}^{2} R''''_{0}\Big)  \epsilon (3I_{o}-4I_{1}I_{A}+I_{1}^{3} ) \ .
\label{betasimp1}
\end{eqnarray}
Finally, we go to the dimensionless renormalized disorder $\tilde R$, defined in Eq.~(\ref{R-rescale}) by
\begin{equation}\label{6.43}
R(u)=:\frac{m^{-4\zeta}}{\epsilon I_1}  \tilde{R}(um^\zeta)\equiv \frac{m^{\epsilon-4\zeta}}{\epsilon \tilde I_1}  \tilde{R}(um^\zeta)
\end{equation}
and group together alike
terms. This yields our final expression
for the 3-loop \(\beta\)-function given in Eq.\ (\ref{betafinal}).
The coefficients ${\cal C}_1$ to ${\cal C}_4$, already given in Eqs.~(\ref{C1})--(\ref{C4}) are constructed from  the  diagrams via
\begin{eqnarray}
{\cal C}_{1} &=& \frac{2I_{A}}{(\epsilon I_{1})^{2}}
-\frac{1}{\epsilon^{2}} -\frac{1}{2\epsilon } = \frac{9+4\pi ^{2}-6
\psi' (\frac{1}{3})}{36} \nn \\
&=& -0.3359768096723647 \\
{\cal C}_{2}&=& \epsilon  \left(\frac{3}{2}
I_{1}+\frac{3}{2}I_{m}+\frac{1}{2}I_{1}^{3}-2 I_{1}I_{A} \right)
(\epsilon I_{1})^{-3}  \nn \\
&=& \frac{3}{4}\zeta (3)+\frac{\pi ^{2}}{18}-\frac{\psi'
(\frac{1}{3})}{12} = 0.6085542725335131\\
{\cal C}_{3}&=& \epsilon (3 I_{j}-2 I_{1}I_{A}) (\epsilon I_{1})^{-3}
= \frac{\psi'
(\frac{1}{3})}{6} -\frac{\pi ^{2}}{9} \nn \\
&=& 0.5859768096723648\\
{\cal C}_{4} &=& \epsilon \left(12 I_l-16I_{1}I_{A}+5I_{1}^{3}+
3I_{o} \right) (\epsilon I_{1})^{-3} \nn \\
&=& 2+\frac{\pi ^{2}}{9}-\frac{\psi' (\frac{1}{3})}{6}= 1.4140231903276352\ .
\end{eqnarray}
These constants
are  closely related to each other analytically.

\section{Reparametrization invariance}\label{newrepara} It is
known in standard field theory, that one can perform a change of
variables, and thus formally change the $\beta$-function, while all
observables remain unchanged. In the context of a functional RG, this
reparametrization invariance is much larger. The function $R (u)$ can
be changed into an arbitrary functional of $f[R]$. The most useful
such reparametrizations involve functionals $f[R]$, which have the
same structure as corrections to $R$, obtained
perturbatively. Especially, when the field $u$ has dimension $\zeta$,
and $R$ times the 1-loop integral has dimension $-4\zeta$, this means
that on dimensional grounds for each additional power of $R$
in $f[R]$, there should be 4 derivatives. Also we do not want $R (u)$
to have different analyticity properties, i.e.\ if $R (u)$ has a
r.h.s.\ Taylor-expansion with a missing linear term (absence of a
super-cusp) then $f[R]$ should have the same properties. The most
suggestive such functional is the 1-loop contribution itself, which we
study now.

The 2-loop RG-equation for the renamed disorder correlator $\tilde
R_{u}$ reads
\begin{eqnarray}\label{1loopRGtilde}
-m \partial_{m}\tilde R_{u} &\equiv& \beta [\tilde R] (u) \nonumber \\
&=& \left(\epsilon-4\zeta \right) \tilde R_{u}+u\zeta {\tilde R}'_{u} + \half
{{\tilde R}}_{u}''{}^{2} -{\tilde R}''_{u} {\tilde
R}''_{0}
 +\frac{1}{2}\left ( \tilde R''_{u} {\tilde R'''_{u}}{}^{2}-\tilde
R''_{0} {\tilde R'''_{u}}{}^{2}-\tilde R''_{u}{\tilde R'''_{0}}{}^{2}
\right)\ .\qquad
\end{eqnarray}
Consider the following change of variables
\begin{equation}\label{changeofvar}
\tilde R_{u} \equiv f[R](u)= R_{u} -\lambda \left(\frac{1}{2}{R''_{u}}^{2}-R''_{u}R''_{0}
\right)+ \ca O (R^{3})\ .
\end{equation}
Varying $m$ yields
\begin{equation}\label{5.3}
-m \partial_{m}\tilde R_{u}  = -m \partial_{m} \left[R_{u} -\lambda
\left(\half {R''_{u}}^{2}-R''_{u}R''_{0} \right) \right]\ .
\end{equation}
This is equivalent to stating that
\begin{eqnarray}\label{5.4}
\beta[\tilde R] (u)  &=&  \beta [R] (u)  -\lambda \Big\{ R'' (u)\beta
[R]'' (u) -R'' (0)\beta [R]'' (u) - R'' (u)\beta [R]'' (0)
\Big\}\ .\qquad
\end{eqnarray}
Solving this equation perturbatively yields the
$\beta$-function for $R_{u}$
\begin{eqnarray}\label{5.5}
\beta[R] (u) &=& (\epsilon-4\zeta ) R_{u}+\zeta uR'_{u} +\left[\half
{R''_{u}}^{2}- R''_{u} R''_{0} \right] (1+\lambda \epsilon )+\frac{1}{2}\left (  R''_{u} {\tilde R'''_{u}}{}^{2}-
R''_{0} { R'''_{u}}{}^{2}- R''_{u}{ R'''_{0}}{}^{2}
\right) \nn\\
&& +\ca O (\epsilon ^{4})
\end{eqnarray}
This equations tells us nothing more than that adding a coefficient of
order $\epsilon$ to the second-order term does not change universal
results at 2-loop order. (The reader may want to verify this
surprising result for the slope of the $\beta$-function at 2-loop
order in a scalar field-theory.)

Suppose now that $\beta [R] (u)=0$. Then this also holds for its
derivatives and multiples thereof.  Therefore, we can add to the fixed-point equation of the $\beta$-function terms of the
form
\begin{equation}\label{5.6}
R'' (u)\beta [R]'' (u) -R'' (0)\beta [R]'' (u) - R'' (u)\beta [R]''
(0)\ .
\end{equation}
Note that these are the same terms, which appeared in
equation (\ref{5.4}).

In the following, we chose $\zeta =0$, since this yields the simplest
relations. We will comment on the more general case later.
Expression (\ref{5.6})  then reads
\begin{eqnarray}\label{5.7}
&&\epsilon \left ( {R''_{u}}^{2}- 2 R''_{u} R''_{0} \right) +
\left(R''_{u} R'''_{u}{}^2 - R''_{0} R'''_{u}{}^2-R''_{u} R'''_{0}{}^2
\right)
+ \left(R''_{u}-R''_{0} \right)^{2}R'''' (u)\ .
\end{eqnarray}
Adding $-1/2$ times (\ref{5.7}) to the $\beta$-function (\ref{5.5})
and choosing there $\lambda =-1/2$ to eliminate the additional 1-loop
order term gives
\begin{eqnarray}\label{5.8}
0 &=& \epsilon R_{u} +\left[\half
{R''_{u}}^{2}- R''_{u} R''_{0} \right]-\frac{1}{2}\left(R''_{u}-R''_{0} \right)^{2}R'''' (u)
+\ca O (\epsilon ^{4})\ .
\end{eqnarray}
In this equation, we have traded the term proportional to $R''R'''{}^2$
for a term of the form  $R'''' R''{}^{2}$. Since the latter is
uniquely defined, this allows us again to fix the anomalous terms
associated to $R'' R'''{}^{2}$.

It would be satisfactory, to have a similar result for the case
$\zeta\neq 0$. The above construction however yields terms of the form
\begin{equation}\label{5.9}
\left(\zeta u R'_{u} \right)'' R''_{u}
\end{equation}
plus the respective anomalous terms. Although one can of course solve
differential equations involving these terms, and thus e.g.\ check the
numerical solution of the fixed point equation to be discussed later,
we have found no way to eliminate these terms, without generating even
more ``unusual'' ones. Our search comprised rescalings of $R_{u}$, of
the field $u$, adding $u \beta[R]'(u)$ to both the variable
transformation and the $\beta$-function itself, and adding multiples
of the $\beta$-function.
On the other hand, one can first write the $\beta$-function without rescaling, then do the non-trivial transformations given above, and finally perform the rescaling. This will simply give the standard rescaling terms.

Let us also comment on the power of reparametrization invariance at
3-loop order. While it proves to be a powerful tool for many diagrams,
it is at least not applicable to fix all anomalous terms.  This can be
anticipated from the difference between diagrams $(o)$ and $(q)$ (see
appendices \ref{s:(o)} and \ref{s:(q)}),
which is proportional to
\begin{equation}
(o)-(q)\sim
{R'''_{0}}^{2}
\left(R''_{u}R^{(4)}_{u}-R''_{0}R^{(4)}_{u}-R''_{u}R^{(4)}_{0} \right)\ .
\end{equation}
While $(o)$ and $(q)$ have the same normal terms,  their difference is proportional to $R'''(0^{+})^{2}$, thus the anomalous terms are different.

\section{Conclusions}
In this article, we have obtained the functional renormalization-group flow equations for the equilibrium properties of   elastic manifolds  in quenched disorder up to 3-loop order.
The analysis of these findings will be given in a separate publication \cite{HusemannWiese2017}: There we will extract the roughness exponent $\zeta$, obtain the fixed-point functions $R$ to 3-loop order, and give the corrrection-to-scaling exponent $\omega$.

An interesting question is how the formalism derived here can be extended to $N>1$ components.  It had been shown in Ref.~\cite{LeDoussalWiese2005a} that there is an ambiguity in the 2-point function already at 1-loop order. While this  allowed the authors of \cite{LeDoussalWiese2005a} to still conclude on the $\beta$-function at 2-loop order, the problem becomes more severe at 3-loop order, and despite considerable efforts in this direction we have not   been able to lift the ambiguities in some of the graphs.

\section*{Acknowledgements}

 We  thank Andreas Ludwig and Boris Kastening for their help in calculating the loop-integrals. We acknowledge a stimulating discussion with Pascal Chauve at an early stage of this work.

\appendix

\section{Loop integrals for  all diagrams up to 3 loops}
\label{a:Diagrams}
\subsection{General formulae,  strategy of calculation, and conventions}
We make use of the Schwinger parameterization
\begin{align}
 \frac{1}{A^n} = \frac{1}{\Gamma(n)} \int_0^\infty \mathrm{d} u \; u^{n-1} e^{-uA}\ ,
\end{align}
and the $d$-dimensional momentum integration
\begin{align}
 \int \frac{\rmd^d p}{(2\pi)^d}\rme^{-ap ^2} \equiv \int_p e^{-ap^2}= \frac{1}{a^{d/2}} \int_p \rme^{-p
^2} = \frac{1}{a^{d/2}}   \frac1{(4\pi)^{d/2}}\ .
\end{align}
In order to avoid cumbersome appearances of factors like \( \frac1{(4\pi)^{d/2}}\), we will write explicitly the last integral, and will only calculate ratios compared to the leading 1-loop diagram \(I_1\), given in the next section.

We will frequently use the   decomposition trick
\begin{align}
\frac{1}{k^2+1} = \frac{1}{k^2} -\frac{1}{k^2(k^2+1)}
\end{align}
which works well for dimension $d\le 4$. The reason for the utility of this decomposition is that it allows one to replace the massive propagator by a massless one, which is easier to integrate over, and a term converging faster for large $k$, which finally renders the integration finite.

Special functions which appear are \begin{eqnarray}
\psi(x) &:=& \frac{\Gamma'(x)}{\Gamma(x)}\ , \\
\psi'(x)&=& \frac{\rmd}{\rmd x}\psi(x)\ .
\end{eqnarray}

\subsection{The 1-loop integral $I_{1}$}\label{app:I1}
The integral $I_{1}$ is defined as
\begin{equation}\label{I1}
I_{1}:= \diagram{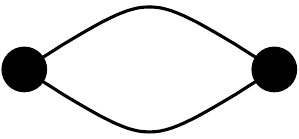}= \int_{k} \frac{1}{(k^{2}+m^{2})^{2}}\ ,
\end{equation}
and is calculated as follows:
\begin{eqnarray}\label{lf14}
I_{1} &=&\int_{k} \int_{0}^{\infty } \rmd \alpha \, \alpha \,\rme^{{-\alpha
(k^{2}+m^{2})} }\nn \\
 &=& \left(\int_{k}\rme^{-k^2} \right)  \int_{0}^{\infty
} \rmd \alpha\,  \alpha^{{1-\frac{d}{2}}} \, \rme^{-\alpha m^{2}}\nn \\
&=&  \left(\int_{k}\rme^{-k^2} \right) m^{-\epsilon } \Gamma
\left(\frac{\epsilon }{2} \right)\ .
\end{eqnarray}
We will also denote the dimensionless integral \begin{equation}
\tilde I_1 = I_1\Big|_{m=1}
\ .\end{equation}
This gives us the normalization-constant for higher-loop calculations
\begin{equation}
\left(\E I_1 \right) = m^{-\E}  \left(\int_{k}\rme^{-k^2} \right) \E
\Gamma \left(\frac{\epsilon }{2} \right) = m^{-\E}  \left(\int_{k}\rme^{-k^2} \right) 2\,
\Gamma \left(1+\frac{\epsilon }{2} \right)\ .
\end{equation}

\subsection{2-loop diagram $I_A$}
The non-trivial 2-loop integral can be written as
\begin{align}
 I_A =\diagram{LH}=\int _{p_1,p_2}g(p_1)g(p_2)^2g(p_1+p_2)
 = \frac{\Gamma(\epsilon)}{m^{2\epsilon}} \tilde{J}_A \ ,
\end{align}
with
\begin{align}
 \tilde{J}_A &=\int_0^\infty \mathrm{d}x\,\mathrm{d}y\, f_A(x,y) = J_1+J_2+J_3\\
 f_A(x,y)&= \frac{y}{(x+y+xy)^{2-\frac{\epsilon}{2}} (1+x+y)^{\epsilon}}
\\
 J_1&= \int_0^1\mathrm{d}y \int_0^\infty \mathrm{d}x f_A(x,y)\\
 J_2&=  \int_1^\infty\mathrm{d}y \int_0^\infty \mathrm{d}x \; \frac{1}{(1+x)^{2-\frac{\epsilon}{2}}}\frac{1}{y^{1+\frac{\epsilon}{2}}}
= \frac{4}{(2-\epsilon)\epsilon}\\
J_3&=  \int_1^\infty\mathrm{d}y \int_0^\infty \mathrm{d}x \; \left[f_A(x,y)
-  \frac{1}{(1+x)^{2-\frac{\epsilon}{2}}}\frac{1}{y^{1+\frac{\epsilon}{2}}}
\right]\ .
\end{align}
The integrals $J_1$ and $I_1$ were solved by expanding the integrand in $\epsilon$
to order $\epsilon$
\begin{align}
 \frac{I_A}{(\epsilon I_1)^2} = \frac{1}{2\epsilon^2} + \frac{1}{4\epsilon}
+ \frac{1}{72} \left[ 9 +4\pi^2  - 6\psi'(\sfrac{1}{3})\right]+\ca{O}(\epsilon)\ .
\end{align}
The result agrees with the one obtained by the  subtraction method.

\subsection{2-loop integral ${I_B}$}
The trivial 2-loop diagram is \begin{equation}
I_B := \parbox{0.15\textwidth}{\includegraphics[width=0.15\textwidth]{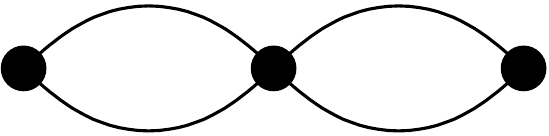}}= I_1^2
\end{equation}

\subsection{$I_i$}
\begin{equation}
I_i = \parbox{0.1\textwidth}{\includegraphics[width=0.1\textwidth]{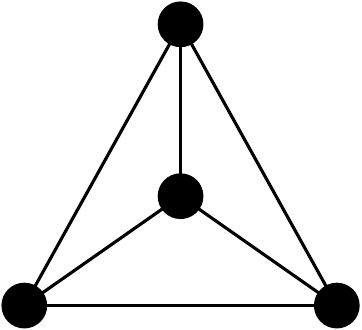}}
\end{equation}

\begin{align}
 \frac{I_i}{(\epsilon I_1)^3} &=\frac{1}{(\epsilon I_1)^3} \int_{p_1,p_2,p_3}  g(p_1)g(p_2)g(p_3)g(p_1+p_3)g(p_2+p_3)g(p_1-p_2) \nn\\
 &= \frac{\zeta(3)}{2\epsilon} + \ca{O}(\epsilon)\ .
\end{align}

\subsection{$I_j$}\begin{equation}
I_j = \parbox{0.1\textwidth}{\includegraphics[width=0.1\textwidth]{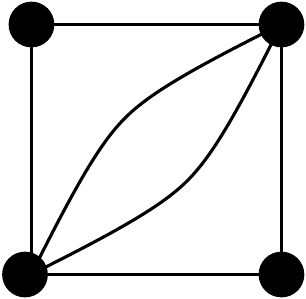}}
\end{equation}
\begin{align}
 \frac{I_j}{(\epsilon I_1)^3} &=\frac{1}{(\epsilon I_1)^3} \int_{p_1,p_2,p_3}  g(p_1)g(p_2)g(p_3)^2g(p_1+p_2+p_3)^2 = \sum_{i=1}^3 I^j_i\\
 I^j_1&= \frac{1}{(\epsilon I_1)^3}\int_{p_1,p_2,p_3}  \frac{1}{p_1^2p_2^2} g(p_3)^2g(p_1+p_2+p_3)^2 = \frac{1}{3\epsilon^3}+\frac{1}{6\epsilon^2}+ \frac{1}{12\epsilon}  +\ca{O}(1)\\
 I^j_2&= -2\frac{1}{(\epsilon I_1)^3}\int_{p_1,p_2,p_3}  \frac{1}{p_1^2p_2^2}g(p_1)g(p_3)^2g(p_1+p_2+p_3)^2 = \ca{O}(1) \\
 I^j_3&= \frac{1}{(\epsilon I_1)^3}\int_{p_1,p_2,p_3}  \frac{1}{p_1^2p_2^2} g(p_1)g(p_2)g(p_3)^2g(p_1+p_2+p_3)^2 = \ca{O}(1)\ .
\end{align}

\subsection{$I_l$}
\begin{equation}
I_l = \parbox{0.1\textwidth}{\includegraphics[width=0.1\textwidth]{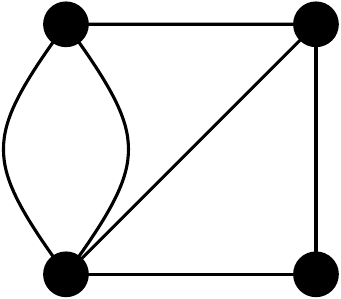}}
\end{equation}
\begin{align}
 \frac{I_l}{(\epsilon I_1)^3} &=\frac{1}{(\epsilon I_1)^3} \int_{p_1,p_2,p_3}  g(p_1)g(p_2)g(p_1+p_2)g(p_3)g(p_1+p_2+p_3)^2 = \sum_{i=1}^4 I^l_i\\
 I^l_1&= \frac{1}{(\epsilon I_1)^3}\int_{p_1,p_2,p_3}  \frac{1}{p_1^2p_2^2(p_1+p_2)^2} g(p_3)g(p_1+p_2+p_3)^2\nn\\
&= \frac{1}{6\epsilon^3}+\frac{1}{4\epsilon^2}+ \frac{7}{24\epsilon}  +\ca{O}(1)
\\
 I^l_2&= \frac{1}{(\epsilon I_1)^3}\int_{p_1,p_2,p_3}  \frac{1}{p_1^2p_2^2(p_1+p_2)^2}g(p_1+p_2)g(p_3)g(p_1+p_2+p_3)^2\nn \\
&= -\frac{4\pi^2+3\psi'(\sfrac{1}{3})-3\psi'(\sfrac{5}{6})}{216\epsilon} +\ca{O}(1) = I^m_2+\ca{O}(1)
\end{align}
\begin{align}
 I^l_3&= \frac{1}{(\epsilon I_1)^3}\int_{p_1,p_2,p_3}  \frac{1}{p_1^2p_2^2} g(p_2)g(p_1+p_2)g(p_3)g(p_1+p_2+p_3)^2 = \ca{O}(1)\\
 I^l_4&= \frac{1}{(\epsilon I_1)^3}\int_{p_1,p_2,p_3}  \frac{1}{p_1^2p_2^2} g(p_1)g(p_2)g(p_1+p_2)g(p_3)g(p_1+p_2+p_3)^2 =\ca{O}(1)
\end{align}
\begin{align}
\frac{I_l}{(\epsilon I_1)^3} &=\frac{1}{6\epsilon^3}+\frac{1}{4\epsilon^2}+ \frac{1}{\epsilon} \left[ -\frac{\pi^2}{54}  +\frac{7}{24} -\frac{1}{72} \left( \psi'(\sfrac{1}{3})-\psi'(\sfrac{5}{6})\right) \right]\ .
\end{align}

\subsection{$I_m$}
\begin{equation}
I_m = \parbox{0.1\textwidth}{\includegraphics[width=0.1\textwidth]{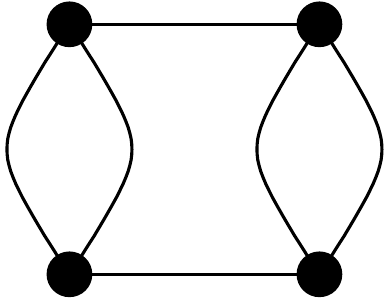}}
\end{equation}
\begin{align}
 \frac{I_m}{(\epsilon I_1)^3}
 &=\frac{1}{(\epsilon I_1)^3} \int_{p_1,p_2,p_3}  g(p_1)g(p_2)g(p_1+p_2+p_3)g(p_3)g(p_1+p_2)^2 = \sum_{i=1}^4 I^m_i\\
 I^m_1&= \frac{1}{(\epsilon I_1)^3}\int_{p_1,p_2,p_3}  \frac{1}{p_1^2p_2^2(p_1+p_2+p_3)^2} g(p_3)g(p_1+p_2)^2\nn \\
 &= I^m_{1,1} +I^m_{1,2}\\
I^m_{1,1}&= \frac{1}{(\epsilon I_1)^3}\int_{p_1,p_2,p_3}  \frac{1}{p_1^2p_2^2(p_1+p_2+p_3)^2} \frac{1}{p_3^2}g(p_1+p_2)^2 = \frac{1}{3\epsilon^3}+ \frac{1}{3\epsilon^2} + \frac{2+\pi^2}{12\epsilon} \\
I^m_{1,2}&= -\frac{1}{(\epsilon I_1)^3}\int_{p_1,p_2,p_3}  \frac{1}{p_1^2p_2^2(p_1+p_2+p_3)^2} \frac{1}{p_3^2}g(p_3)g(p_1+p_2)^2 = - \frac{\pi^2}{24\epsilon}
\\ \nn
 I^m_2&=-\frac{1}{(\epsilon I_1)^3}\int_{p_1,p_2,p_3}  \frac{1}{p_1^2p_2^2(p_1+p_2+p_3)^2}g(p_1+p_2+p_3)g(p_3)g(p_1+p_2)^2 \\
&= -\frac{4\pi^2+3\psi'(\sfrac{1}{3})-3\psi'(\sfrac{5}{6})}{216\epsilon} +\ca{O}(1)\\
 I^m_3&= -2 \frac{1}{(\epsilon I_1)^3}\int_{p_1,p_2,p_3}  \frac{1}{p_1^2p_2^2} g(p_2)g(p_1+p_2+p_3)g(p_3)g(p_1+p_2)^2 = -\frac{\pi^2}{12\epsilon}+\ca{O}(1)\ .
\end{align}
Finally,
\begin{align}
 I^m_4&= \frac{1}{(\epsilon I_1)^3}\int_{p_1,p_2,p_3}  \frac{1}{p_1^2p_2^2} g(p_1)g(p_2)g(p_1+p_2+p_3)g(p_3)g(p_1+p_2)^2
      = I^m_{4,0}+I^m_{4,1}\\
 I^m_{4,0}&= I_1(0) \frac{1}{(\epsilon I_1)^3}\int\mathrm{d}^2p  \frac{1}{p_1^2p_2^2} g(p_1)g(p_2) g(p_1+p_2)^2  \nn \\ &=\frac{5\pi^2 -3\psi'(\sfrac{1}{3})+3\psi'(\sfrac{5}{6})}{216\epsilon} + \ca{O}(1) \\
 I^m_{4,1}&= \frac{1}{(\epsilon I_1)^3}\int\mathrm{d}^2p  (I_1(p_1+p_2)-I_1(0)) \frac{1}{p_1^2p_2^2} g(p_1)g(p_2) g(p_1+p_2)^2  =\ca{O}(1)\ .
\end{align}
All in all
\begin{align}
\frac{I_m}{(\epsilon I_1)^3} &=\frac{1}{3\epsilon^3}+\frac{1}{3\epsilon^2}-   \frac{4\pi^2 -18 +\psi'(\sfrac{1}{3})-\psi'(\sfrac{5}{6})}{108\epsilon}  +\ca{O}(1) \nn \\
&=\frac{1}{3\epsilon^3}+\frac{1}{3\epsilon^2} +   \frac{3+2\pi^2 - 3\psi'(\sfrac{1}{3})}{18\epsilon}+\ca{O}(1) \ ,\label{A.48}
\end{align}
where two PolyGamma-identities were used,
\begin{align}
 \psi'(\sfrac{1}{3})+\psi'(\sfrac{5}{6})&=4\psi'(\sfrac{2}{3})\\
\psi'(\sfrac{1}{3})+\psi'(\sfrac{2}{3}) &=\frac{4\pi^2}{3}\ .
\end{align}

\section{Complimentary Material for Section~\ref{sec:ERG}}
\subsection{Functional RG Equations for $S^{(3)}$ and $S^{(4)}$} \label{app:ReplicaExpansionRGE}
The flow equation of the third $\Gamma$-cumulant in the ERG hierachy is given by
\begin{align} \label{eq:dotS3}\nonumber
\dot{S}^{(3)}[u_{abc}] &=  \int_{x_1,x_2} \;\dot{g}(x_1,x_2)  \left\{ -3T S_{110}^{(3)}[u_{abc}](x_1,x_2) + \frac 32 S_{1100}^{(4)}[u_{aabc}](x_1,x_2) \right\} \\ \nonumber
&\;\; +  \int_{x_1,...,x_4} \left[ \frac{\mathrm{d}}{\mathrm{d}m} g(x_1,x_2)g(x_3,x_4) \right] \left\{ \frac{3T}{2} \ca{R}''[u_{ab}](x_2,x_3) \ca{R}''[u_{ac}](x_4,x_1)  \right. \\ \nonumber &\hspace{2cm} \left.+3 \ca{R}''[u_{ab}](x_2,x_3) \left[S_{110}^{(3)}[u_{aac}](x_4,x_1) - S_{110}^{(3)}[u_{abc}](x_4,x_1)\right] \right\} \\ \nonumber
&\;\; +  \int_{x_1,...,x_6} \left[ \frac{\mathrm{d}}{\mathrm{d}m} g(x_1,x_2)g(x_3,x_4)g(x_5,x_6) \right] \\ \nonumber & \hspace{3cm} \left\{ 3 \ca{R}''[u_{ab}](x_2,x_3) \ca{R}''[u_{ac}](x_4,x_5) \ca{R}''[u_{ac}](x_6,x_1) \right. \\  & \hspace{3cm}\; \left. - \ca{R}''[u_{ab}](x_2,x_3) \ca{R}''[u_{bc}](x_4,x_5) \ca{R}''[u_{ac}](x_6,x_1)\right\}
\end{align}
We split the flow equation for the fourth $\Gamma$-cumulant
\begin{align}\label{eq:dotS4}
\dot{S}^{(4)}[u_{abcd}] = \dot{S}^{(4)}_1[u_{abcd}] +\dot{S}^{(4)}_2[u_{abcd}] +\dot{S}^{(4)}_3[u_{abcd}] +\dot{S}^{(4)}_4[u_{abcd}]
\end{align}
into four parts
\begin{align}
S_1^{(4)}[u_{abcd}]&= 2 \int_{x_1,x_2} \; \dot{g}(x_1,x_2) \left\{ -3T S^{(4)}[u_{abcd}](x_2,x_1)
 + S^{(5)}[u_{aabcd}](x_2,x_1) \right\}
 \\
\nn
 S_2^{(4)}[u_{abcd}]&= 6T \int_{x_1,...,x_4} \left[ \frac{\mathrm{d}}{\mathrm{d}m} g(x_1,x_2)g(x_3,x_4) \right] \Big\{ \ca{R}''[u_{ab}](x_2,x_3) S^{(3)}_{200}[u_{acd}](x_4,x_1) \Big\} \\ \nonumber
 & + 6 \int_{x_1,...,x_4} \left[ \frac{\mathrm{d}}{\mathrm{d}m} g(x_1,x_2)g(x_3,x_4) \right] \Big\{ \ca{R}''[u_{ab}](x_2,x_3) S^{(4)}_{1100}[u_{aacd}](x_4,x_1)  \\ \nonumber
 & \hspace{6cm} - \ca{R}''[u_{ab}](x_2,x_3) S^{(4)}_{1100}[u_{abcd}](x_4,x_1) \\
 & + S^{(3)}_{200}[u_{abc}](x_2,x_3) S^{(3)}_{110}[u_{aad}](x_4,x_1) + S^{(3)}_{110}[u_{abc}](x_2,x_3) S^{(3)}_{110}[u_{bad}](x_4,x_1) \Big\}
\\ \nn
%
%
 S_3^{(4)}[u_{abcd}]&= 4T \int_{x_1,...,x_6} \left[ \frac{\mathrm{d}}{\mathrm{d}m} g(x_1,x_2)g(x_3,x_4)g(x_5,x_6) \right] \\ \nonumber & \hspace{4cm} \Big\{ \ca{R}''[u_{ab}](x_2,x_3) \ca{R}''[u_{ac}](x_4,x_5) \ca{R}''[u_{ad}](x_6,x_1) \Big\} \\ \nonumber & +6 \int_{x_1,...,x_6} \left[ \frac{\mathrm{d}}{\mathrm{d}m} g(x_1,x_2)g(x_3,x_4)g(x_5,x_6) \right] \\ \nonumber &\hspace{4cm}\Big\{
2 \ca{R}''[u_{ab}](x_2,x_3) \ca{R}''[u_{ac}](x_4,x_5) S^{(3)}_{110}[u_{aad}](x_6,x_1)  \\ \nonumber &\hspace{4cm}
-2 \ca{R}''[u_{ab}](x_2,x_3) \ca{R}''[u_{ac}](x_4,x_5) S^{(3)}_{110}[u_{acd}](x_6,x_1) \\ \nonumber &\hspace{4cm}
-2 \ca{R}''[u_{ac}](x_2,x_3) \ca{R}''[u_{ab}](x_4,x_5) S^{(3)}_{110}[u_{acd}](x_6,x_1) \\ \nonumber &\hspace{4cm}
+ 2 \ca{R}''[u_{bc}](x_2,x_3) \ca{R}''[u_{ab}](x_4,x_5) S^{(3)}_{110}[u_{acd}](x_6,x_1) \\   &\hspace{4cm}
+ \ca{R}''[u_{ab}](x_2,x_3) \ca{R}''[u_{ab}](x_4,x_5) S^{(3)}_{110}[u_{acd}](x_6,x_1)   \Big\}
\end{align}
\begin{align}\nn
 S_4^{(4)}[u_{abcd}]&= 3 \int_{x_1,...,x_8} \left[ \frac{\mathrm{d}}{\mathrm{d}m} g(x_1,x_2)g(x_3,x_4)g(x_5,x_6) g(x_7,x_8) \right]
\\ \nonumber &  \hspace{2cm} \Big\{
4 \ca{R}''[u_{ab}](x_2,x_3) \ca{R}''[u_{ac}](x_4,x_5) \ca{R}''[u_{ad}](x_6,x_7) \ca{R}''[u_{ad}](x_8,x_1) \\ \nonumber &\hspace{2cm}
+2 \ca{R}''[u_{ab}](x_2,x_3) \ca{R}''[u_{ac}](x_4,x_5) \ca{R}''[u_{cd}](x_6,x_7) \ca{R}''[u_{ac}](x_8,x_1)\\ \nonumber &\hspace{2cm}
-4 \ca{R}''[u_{ab}](x_2,x_3) \ca{R}''[u_{ac}](x_4,x_5) \ca{R}''[u_{cd}](x_6,x_7) \ca{R}''[u_{ad}](x_8,x_1)\\   &\hspace{2cm}
+\ca{R}''[u_{ab}](x_2,x_3) \ca{R}''[u_{bc}](x_4,x_5) \ca{R}''[u_{cd}](x_6,x_7) \ca{R}''[u_{ad}](x_8,x_1) \Big\}
\end{align}

\subsection{Third $\Gamma$-cumulant $S^{(3)}$ to 3-loop order\label{app:S33loop}}
In total there are four contributions to the flow of $S^{(3)}$ in 3-loop order
\begin{align}\label{eq:flowS3}
 \dot{S}^{(3)}[u_{abc}] = \frac{\tilde{\mathrm{d}}}{\mathrm{d}m_g} \sum_{i=1}^{4} \uu_i[u_{abc}] + \ca{O}(\epsilon^5)\ ,
\end{align}
where the first contribution is known from the  2-loop calculation and reads
\begin{align}
 \uu_1 &= \frac{1}{2}(A_1+A_2+A_3) \sim \ca{O}(\epsilon^3)\ ,
\end{align}
where only the local part of $R[v]$ is inserted, so $\uu_1$ is of order $\epsilon^3$. The second contribution comes from inserting the non-local part of $R[v]$ to second order, that is Eq.~(\ref{eq:OneLoopNonLocal}), into $\frac12 (A_1+A_2+A_3)$.
\begin{align}
 \uu_2 &= \uu_{2,1} + \uu_{2,2} +\uu_{2,3}\sim \ca{O}(\epsilon^4)\ ,
\end{align}
where we split the contributions according to different types of integrals. This is not the shortest way to write but better comprehensible. The same is done in the contributions from the $R S^{(3)}$ term
\begin{align}
 \uu_3 &= \uu_{3,1} + \uu_{3,2} \sim \ca{O}(\epsilon^4)\ ,
\end{align}
where $\uu_1$ was used for $S^{(3)}$ on the right-hand side. Finally
\begin{align}
 \uu_4 &= \uu_{4,1} + \uu_{4,2} \sim \ca{O}(\epsilon^4)
\end{align}
is the feeding term from $S^{(4)}$, where we insert Eq.~(\ref{eq:S4threeloop}).

Eq. (\ref{eq:flowS3}) integrates to
\begin{align}
 S^{(3)}[u_{abc}] = \sum_{i=1}^{5} \uu^{(3),i}[u_{abc}] + \ca{O}(\epsilon^5)
\end{align}
with the 2-loop result $\uu^{(3),1}[u_{abc}] = \uu_1[u_{abc}]$ and
\begin{align}\nn
 \uu^{(3),2}[u_{abc}] &= \int^m(\uu_{2,1} - T) = -\frac 12 I_1 \int_{x_1,x_2,x_3} g(x_1,x_2)g(x_2,x_3) g(x_3,x_1) \\ \nonumber
& \quad \times \Bigg(\left( 2 \ca{R}''_{ab}(x_1) \Big[\ca{R}''_{ac}(x_2)+ \ca{R}''_{bc}(x_2)\Big] + \Big[\ca{R}''_{ac}(x_1)- \ca{R}''_{bc}(x_1)\Big]\Big[\ca{R}''_{ac}(x_2)- \ca{R}''_{bc}(x_2)\Big]\right) \\ \nonumber
& \hspace{1cm} \times \Big[ R_{ab}''''(x_3) \ca{R}''_{ab}(x_3)  +R_{ab}'''(x_3)^2 - R'''(0^+)^2 \Big] \\ \nonumber
& \hspace{0.5cm} +\left( \ca{R}''_{ab}(x_1) \ca{R}''_{ab}(x_2) + \Big[\ca{R}''_{ac}(x_1)+ \ca{R}''_{bc}(x_1)\Big]\Big[\ca{R}''_{ac}(x_2)+ \ca{R}''_{bc}(x_2)\Big]\right) \\ \nonumber
& \hspace{1cm} \times \Big[ R_{ac}''''(x_3) \ca{R}''_{ac}(x_3)+ R_{bc}''''(x_3) \ca{R}''_{bc}(x_3) + R_{ac}'''(x_3)^2+ R_{bc}'''(x_3)^2 - 2R'''(0^+)^2 \Big] \\ \nonumber
& \hspace{0.5cm} + 2 \ca{R}''_{ab}(x_1) \Big[\ca{R}''_{ac}(x_2)- \ca{R}''_{bc}(x_2)\Big]\Big[ R_{ac}''''(x_3) \ca{R}''_{ac}(x_3)- R_{bc}''''(x_3) \ca{R}''_{bc}(x_3) \\   &\hspace{8cm} + R_{ac}'''(x_3)^2- R_{bc}'''(x_3)^2 \Big] \Bigg)
\end{align}
\begin{align}\nn
 \uu^{(3),3}[u_{abc}] &= \int^m(\uu_{2,2} +\uu_{3,1}) = \frac 12 \int_{x_1,x_2,x_3,x_4} g(x_1,x_2)g(x_1,x_4) g(x_2,x_4)g(x_3,x_4)^2 \\ \nonumber
& \quad \times \Bigg(\left( 2 \ca{R}''_{ab}(x_1) \Big[\ca{R}''_{ac}(x_2)+ \ca{R}''_{bc}(x_2)\Big] + \Big[\ca{R}''_{ac}(x_1)- \ca{R}''_{bc}(x_1)\Big]\Big[\ca{R}''_{ac}(x_2)- \ca{R}''_{bc}(x_2)\Big]\right)
\\ \nonumber& \hspace{2cm} \times
\ca{R}''_{ab}(x_3) R_{ab}''''(x_4) \\ \nonumber
& \hspace{1cm} +\left(  \ca{R}''_{ab}(x_1) \ca{R}''_{ab}(x_2) + \Big[\ca{R}''_{ac}(x_1)+ \ca{R}''_{bc}(x_1)\Big]\Big[\ca{R}''_{ac}(x_2)+ \ca{R}''_{bc}(x_2)\Big]\right) \\ \nonumber
& \hspace{2cm} \times \Big[ \ca{R}''_{ac}(x_3)R_{ac}''''(x_4) + \ca{R}''_{bc}(x_3) R_{bc}''''(x_4)\Big] \\
& \hspace{1cm} + 2 \ca{R}''_{ab}(x_1) \Big[\ca{R}''_{ac}(x_2)- \ca{R}''_{bc}(x_2)\Big]\Big[ \ca{R}''_{ac}(x_3)R_{ac}''''(x_4) - \ca{R}''_{bc}(x_3)R_{bc}''''(x_4)  \Big] \Bigg)
\end{align}
\begin{align}
\nn
 \uu^{(3),4}[u_{abc}] &= \int^m(\uu_{2,3} +\uu_{4,2}) = \int _{x_1,x_2,y_1,y_2}y \;g(y_1,y_2)^2g(x_1,x_2) g(x_1,y_1)g(x_2,y_2) \\ \nonumber
& \quad \times \Bigg\{\left\{  \ca{R}''_{ab}(x_1) \Big[\ca{R}''_{ac}(x_2)+ \ca{R}''_{bc}(x_2)\Big] +\frac 12  \Big[\ca{R}''_{ac}(x_1)- \ca{R}''_{bc}(x_1)\Big]\Big[\ca{R}''_{ac}(x_2)- \ca{R}''_{bc}(x_2)\Big]\right\}
 \nn \\ &\hspace{2cm}\times \Big[ R_{ab}'''(y_1) R_{ab}'''(y_2) -R'''(0^+)^2\Big] \nn \\
\nonumber
& + \frac 12 \left\{\ca{R}''_{ab}(x_1) \ca{R}_{ab}''(x_2) +\Big[\ca{R}''_{ac}(x_1) + \ca{R}''_{bc}(x_1)\Big] \Big[\ca{R}''_{ac}(x_2)+ \ca{R}''_{bc}(x_2)\Big]\right\}
\nn \\
&\hspace{2cm}\times \Big[ R_{ac}'''(y_1) R_{ac}'''(y_2)+  R_{bc}'''(y_1) R_{bc}'''(y_2)-2 R'''(0^+)^2\Big] \nn \\
&+ \ca{R}''_{ab}(x_1) \Big[\ca{R}''_{ac}(x_2)- \ca{R}''_{bc}(x_2)\Big] \Big[ R_{ac}'''(y_1) R_{ac}'''(y_2)-  R_{bc}'''(y_1) R_{bc}'''(y_2)\Big] \Bigg\}
\end{align}
\begin{align}
\nn
 \uu^{(3),5}[u_{abc}] &= \int^m(\uu_{3,2} +\uu_{4,1}) = \frac 12 \int_{x_1,x_2,y_1,y_2} \;g(y_1,y_2)g(x_1,y_1) g(x_1,y_2)g(x_2,y_1)g(x_2,y_2)  \\ \nonumber
& \quad \times \Bigg\{\Big[  \ca{R}''_{ab}(x_1) \ca{R}''_{ab}(x_2) + \ca{R}''_{bc}(x_1) \ca{R}''_{bc}(x_2)  +\ca{R}''_{ac}(x_1) \ca{R}''_{ac}(x_2) \Big]\\ \nonumber
& \hspace{2cm} \times \Big[  R_{ab}'''(y_1) R_{ac}'''(y_2)+  R_{ac}'''(y_1) R_{bc}'''(y_2) -R_{ab}'''(y_1) R_{bc}'''(y_2) \Big]
\nn\\ \nonumber
& +2 \ca{R}''_{ab}(x_1) \ca{R}''_{ac}(x_2) \Big[R_{ab}'''(y_1) R_{ab}'''(y_2)+ R_{ab}'''(y_1) R_{ac}'''(y_2)+ R_{ab}'''(y_1) R_{bc}'''(y_2) \nn \\ \nonumber & \hspace{4cm} - R_{ac}'''(y_1) R_{bc}'''(y_2)+R_{ac}'''(y_1) R_{ac}'''(y_2) -R'''(0^+)^2 \Big] \nn\\
& +2 \ca{R}''_{ab}(x_1) \ca{R}''_{bc}(x_2) \Big[R_{ab}'''(y_1) R_{ab}'''(y_2)- R_{ab}'''(y_1) R_{ac}'''(y_2)- R_{ab}'''(y_1) R_{bc}'''(y_2) \nn \\ \nonumber  & \hspace{4cm} - R_{ac}'''(y_1) R_{bc}'''(y_2)+R_{bc}'''(y_1) R_{bc}'''(y_2) -R'''(0^+)^2 \Big] \nn\\
& +2 \ca{R}''_{ac}(x_1) \ca{R}''_{bc}(x_2) \Big[R_{bc}'''(y_1) R_{bc}'''(y_2)- R_{ab}'''(y_1) R_{ac}'''(y_2)+ R_{ab}'''(y_1) R_{bc}'''(y_2) \nn \\  & \hspace{4cm} + R_{ac}'''(y_1) R_{bc}'''(y_2)+R_{ac}'''(y_1) R_{ac}'''(y_2) -R'''(0^+)^2 \Big] \Bigg\}
\end{align}

\section{Systematic treatment of diagrams up to 3 loops: sloops and recursive construction}\label{app:C}
We present a systematic procedure to obtain (relatively)
simple results for diagrams at up to 3 loops. The idea is
to write the diagram, and then to consider all possible
sloops which lead to the same diagram. Subtracting them
with the right weight leads to results which are much simpler
than those obtained by trying to reduce expressions term by term.
The notation used throughout this section is
\begin{align}\label{A.1}
h_{ab}&:= R''_{ab} (1-\delta _{ab})\ ,\qquad \,g_{ab}:= R'''_{ab} (1-\delta
_{ab}) \ ,\qquad f_{ab}:= R''''_{ab} (1-\delta _{ab})\ ,\nn\\
p_{ab}&:=R^{(5)}_{ab} (1-\delta _{ab})\ , \qquad s_{ab}:=R^{(6)}_{ab}
(1-\delta _{ab})
\ .
\end{align}
We also use $h_{0}:=R''_{aa}$ a.s.o.
The notation is such that all summations (which are implicit) are
restricted. An example is
\begin{equation}
h_{ab}:=\sum_{a b}h_{ab} \equiv \sum_{a\neq b}h_{ab} = \sum_{a b}R''_{ab} -\sum_a R''_{aa}\ .
\end{equation}
We will  write rather instistinguishably, in a little abuse of notation, $R(u_a-u_b)\equiv R_u\equiv R_{ab}$, whatever is more convenient or suggestive.
Below, we will give all diagrams.

There is always an additional combinatorial
factor. At \(n\)-loop order, denote the number of propagators between points $i$ and $j$ as $n_{i,j}$.  Further denote the number of symmetries ${\cal S}$ as $N_{\cal S}$. Then the combinatorial factor for the contribution to $R$ is
\begin{equation}
\mbox{Comb} = \left(\frac12\right)^{\!\!\!n} \times \frac1{N_{\cal S}} \times \prod_{i,j} \frac1{n_{i,j}!}
\end{equation}
 at \(n\)-loop order, {\em   written apart from  the diagram}. We will give this factor at the beginning of each diagram with the same conventions as above.

\subsection{1 loop}
Here we give the 1-loop diagram.  A (closed) dashed line represents a sloop. ${\rm Comb}=\frac12 \times \frac12 \times \frac12$.
\begin{eqnarray}
\diagram{1loop} &=& 4\,{{h_{ab}}^2} + 4\,h_{ab}\,h_{ac}\\
\diagram{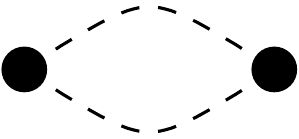} &=&\sum_{a,b} 4\,R''_{ab}\,R''_{ac}=4\,{{h_0}^2} + 8\,h_0\,h_{ab} +
  4\,h_{ab}\,h_{ac} \\
\diagram{1loop}
-\diagram{1loopsloop}&=&
  -4\,{{h_0}^2} - 8\,h_0\,h_{ab} +
  4\,{{h_{ab}}^2} = 4 \sum_{a,b} \left[ {{R''_{ab}}^2}
-2\,R''_0\,R''_{ab} \right]
  + \mbox{const}\qquad
\ .
\end{eqnarray}

\subsection{2 loops}
\subsubsection{The hat-diagram}\label{s:LH}
${\rm Comb}=\frac1{2^2} \times \frac12 \times \frac12$
\begin{eqnarray}
\diagram{LH}
&=& 8\,\left( 2\,{g_{ab}^2}\,h_{ab} +
    3\,{g_{ab}^2}\,h_{bc} - g_{ab}\,g_{ac}\,h_{bc} +
    2\,g_{ac}\,g_{bc}\,h_{bc} + g_{ac}\,g_{bc}\,h_{cd}
     \right)\\
\diagram{LHsloop2}
&=&8\,\left( {g_{ab}^2}\,h_{bc} - g_{ab}\,g_{ac}\,h_{bc} +
    2\,g_{ac}\,g_{bc}\,h_{bc} + g_{ac}\,g_{bc}\,h_{cd}
     \right)\\
\diagram{LHsloop3} &=&8\,\left( h_0\,{g_{ab}^2} +
    h_0\,g_{ab}\,g_{ac} +
    {g_{ab}^2}\,h_{bc} + g_{ac}\,g_{bc}\,h_{cd} \right)\ .
\end{eqnarray}
The simplest combination is
\begin{equation}
\diagram{LH}-\diagram{LHsloop2}
 = 16\,{g_{ab}^2}\,\left( h_{ab} + h_{bc} \right)
\ .
\end{equation}
\subsubsection{The bubble-chain}\label{s:banana}
The bubble-chain has ${\rm Comb}=\frac1{2^2} \times \frac12 \times (\frac12)^2$, and reads
\begin{eqnarray}\label{lf6-bis}
\diagram{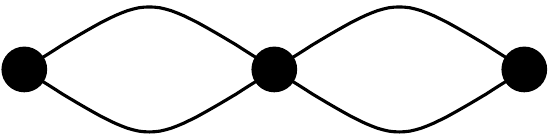} &=& 16\, f_{ab}\,h_{ab}^{2} +
32\,f_{ab}\,h_{ab}\,h_{bc}+8\,f_{ac}\,h_{ab}\,h_{cd}+8\,f_{ac\,}h_{bc}\,h_{cd}
\\
\diagram{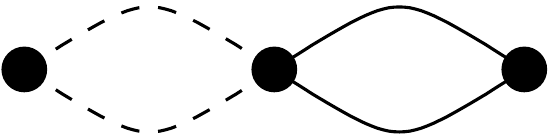} &=& 16\, h_{0}\, f_{ab}\,h_{ab} + 16\, h_{0}\,
f_{ab}\,h_{bc}
+16\,f_{bc}\,h_{ab}\,h_{bc}+8\, f_{ac}\,h_{ab}\,h_{cd}+ 8\,
f_{ac}\,h_{bc}\, h_{cd}\ .\qquad ~~~ \
\end{eqnarray}
Now two sloops are a little bit more complicated, and in fact to be
specific, we set
\begin{equation}
\diagram{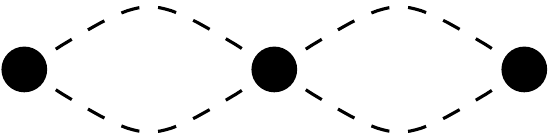}:=\frac{1}{2}\left[ \diagram{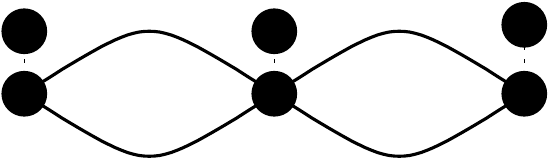} +
\diagram{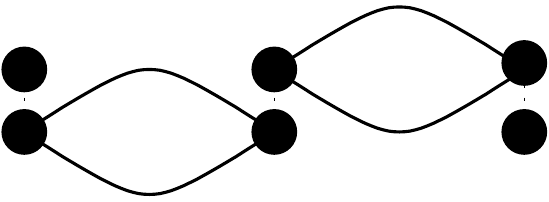}\right]
\end{equation}
We have
\begin{eqnarray}
 \diagram{2banana2sloopsa} &=&
16\,h_{0}^{2}\,f_{ab}+32\,h_{0}\,f_{ab}\,h_{ac}+16\, f_{ab}\,h_{ac}\,h_{ad} \\
 \diagram{2banana2sloopsb} &=&
16\,h_{0}^{2}\,f_{ab}+32\,h_{0}\,f_{ab}\,h_{ac}+16\, f_{ab}\,h_{ac}\,h_{bd} \\
 \diagram{2banana2sloops}
&=&16\,h_{0}^{2}\,f_{ab}+32\,h_{0}\,f_{ab}\,h_{ac}+8\,
f_{ab}\,h_{ac}\,h_{ad}+8\, f_{ab}\,h_{ac}\,h_{bd}\ .
\end{eqnarray}
Note that we have dropped the term $f_{0}$, which naively would be
there in the calculations. This can be done, since $f_{0}\sum
_{abc}R''_{ab} R''_{ac}$ is itself a 3-replica-term.

Then the simplest combination is
\begin{equation}
\diagram{banana}-2 \diagram{banana2sloop} + \diagram{2banana2sloops}
= 16 f_{ab} \left(h_{ab}-h_{0} \right)^{2} = 16 \sum _{a,b}
R''''_{ab}\left(R''_{ab}-R''_{0} \right)^{2}\ .
\end{equation}

\subsection{3 loops}

\subsubsection{Diagram ({\it h})}\label{s:(h)}
Diagram (\textit{h)} has ${\rm Comb}=\frac1{2^3} \times \frac12 \times (\frac12)^3$.
\begin{eqnarray}
\diagram{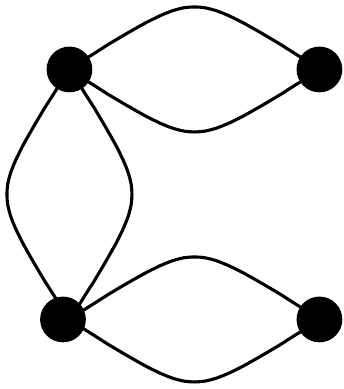}  &=&64\,f_{ab}^2\,h_{ab}^2 +
  128\,{f_{ab}}^2\,h_{ab}\,h_{bc} +
  64\,f_{ab}\,f_{bc}\,h_{ab}\,h_{bc} +
  64\,f_{ab}\,f_{bc}\,h_{ab}\,h_{cd} +
  32\,{f_{bc}}^2\,h_{ab}\,h_{cd}\nn \\
&& +
  64\,f_{ac}\,f_{bc}\,h_{ac}\,h_{cd} +
  32\,{f_{bc}}^2\,h_{ac}\,h_{cd} +
  16\,f_{ac}\,f_{cd}\,h_{ab}\,h_{de} +
  32\,f_{ac}\,f_{cd}\,h_{bc}\,h_{de} \nn\\
  &&+
  16\,f_{ad}\,f_{cd}\,h_{bd}\,h_{de} \\
\diagram{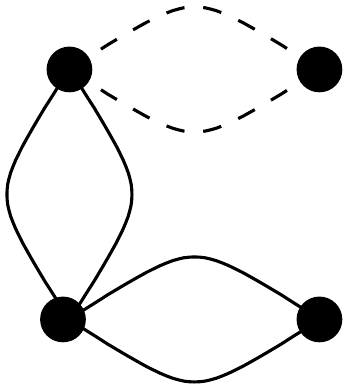}  &=&64\,h_{0}\,{f_{ab}}^2\,h_{ab} +
  64\,h_{0}\,{f_{ab}}^2\,h_{bc} +
  64\,h_{0}\,f_{ab}\,f_{bc}\,h_{bc} +
  64\,{f_{bc}}^2\,h_{ab}\,h_{bc} +
  32\,h_{0}\,f_{ab}\,f_{bc}\,h_{cd}\nn \\
&& +
  32\,h_{0}\,f_{ac}\,f_{bc}\,h_{cd} +
  32\,{f_{bc}}^2\,h_{ab}\,h_{cd} +
  32\,f_{ac}\,f_{cd}\,h_{ab}\,h_{cd} +
  32\,{f_{bc}}^2\,h_{ac}\,h_{cd} \nn \\
&&+
  32\,f_{ac}\,f_{cd}\,h_{bc}\,h_{cd} +
  16\,f_{ac}\,f_{cd}\,h_{ab}\,h_{de} +
  32\,f_{ac}\,f_{cd}\,h_{bc}\,h_{de} +
  16\,f_{ad}\,f_{cd}\,h_{bd}\,h_{de} \\
\diagram{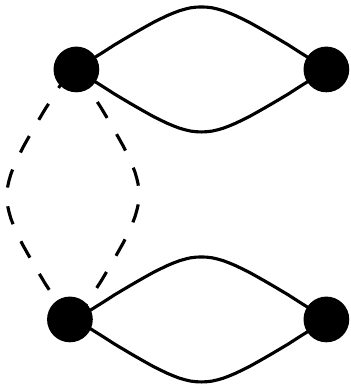}  &=&64\,f_{ab}\,f_{bc}\,h_{ab}\,h_{bc} +
  64\,f_{ab}\,f_{ac}\,h_{ab}\,h_{cd} +
  64\,f_{ac}\,f_{bc}\,h_{ac}\,h_{cd} +
  16\,f_{ab}\,f_{ad}\,h_{bc}\,h_{de}\nn \\
&& +
  32\,f_{ab}\,f_{bd}\,h_{bc}\,h_{de}  +
  16\,f_{ad}\,f_{cd}\,h_{bd}\,h_{de} \\
\diagram{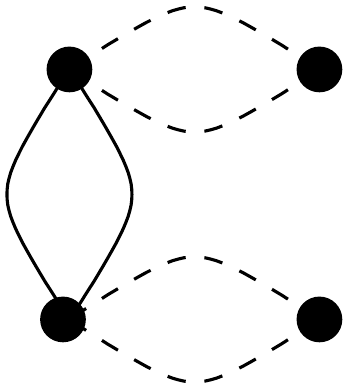}&=& 64\,{h_{0}}^2\,{f_{ab}}^2 +
  64\,{h_{0}}^2\,f_{ab}\,f_{ac} +
  128\,h_{0}\,{f_{ab}}^2\,h_{ac} +
  64\,h_{0}\,f_{ab}\,f_{ac}\,h_{ad} +
  32\,{f_{ab}}^2\,h_{ac}\,h_{ad} \nn \\
&&+
  16\,f_{ab}\,f_{ad}\,h_{ac}\,h_{ae} +
  32\,{f_{ab}}^2\,h_{ac}\,h_{bd} +
  64\,h_{0}\,f_{ab}\,f_{ac}\,h_{cd} +
  32\,f_{ab}\,f_{ad}\,h_{ac}\,h_{de} \nn \\
&&+
  16\,f_{ab}\,f_{bd}\,h_{ac}\,h_{de}\ .
\end{eqnarray}
For two intersecting 2-loops, there are 2 possibilities, and we define:
\begin{equation}\label{lf7}
\diagram{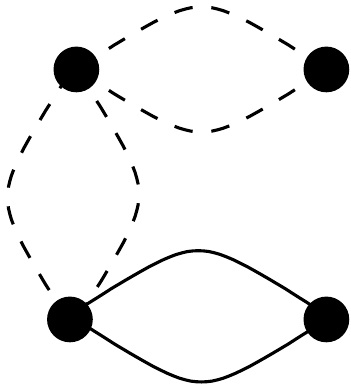} =\half \left[\diagram{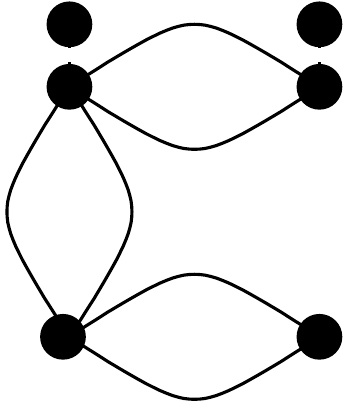}
+\diagram{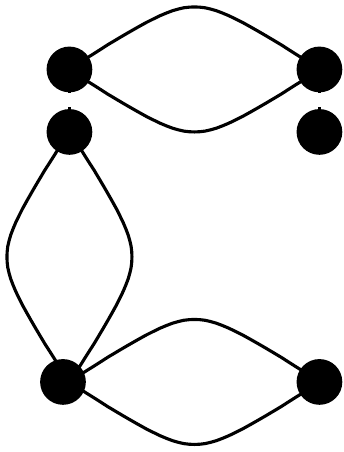} \right]
\end{equation}
The terms are
\begin{eqnarray}
\diagram{3looph2sloops2a}&=&64\,h_{0}\,f_{ab}\,f_{bc}\,h_{bc} +
  32\,h_{0}\,f_{ab}\,f_{ac}\,h_{cd} +
  32\,h_{0}\,f_{ac}\,f_{bc}\,h_{cd} +
  64\,f_{ac}\,f_{cd}\,h_{bc}\,h_{cd} \nn \\
&&+
  32\,f_{ab}\,f_{ad}\,h_{ac}\,h_{de} +
  32\,f_{ad}\,f_{cd}\,h_{bd}\,h_{de}
\eea
\bea\diagram{3looph2sloops2b}&=&64\,h_{0}\,f_{ab}\,f_{bc}\,h_{bc} +
  32\,h_{0}\,f_{ab}\,f_{bc}\,h_{cd} +
  32\,h_{0}\,f_{ac}\,f_{bc}\,h_{cd} +
  64\,f_{ac}\,f_{cd}\,h_{ab}\,h_{cd} \nn \\
&&+
  32\,f_{ad}\,f_{cd}\,h_{ab}\,h_{de} +
  32\,f_{ab}\,f_{bd}\,h_{ac}\,h_{de} \\
\diagram{3looph2sloops2} &=& 64\,h_{0}\,f_{ab}\,f_{bc}\,h_{bc} +
  32\,h_{0}\,f_{ab}\,f_{ac}\,h_{cd} +
  32\,h_{0}\,f_{ac}\,f_{bc}\,h_{cd} +
  32\,f_{ac}\,f_{cd}\,h_{ab}\,h_{cd} \nn \\
&&+
  32\,f_{ac}\,f_{cd}\,h_{bc}\,h_{cd}+
  32\,f_{ab}\,f_{ad}\,h_{ac}\,h_{de} +
  16\,f_{ab}\,f_{bd}\,h_{ac}\,h_{de} +
  16\,f_{ad}\,f_{cd}\,h_{bd}\,h_{de}\ .\qquad \qquad
\end{eqnarray}
Now 3 intersecting sloops. They can intersect in 3 different manners,
and we take the average, with the weight proportional to their
combinatorial factor,
\begin{eqnarray}
\diagram{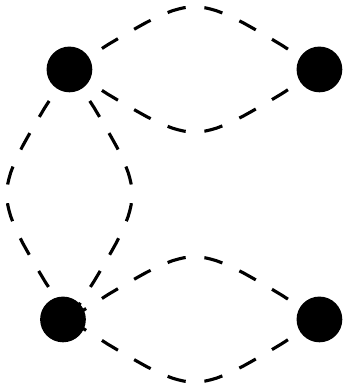} = \frac{1}{4}\left[\diagram{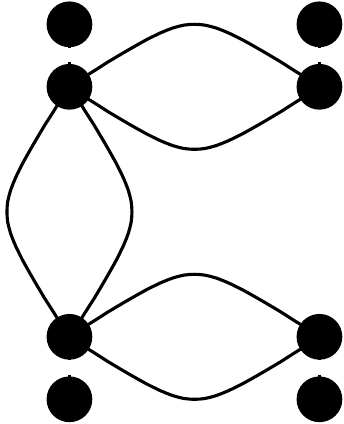}
+ {2} \diagram{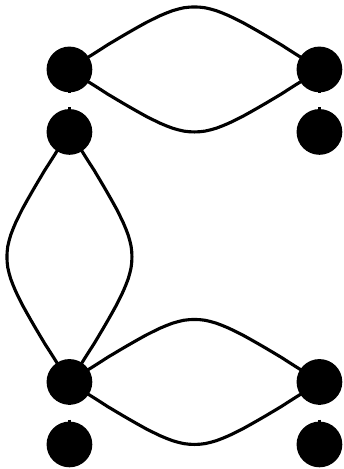}+
\diagram{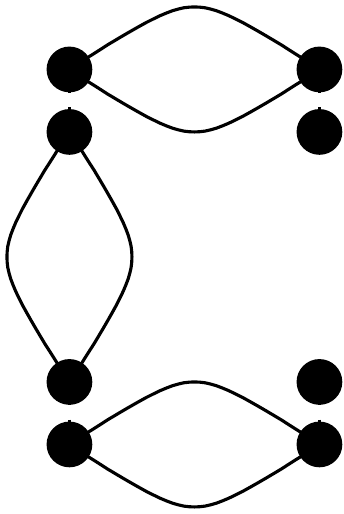}  \right]
.\end{eqnarray}
The respective contributions are:
\begin{eqnarray}
\diagram{3looph3sloopsa}  &=& 64\,{h_{0}}^2\,f_{ab}\,f_{ad} +
  128\,h_{0}\,f_{ab}\,f_{ad}\,h_{ac} +
  64\,f_{ab}\,f_{ad}\,h_{ac}\,h_{ae} \\
\diagram{3looph3sloopsb}  &=&64\,{h_{0}}^2\,f_{ab}\,f_{ad} +
  64\,h_{0}\,f_{ab}\,f_{ad}\,h_{ae} +
  64\,h_{0}\,f_{ab}\,f_{ad}\,h_{bc} +
  64\,f_{ab}\,f_{ad}\,h_{ae}\,h_{bc} ~~~~~~~~~\\
\diagram{3looph3sloopsc}  &=&64\,{h_{0}}^2\,f_{ab}\,f_{ad} +
  128\,h_{0}\,f_{ab}\,f_{ad}\,h_{bc} +
  64\,f_{ab}\,f_{ad}\,h_{bc}\,h_{de}\\
\diagram{3looph3sloops} &=& 64\,{h_{0}}^2\,f_{ab}\,f_{ad} +
  64\,h_{0}\,f_{ab}\,f_{ad}\,h_{ac} +
  16\,f_{ab}\,f_{ad}\,h_{ac}\,h_{ae} +
  64\,h_{0}\,f_{ab}\,f_{ad}\,h_{bc} \nn \\
&&+
  32\,f_{ab}\,f_{ad}\,h_{ae}\,h_{bc} +
  16\,f_{ab}\,f_{ad}\,h_{bc}\,h_{de}\ .
\end{eqnarray}
The final combination is
\begin{eqnarray}
&&\!\!\!\diagram{3looph}-2 \diagram{3looph12sloop1}-\diagram{3looph12sloop2}+
\diagram{3looph2sloops1} +2 \diagram{3looph2sloops2} -
\diagram{3looph3sloops}  \nn \\
&&\qquad = 64 f_{ab}^{2} \left(h_{ab}-h_{0}
\right)^{2}= 64 \sum _{a,b} f_{ab}^{2} \left(h_{ab}-h_{0}
\right)^{2} \ .
\end{eqnarray}
Note that each sloop comes with a factor of $(-1)$ and furthermore
one has taken into account the proper combinatorial factor. This result is confirmed by the recursive-construction algorithm.

\subsubsection{Diagram (\textit{i})}\label{s:(i)}
 ${\rm Comb}=\frac1{2^3} \times \frac1{4!} \times 1$.
For a given order of the contractions, we have:
\begin{eqnarray}
\parbox{1cm}{\fig{1cm}{3loopi}} &=& 16 \left (6\,{g_{ab}^4} + 16\,{g_{ab}^3}\,g_{ac} +
  3\,{g_{ab}^2}\,{g_{ac}^2} +
  6\,{g_{ab}^2}\,g_{ac}\,g_{ad} +
  g_{ab}\,g_{ac}\,g_{ad}\,g_{ae} +
  12\,{g_{ab}^2}\,g_{ac}\,g_{bc} \right) \qquad \\
\parbox{1cm}{\fig{1cm}{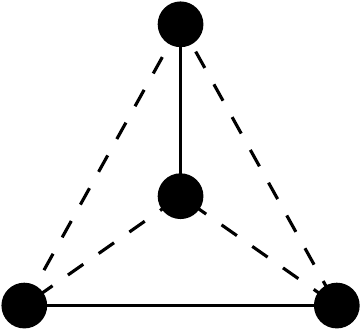}} &=&
16\left( {g_{ab}^2}\,{g_{ac}^2} +
    2\,{g_{ab}^2}\,g_{ac}\,g_{ad} +
    g_{ab}\,g_{ac}\,g_{ad}\,g_{ae} \right)\\
\parbox{1cm}{\fig{1cm}{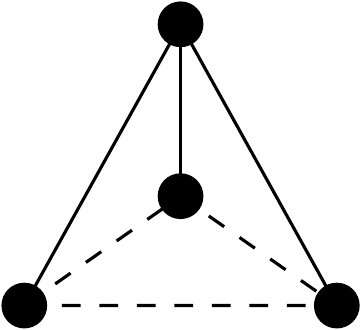}} &=&
16\left( 4\,{g_{ab}^3}\,g_{ac} +
    3\,{g_{ab}^2}\,g_{ac}\,g_{ad} +
    g_{ab}\,g_{ac}\,g_{ad}\,g_{ae} +
    3\,{g_{ab}^2}\,g_{ac}\,g_{bc} \right)
\ .
\end{eqnarray}
The simplest combination is
\begin{equation}
\parbox{1cm}{\fig{1cm}{3loopi}}
+3 \parbox{1cm}{\fig{1cm}{3loopisloop4}}
-4 \parbox{1cm}{\fig{1cm}{3loopisloop3}}
=96 \left( {g_{ab}^4} + {g_{ab}^2}\,{g_{ac}^2} \right)
=\mbox{1-rep} + 96 \sum_{a,b}\left( {R'''_{ab}}^4 -2 {R'''_{ab}}^2 {R'''_0}^2\right) + \mbox{3-reps}
\ .
\end{equation}
Note that the factors are combinatorial factors for the number
of possibilities to chose the sloop, while the signs are less intuitive.
The diagram is supercusp-free.

\subsubsection{Diagram (\textit{j})}\label{s:(j)}
Diagram (\textit{j}) has ${\rm Comb}=\frac1{2^3} \times \frac14 \times  \frac12$.
We number 1 to 4 for points $x_1$ to $x_4$,
\begin{equation}
{\parbox{0cm}{\rule{0mm}{7mm}}^1_3}
\diagram{3loopj}
{\parbox{0cm}{\rule{0mm}{7mm}}^2_4}
\end{equation}
We have performing, the contractions in the order (23)(23)(13)(12)(34)(24)
or (13)(12)(34)(24)(23) (23)
\begin{eqnarray}
\diagram{3loopj}&=&
16\big( 4\,{f_{ab}^2}\,{h_{ab}^2} +
    12\,{f_{ab}^2}\,h_{ab}\,h_{bc} -
    4\,f_{ab}\,f_{ac}\,h_{ab}\,h_{bc} +
    2\,f_{ab}\,f_{bc}\,h_{ab}\,h_{bc} +
    f_{ab}\,f_{ac}\,{h_{bc}^2} \nn \\
&&+
    2\,f_{ab}\,f_{bc}\,{h_{bc}^2} +
    4\,f_{ab}\,f_{bc}\,h_{ab}\,h_{bd} +
    3\,{f_{ab}^2}\,h_{bc}\,h_{bd} +
    4\,{f_{ac}^2}\,h_{ab}\,h_{cd} -
    2\,f_{ac}\,f_{ad}\,h_{ab}\,h_{cd} \nn \\
&&+
    f_{ad}\,f_{cd}\,h_{bd}\,h_{de} \big)\ .
\end{eqnarray}
Sloops: The 2-sloop contracted as (23)(23)(13)(12)(34)(24)
gives
\begin{eqnarray}
\diagram{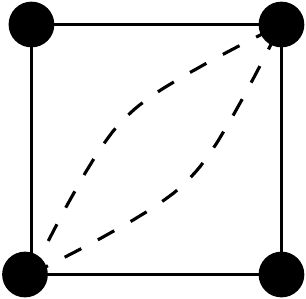}&=&
16\big( 4\,{f_{ab}^2}\,h_{ab}\,h_{bc} -
    4\,f_{ab}\,f_{ac}\,h_{ab}\,h_{bc} +
    2\,f_{ab}\,f_{bc}\,h_{ab}\,h_{bc} +
    f_{ab}\,f_{ac}\,{h_{bc}^2} +
    2\,f_{ab}\,f_{bc}\,{h_{bc}^2} \nn \\
&&+
    4\,f_{ab}\,f_{bc}\,h_{ab}\,h_{bd}+
    {f_{ab}^2}\,h_{bc}\,h_{bd} +
    2\,{f_{ac}^2}\,h_{ab}\,h_{cd} -
    2\,f_{ac}\,f_{ad}\,h_{ab}\,h_{cd} +
    f_{ad}\,f_{cd}\,h_{bd}\,h_{de} \big)\ .\qquad\ \ \
\end{eqnarray}
The 3-sloop is
\begin{eqnarray}
\diagram{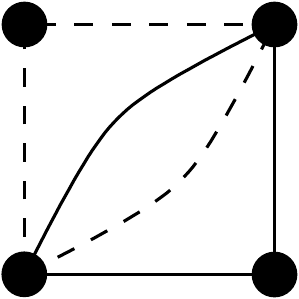}
&=& 16\big( 2\,h_0\,{f_{ab}^2}\,h_{ab} +
    h_0\,f_{ac}\,f_{ad}\,h_{ab} +
    3\,h_0\,{f_{ab}^2}\,h_{bc} -
    h_0\,f_{ab}\,f_{ac}\,h_{bc} +
    2\,h_0\,f_{ac}\,f_{bc}\,h_{bc} \nn \\
&&+
    2\,{f_{ab}^2}\,h_{ab}\,h_{bc}+
    2\,f_{ab}\,f_{bc}\,h_{ab}\,h_{bd} +
    {f_{ab}^2}\,h_{bc}\,h_{bd} +
    2\,{{f_{ac}}^2}\,h_{ab}\,h_{cd} -
    f_{ac}\,f_{ad}\,h_{ab}\,h_{cd} \nn \\
&&+
    f_{ad}\,f_{cd}\,h_{bd}\,h_{de} \big).
\end{eqnarray}
The 4-sloop(13)(12)(34)(24), then contracted (23)(23)
gives
\begin{equation}
\diagram{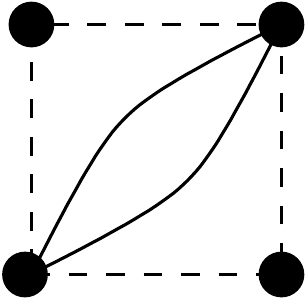}=
16 \left( 3\,{h_0^2}\,{f_{ab}^2} +
    {h_0^2}\,f_{ab}\,f_{ac} +
    2\,h_0\,f_{ac}\,f_{ad}\,h_{ab} +
    6\,h_0\,{f_{ab}^2}\,h_{bc} +
    3\,{f_{ab}^2}\,h_{bc}\,h_{bd} +
    f_{ad}\,f_{cd}\,h_{bd}\,h_{de} \right).
\end{equation}
We can study another configuration, which we do not know   how
to draw, so call it $S$
\begin{equation}
S=16(h_{ab} (x_1)+h_{0}) g_{ac} (x_2)g_{ac} (x_3) R_{de} (x_4)\ ,
\end{equation}
where we have already dropped the term $a=c$, which will disappear
after the next contraction.
Contracting (34) and then (24) gives
\begin{equation}
64\,h_0\,{f_{ab} }^2\,h_{ab} +64\,h_0\,{f_{ab} }^2\,h_{bc}
+64\,{f_{bc} }^2\,h_{ab} \,h_{bc} +32\,{f_{ac} }^2\,h_{ab}\,h_{cd}
+32\,{f_{bc} }^2\,h_{ac} \,h_{cd} \ .
\end{equation}
A simple combination seems to be
\begin{equation}
\diagram{3loopj}
-\diagram{3loopjsloop2}
= 32\,\left( 2\,{f_{ab}^2}\,{h_{ab}^2} +
    4\,{f_{ab}^2}\,h_{ab}\,h_{bc} +
    {f_{ab}^2}\,h_{bc}\,h_{bd} +
    {f_{ac}^2}\,h_{ab}\,h_{cd} \right)\ .
\end{equation}
A still simpler configuration is
\begin{eqnarray}
\diagram{3loopj}
-\diagram{3loopjsloop2}- S &=&
64(-h_0{f_{ab}}^2h_{ab}+{f_{ab}}^2{h_{ab}}^2-h_0{f_{ab}}^2
  h_{bc}+{f_{ab}}^2h_{ab}h_{bc}) \nonumber \\
&=&64[{f_{ab}}^2{h_{ab}} (h_{ab}-h_{0})+{f_{ab}}^2h_{bc} (
h_{ab}-h_{0})]\nonumber \\
 &=& 64[{f_{ab}}^2 (h_{ab}-h_{0})] ({h_{ab}}+h_{bc})\ .
\end{eqnarray}
The trivial de-slooping gives (confirmed by the recursive-construction algorithm)
\begin{equation}
\diagram{3loopj}
-\diagram{3loopjsloop2} = 64 f_{ab}^{2} (h_{ab}-h_{0})^{2} =64 \sum
_{ab}{R''''_{ab}}^{2}\left(R''_{ab}-R''_{0} \right)^{2}\ .
\end{equation}

\subsubsection{Diagram (\textit{k})}\label{s:(k)}
Next is diagram (\textit{k}). It has ${\rm Comb}=\frac1{2^3} \times \frac12 \times \frac1{3!}$.  With contractions (13)(13)(13)(12)(34)(24) we have
\begin{eqnarray}
{\parbox{0cm}{\rule{0mm}{7mm}}^1_3}
\diagram{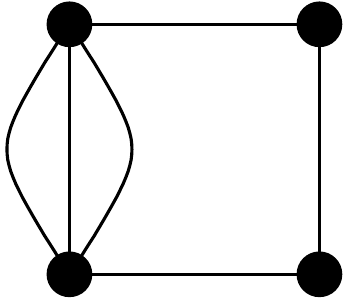}
\,{\parbox{0cm}{\rule{0mm}{7mm}}^2_4}
&=&
16\,\big( 12\,{f_{ab}^2}\,h_{ab}\,h_{bc} -
    6\,{{f_{ac}}^2}\,h_{ab}\,h_{bc} +
    6\,{{f_{ac}}^2}\,{h_{bc}^2} +
    2\,f_{ab}\,f_{bc}\,h_{ab}\,h_{bd} +
    7\,{f_{ab}^2}\,h_{bc}\,h_{bd} \nn \\
&&+
    f_{ad}\,f_{cd}\,{h_{bd}^2} -
    2\,f_{ab}\,f_{ac}\,h_{bc}\,h_{cd} +
    f_{ab}\,f_{ad}\,h_{bc}\,h_{cd} +
    2\,f_{ab}\,f_{bc}\,h_{bc}\,h_{cd} -
    2\,f_{ab}\,f_{bd}\,h_{bc}\,h_{cd} \nn \\
&&+
    f_{ad}\,f_{cd}\,h_{bd}\,h_{de} \big)
\end{eqnarray}
The  2-sloop (13)(13), then (13)(12)(34)(24)
\begin{eqnarray}
\diagram{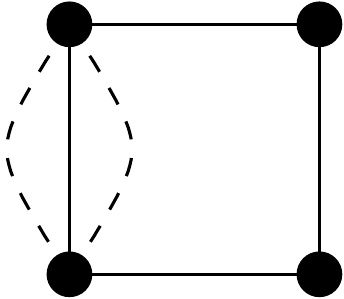}&=&
16\,\big( 4\,{f_{ab}^2}\,h_{ab}\,h_{bc} -
    2\,{{f_{ac}}^2}\,h_{ab}\,h_{bc} +
    2\,{{f_{ac}}^2}\,{h_{bc}^2} +
    2\,f_{ab}\,f_{bc}\,h_{ab}\,h_{bd} +
    3\,{f_{ab}^2}\,h_{bc}\,h_{bd} \nn \\
&&+
    f_{ad}\,f_{cd}\,{h_{bd}^2} -
    2\,f_{ab}\,f_{ac}\,h_{bc}\,h_{cd} +
    f_{ab}\,f_{ad}\,h_{bc}\,h_{cd} +
    2\,f_{ab}\,f_{bc}\,h_{bc}\,h_{cd} -
    2\,f_{ab}\,f_{bd}\,h_{bc}\,h_{cd} \nn \\
&&+
    f_{ad}\,f_{cd}\,h_{bd}\,h_{de} \big)
\end{eqnarray}
We find that the difference is
\begin{equation}
\diagram{3loopk}
-\diagram{3loopksloop2}
= 64\,\left( 2\,{f_{ab}^2}\,h_{ab}\,h_{bc} -
    {f_{ac}^2}\,h_{ab}\,h_{bc} +
    {f_{ac}^2}\,{h_{bc}^2} +
    {f_{ab}^2}\,h_{bc}\,h_{bd} \right)
\end{equation}
Trivial deslooping gives
\begin{equation}
0
\end{equation}
This is important since there is no counter-term in the theory for the divergence between the two leftmost vertices.

\subsubsection{Diagram (\textit{l})}\label{s:(l)}
Diagram (\textit{l}) has ${\rm Comb}=\frac1{2^3} \times 1 \times \frac12$. We use the  notation
\begin{equation}
{\parbox{0cm}{\rule{0mm}{7mm}}^1_3}
\diagram{3loopl}
{\parbox{0cm}{\rule{0mm}{7mm}}^2_4}
\end{equation}
Diagram $(l)$ is
\begin{eqnarray}
\diagram{3loopl}&=&16\Big(
4f_{ab}g_{ab}^2h_{ab}+8f_{ab}g_{ab}^2h_{bc}-3f_{ab}g_{ab}g_{ac}h_{bc}
+f_{bc}g_{ab}g_{ac}h_{bc}-f_{ab}g_{ac}^2h_{bc}+f_{bc}g_{ac}^2h_{bc}
\nn \\
&&-3f_{ab}g_{ab}g_{bc}h_{bc}-5f_{bc}g_{ab}g_{bc}h_{bc}+f_{ab}g_{bc}^2h_{bc}
+f_{bc}g_{ac}^2h_{cd}-f_{ac}g_{ab}g_{bc}h_{cd}-2f_{bc}g_{ab}g_{bc}h_{cd}
\nn \\
&&+4f_{ac}g_{ac}g_{bc}h_{cd}
+f_{bc}g_{ab}g_{bd}h_{cd}+f_{cd}g_{ad}g_{bd}h_{cd}-f_{bc}g_{ac}g_{cd}h_{cd}+f_{cd}g_{ad}g_{bd}h_{de}\Big)\ .
\end{eqnarray}
The 2-sloop is
\begin{eqnarray}
\diagram{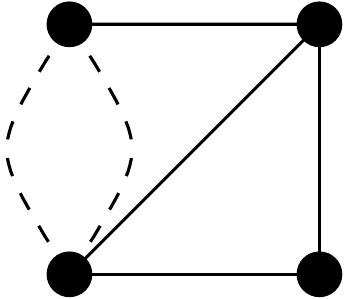}&=& 16 \Big( 2f_{ab} g_{ab}^2h_{bc} -f_{ab}
g_{ab} g_{ac} h_{bc} +f_{bc} g_{ab} g_{ac} h_{bc} -f_{ab} g_{ac}
^2h_{bc}+f_{bc} g_{ac}^2h_{bc} -f_{ab} g_{ab} g_{bc} h_{bc}
\nn \\
&&-3f_{bc} g_{ab} g_{bc} h_{bc}  +f_{ab} g_{bc} ^2h_{bc} +f_{bc}
g_{ac}^2h_{cd}-f_{ac} g_{ab} g_{bc} h_{cd}-2f_{bc} g_{ab} g_{bc}
h_{cd}+2f_{ac} g_{ac} g_{bc} h_{cd}\nn \\
&&+f_{bc} g_{ab}
g_{bd}h_{cd} +f_{cd}g_{ad} g_{bd}h_{cd}-f_{bc} g_{ac}
g_{cd}h_{cd}+f_{cd}g_{ad} g_{bd}h_{de} \Big)\ .
\end{eqnarray}
There is the special sloop configuration, which is obtained by
starting from $16 g_{ab} (1)g_{ab} (2)h_{ac} (2)R_{de} (4)$. It is
denoted and reads
\begin{equation}\diagram{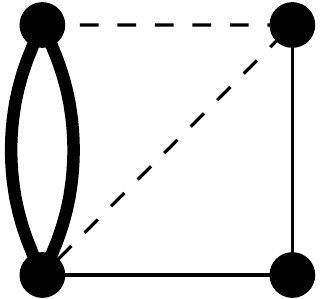}=
32\Big(-f_{ab} \,g_{ab} \,g_{ac}
\,h_{bc} -f_{ab} \,g_{ab} \,g_{bc} \,h_{bc} +f_{bc} \,g_{ac} \,g_{bc}
\,h_{bc} +f_{ac} \,g_{ac} \,g_{bc} \,h_{cd} \Big)\ .
\end{equation}
Now diagram ($l$) is with the 2-sloop subtracted
\begin{eqnarray}
\diagram{3loopl}
-\diagram{3looplsloop2}
&=&32\,\Big( 2\,f_{ab}\,{g_{ab}^2}\,h_{ab} +
    3\,f_{ab}\,{g_{ab}^2}\,h_{bc} -
    f_{ab}\,g_{ab}\,g_{ac}\,h_{bc} -
    f_{ab}\,g_{ab}\,g_{bc}\,h_{bc} \nn \\
&&+
    f_{bc}\,g_{ac}\,g_{bc}\,h_{bc} +
    f_{ac}\,g_{ac}\,g_{bc}\,h_{cd} \Big)\ .
\end{eqnarray}
An even simpler configuration is
\begin{equation}
\diagram{3loopl} -\diagram{3looplsloop2}-\diagram{3looplhatsloop}=
64 \Big( f_{ab} \,g_{ab}^2\,h_{ab} +f_{ab} \, g_{ab}^2\,h_{bc}\Big)\ .
\end{equation}
There are of course much more possible sloops, involving three or four
vertices. However, we did not use them here, and thus do not display
them.

The recursive-construction algorithm gives, consistent with the above
\begin{equation}
\diagram{3loopl} = 64
\left[ R''_{u} ( R'''_{u})^{2}R''''_{u}-R''_{0} ( R'''_{u})^{2}R''''_{u}
-R''_{u} ( R'''_{0})^{2} R''''_{0}  \right]\ .
\end{equation}

\subsubsection{Diagram (\textit{m})}\label{s:(m)}
Diagram (\textit{m})  has ${\rm Comb}=\frac1{2^3} \times \frac14 \times (\frac12)^2$. It is not independent of the path of contractions. We number
\begin{equation}
{\parbox{0cm}{\rule{0mm}{9mm}}^1_2}{\diagram{3loopm}} {\parbox{0cm}{\rule{0mm}{9mm}}^3_4}
\end{equation}
The simplest
result is obtained by using contractions (12)(12)(34)(34)(13)(24)
\begin{eqnarray}
\!\!\diagram{3loopm}
&=&16\Big( 4\,{g_{ab}^4} + 8\,{g_{ab}^3}\,g_{ac} +
    8\,{g_{ab}^2}\,{g_{ac}^2} +
    8\,{g_{ab}^2}\,g_{ac}\,g_{ad} +
    g_{ab}\,g_{ac}\,g_{ad}\,g_{ae} -
    4\,{g_{ab}^2}\,g_{ac}\,g_{bc} \nn \\
&&-
    2\,{g_{ab}^2}\,g_{ac}\,g_{bd} -
    g_{ab}\,g_{ad}\,g_{bc}\,g_{cd} \Big)\ .
\end{eqnarray}
Another result is obtained using (12)(12)(13)(24)(34)(34) instead
of (12)(12)(34)(34)(13)(24). The difference is
\begin{equation}
(12)(12)(13)(24)(34)(34)-(12)(12)(34)(34)(13)(24)
=32\,g_{ab}\,g_{ac}\,g_{bc}\,g_{cd}
\end{equation}
We check that this projects to 0.

Now the sloops give
\begin{eqnarray}
\diagram{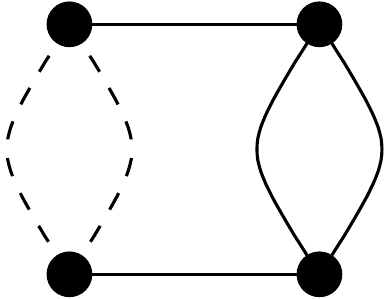}
&=& 16 \Big( 4\,{g_{ab}^3}\,g_{ac} +
    2\,{g_{ab}^2}\,{g_{ac}^2} +
    6\,{g_{ab}^2}\,g_{ac}\,g_{ad} +
    g_{ab}\,g_{ac}\,g_{ad}\,g_{ae} -
    2\,{g_{ab}^2}\,g_{ac}\,g_{bc} -
    2\,{g_{ab}^2}\,g_{ac}\,g_{bd} \nn \\
&&-
    g_{ab}\,g_{ad}\,g_{bc}\,g_{cd} \Big)\\
\diagram{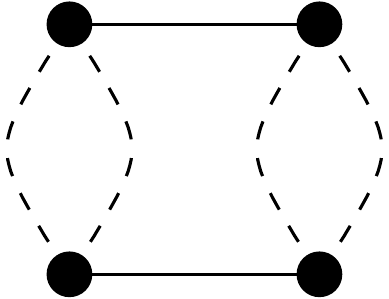}
&=& 16\left( 4\,{g_{ab}^2}\,g_{ac}\,g_{ad} +
    g_{ab}\,g_{ac}\,g_{ad}\,g_{ae} -
    2\,{g_{ab}^2}\,g_{ac}\,g_{bd} -
    g_{ab}\,g_{ad}\,g_{bc}\,g_{cd} \right)\ .
\end{eqnarray}
The following combination is simple
\begin{eqnarray}
&&\diagram{3loopm}
 -2\,\diagram{3loopm1sloop2}
+\diagram{3loopm2sloop2}
=64\,{g_{ab}^4} + 64\,{g_{ab}^2}\,{g_{ac}^2}
\nn \\
&&\qquad =\mbox{1-rep} + 64 \sum_{a,b}\left( {R'''_{ab}}^4 -2 {R'''_{ab}}^2
{R'''_{0}}^{2} \right) + \mbox{3-reps}\ .
\end{eqnarray}

\subsubsection{Diagram (\textit{n})}\label{s:(n)}
${\rm Comb}=\frac1{2^3} \times \frac1{3!} \times (\frac12)^3$.
We have with the choice of contractions (13)(13)(23)(23)(34)(34)
\begin{eqnarray}
\diagram{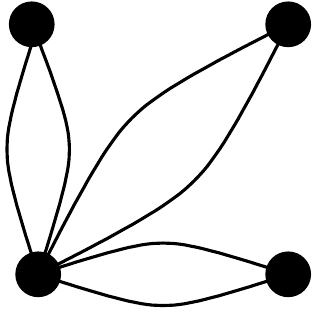}&=&
16(4{h_{ab}}^3s_{ab}+12{h_{ab}}^2h_{bc}s_{ab}+6h_{ab}h_{ac}h_{cd}s_{ac}+6h_{ac}h_{bc}h_{cd}s_{ac}+3h_{ab}h_{cd}h_{de}s_{ad}\nn \\
&&+h_{ad}h_{bd}h_{de}s_{cd})\ .\qquad
\end{eqnarray}
A single sloop is
\begin{eqnarray}\label{lf8}
\diagram{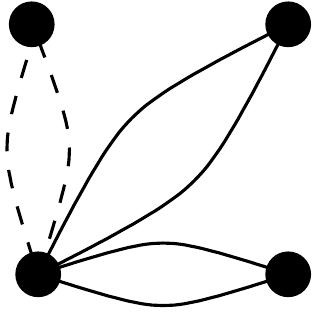}&=&
16(4h_0{h_{ab}}^2s_{ab}+8h_0h_{ab}h_{bc}s_{ab}+2h_0h_{ab}h_{cd}s_{ac}+
2h_0h_{bc}h_{cd}s_{ac}+3h_{ab}h_{cd}h_{de}s_{ad}\nn \\
&&+4h_{ab}{h_{bc}}^2s_{bc}+4h_{ab}h_{bc}h_{cd}s_{bc}+4h_{ac}h_{bc}h_{cd}s_{bc}+h_{ad}h_{bd}h_{de}
s_{cd}) \ .\nn \\
\end{eqnarray}
Double and triple sloops yield
\begin{eqnarray}\label{lf9}
 \diagram{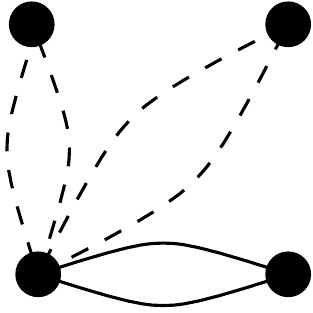}&=& \frac{1}{2}\left[\diagram{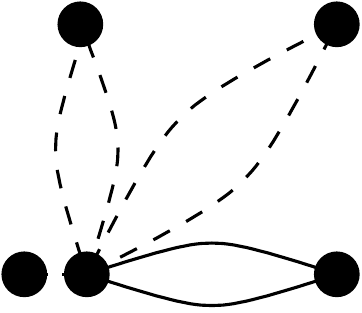}+
\diagram{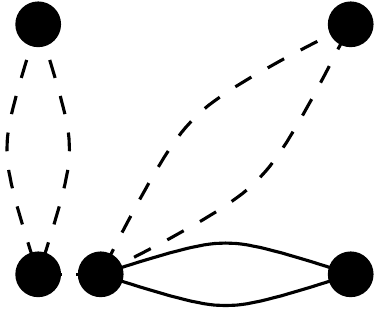} \right]   \\
 \diagram{3loopnC1}&=&
32(2h_0^2h_{ab}s_{ab}+2h_0^2h_{bc}s_{ab}+2h_0h_{ab}h_{cd}s_{ac}+
h_{ab}h_{ac}h_{de}s_{ad}+4h_0h_{ab}h_{bc}s_{bc}\nn \\
&&+2h_0h_{ac}h_{cd}s_{bc}+
2h_{ac}h_{bc}h_{cd}s_{cd}+h_{ad}h_{bd}h_{de}s_{cd})  \\
 \diagram{3loopnC2}&=&
32(2h_0^2h_{ab}s_{ab}+2h_0^2h_{bc}s_{ab}+2h_0h_{ab}h_{cd}s_{ac}+
2h_0h_{bc}h_{cd}s_{ac}+2h_{ab}h_{cd}h_{de}s_{ad}\nn \\
&&+4h_0h_{ab}h_{bc}s_{bc}+
2h_{ad}h_{bc}h_{cd}s_{cd})  \\
 \diagram{3loopnC}&=&
16(4h_0^2h_{ab}s_{ab}+4h_0^2h_{bc}s_{ab}+4h_0h_{ab}h_{cd}s_{ac}+
4h_0h_{bc}h_{cd}s_{ac}+3h_{ab}h_{ac}h_{de}s_{ad}\nn \\
&&+8h_0h_{ab}h_{bc}s_{bc} +
2h_{ac}h_{bc}h_{cd}s_{cd}+2h_{ad}h_{bc}h_{cd}s_{cd}+h_{ad}h_{bd}h_{de}s_{cd})\ .
\\
\label{lf10}
 \diagram{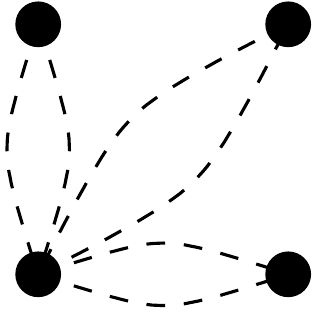}&=& \frac{1}{4}\left[\diagram{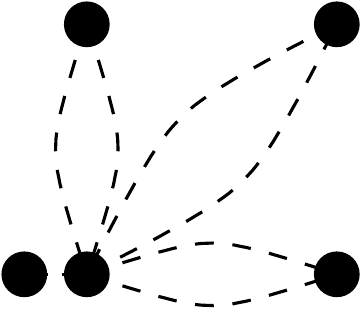}+
3\diagram{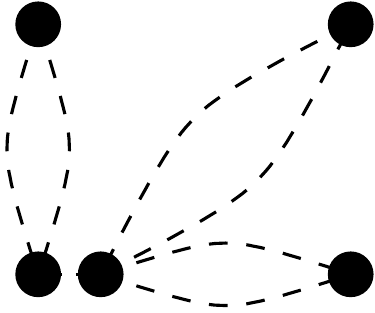} \right] \\
 \diagram{3loopnD1}&=&
16(h_0^3s_{ab}+3h_0^2h_{ab}s_{ac}+3h_0h_{ab}h_{ac}s_{ad}+
h_{ab}h_{ac}h_{ae}s_{ad}) \\
 \diagram{3loopnD2}&=&
16(h_0^3s_{ab}+3h_0^2h_{ab}s_{ac}+2h_0h_{ab}h_{cd}s_{ac}+h_0h_{ab}h_{ac}s_{ad}+h_{ab}h_{ac}h_{de}s_{ad})\\
\diagram{3loopnD}&=&
4(4h_0^3s_{ab}+12h_0^2h_{ab}s_{ac}+6h_0h_{ab}h_{cd}s_{ac}+
6h_0h_{ab}h_{ac}s_{ad}+h_{ab}h_{ac}h_{ae}s_{ad}\nn \\
&&+3h_{ab}h_{ac}h_{de}s_{ad}).
\end{eqnarray}
The final result is
\begin{equation}
 \diagram{3loopn}-3 \diagram{3loopnB}+3 \diagram{3loopnC}-4
\diagram{3loopnD} =64 (h_{ab}-{h_{0}})^3s_{ab}\ .
\end{equation}
This is confirmed by the recursive-construction algorithm.

\subsubsection{Diagram (\textit{o})}\label{s:(o)}
${\rm Comb}=\frac1{2^3} \times \frac12 \times (\frac12)^2$.
\begin{eqnarray}
\diagram{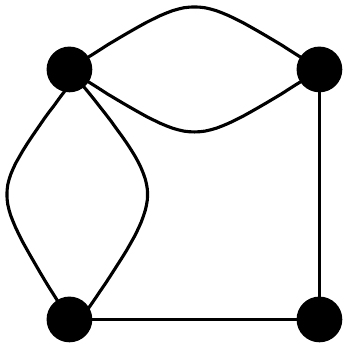} &=& 64\,f_{ab}\,{g_{ab}}^2\,h_{ab} +
  128\,f_{ab}\,{g_{ab}}^2\,h_{bc} -
  64\,f_{ab}\,g_{ab}\,g_{ac}\,h_{bc} -
  64\,f_{ab}\,g_{ab}\,g_{bc}\,h_{bc} \nn \\
&&-
  64\,f_{bc}\,g_{ab}\,g_{bc}\,h_{bc} +
  16\,f_{ab}\,g_{ac}\,g_{bc}\,h_{cd} +
  96\,f_{bc}\,g_{ac}\,g_{bc}\,h_{cd} -
  16\,f_{ab}\,g_{ad}\,g_{bc}\,h_{cd} \nn \\
&&-
  32\,f_{ac}\,g_{ad}\,g_{bc}\,h_{cd} -
  16\,f_{cd}\,g_{ad}\,g_{bc}\,h_{cd} +
  16\,f_{ab}\,{g_{bc}}^2\,h_{cd} -
  16\,f_{ab}\,g_{bc}\,g_{bd}\,h_{cd} \nn \\
&&-
  32\,f_{ac}\,g_{bc}\,g_{cd}\,h_{cd} +
  16\,f_{ad}\,g_{bd}\,g_{cd}\,h_{de}\\
\diagram{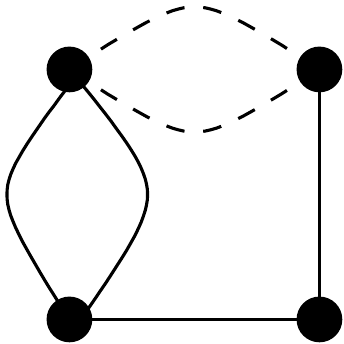}&=&32\,f_{ab}\,{g_{ab}}^2\,h_{bc} -
  32\,f_{ab}\,g_{ab}\,g_{ac}\,h_{bc} -
  32\,f_{ab}\,g_{ab}\,g_{bc}\,h_{bc} +
  32\,f_{bc}\,g_{ac}\,g_{bc}\,h_{bc} \nn \\
&&+
  16\,f_{ab}\,g_{ac}\,g_{bc}\,h_{cd}  +
  64\,f_{bc}\,g_{ac}\,g_{bc}\,h_{cd} -
  16\,f_{ab}\,g_{ad}\,g_{bc}\,h_{cd} -
  32\,f_{ac}\,g_{ad}\,g_{bc}\,h_{cd} \nn \\
&&-
  16\,f_{cd}\,g_{ad}\,g_{bc}\,h_{cd} +
  16\,f_{ab}\,{g_{bc}}^2\,h_{cd}  -
  16\,f_{ab}\,g_{bc}\,g_{bd}\,h_{cd} -
  32\,f_{ac}\,g_{bc}\,g_{cd}\,h_{cd} \nn \\
&&+
  16\,f_{ad}\,g_{bd}\,g_{cd}\,h_{de}
\eea
\bea
\diagram{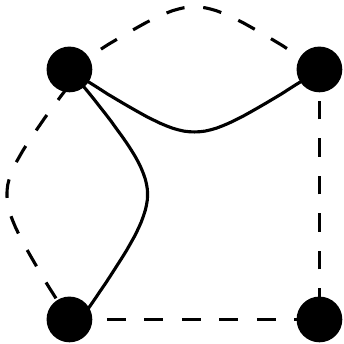}&=&4\,h_{0}\,f_{ab}\,{g_{ab}}^2 +
  8\,{h_0}\,f_{ab}\,g_{ab}\,g_{ac} +
  4\,{h_0}\,f_{ab}\,g_{ac}\,g_{ad} +
  4\,f_{ab}\,{g_{ab}}^2\,h_{ac} +
  8\,f_{ab}\,g_{ab}\,g_{ac}\,h_{ad} \nn \\
&&+
  4\,f_{ab}\,g_{ac}\,g_{ad}\,h_{ae}\ .
\end{eqnarray}
For 2 touching loops, intersections are possible:
\begin{eqnarray}
\diagram{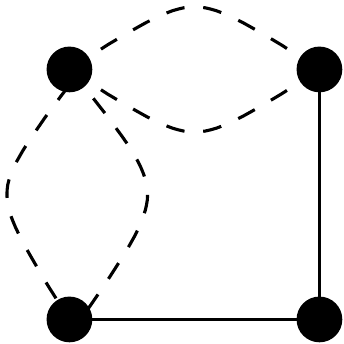} =\frac{1}{2}\left[ \diagram{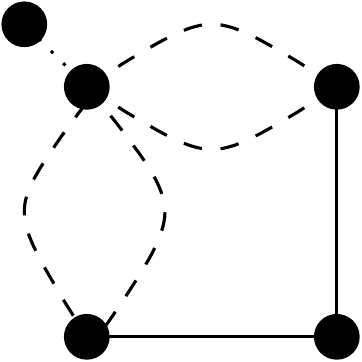}+
\diagram{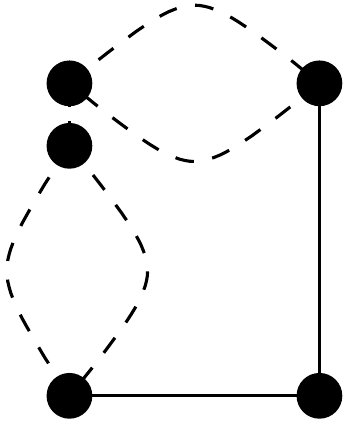} \right]
\end{eqnarray}
The terms are
\begin{eqnarray}
 \diagram{3loopoD1}&=&32\,f_{ab}\,{g_{ac}}^2\,h_{cd} -
  32\,f_{ab}\,g_{ac}\,g_{ad}\,h_{cd} -
  64\,f_{ac}\,g_{bc}\,g_{cd}\,h_{cd} +
  32\,f_{ad}\,g_{bd}\,g_{cd}\,h_{de}d1 \\
 \diagram{3loopoD2}&=&32\,f_{ab}\,g_{ac}\,g_{bc}\,h_{cd} +
  64\,f_{ac}\,g_{ac}\,g_{bc}\,h_{cd} -
  64\,f_{ac}\,g_{ad}\,g_{bc}\,h_{cd} -
  32\,f_{ab}\,g_{ac}\,g_{bd}\,h_{cd} \nn \\
&&-
  32\,f_{cd}\,g_{ac}\,g_{bd}\,h_{cd}\qquad \qquad \\
 \diagram{3loopoD}&=&16\,f_{ab}\,{g_{ac}}^2\,h_{cd} -
  16\,f_{ab}\,g_{ac}\,g_{ad}\,h_{cd} +
  16\,f_{ab}\,g_{ac}\,g_{bc}\,h_{cd} +
  32\,f_{ac}\,g_{ac}\,g_{bc}\,h_{cd} \nn \\
&&-
  32\,f_{ac}\,g_{ad}\,g_{bc}\,h_{cd} -
  16\,f_{ab}\,g_{ac}\,g_{bd}\,h_{cd} -
  16\,f_{cd}\,g_{ac}\,g_{bd}\,h_{cd} -
  32\,f_{ac}\,g_{bc}\,g_{cd}\,h_{cd} \nn \\
&&+
  16\,f_{ad}\,g_{bd}\,g_{cd}\,h_{de}\ .
\end{eqnarray}
The simplest combination is
\begin{eqnarray}
&&\diagram{3loopo}- 2 \diagram{3loopoB}+\diagram{3loopoD} = 64
f_{ab}g_{ab}^{2}\left(h_{ab}+h_{bc} \right)  \nn \\
&&\qquad = 64\sum _{a,b}
R^{(4)}_{ab}\,{R'''_{ab}}^2\,R''_{ab}
- R''_0\,R''''_{ab}\,{R'''_{ab}}^2- R''''_0\,{R'''_{0}}^{2}\,R''_{ab} \label{C77}\ .
\end{eqnarray}
This is  confirmed by recursive construction.

\subsubsection{Diagram (\textit{p})}\label{s:(p)}The diagram (\textit{p})  has ${\rm Comb}=\frac1{2^3} \times 1 \times (\frac12)^2$ and is
\begin{eqnarray}
\diagram{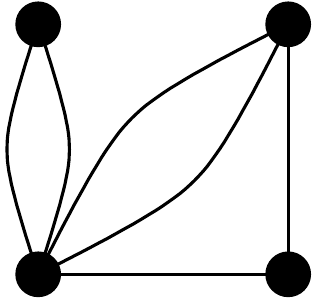}&=&
16(4g_{ab}{h_{ab}}^2p_{ab}+10g_{ab}h_{ab}h_{bc}p_{ab}-
2g_{ac}h_{ab}h_{bc}p_{ab}-2g_{bc}h_{ab}h_{bc}p_{ab}-g_{cd}h_{ab}h_{cd}p_{ac}\nn \\
&&+
3g_{bc}h_{ac}h_{cd}p_{ac}+g_{cd}h_{ab}h_{de}p_{ad}+2g_{ac}{h_{bc}}^2p_{bc}+3g_{bc}h_{ab}h_{cd}p_{bc}
-g_{bd}h_{ab}h_{cd}p_{bc}\nn \\
&&+3g_{bc}h_{ac}h_{cd}p_{bc}-
g_{bd}h_{ac}h_{cd}p_{bc}-g_{cd}h_{ac}h_{cd}p_{bc}+g_{bd}h_{ac}h_{cd}p_{cd}+g_{bd}h_{ad}h_{de}p_{cd})\ .
\qquad \ \ \ \ \
\end{eqnarray}
The two 1-sloop terms are
\begin{eqnarray}
\diagram{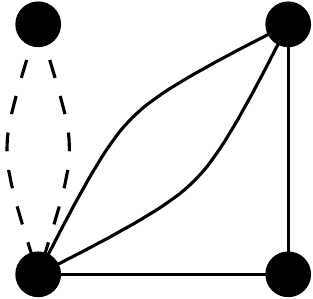}&=&
16(4h_0g_{ab}h_{ab}p_{ab}+6h_0g_{ab}h_{bc}p_{ab}-2h_0g_{ac}h_{bc}p_{ab}-
2h_0g_{bc}h_{bc}p_{ab}+h_0g_{bc}h_{cd}p_{ac}
\nn \\
&&-g_{cd}h_{ab}h_{cd}p_{ac}+
g_{cd}h_{ab}h_{de}p_{ad}+2h_0g_{ac}h_{bc}p_{bc}+4g_{bc}h_{ab}h_{bc}p_{bc}
+h_0g_{ac}h_{cd}p_{bc}\nn \\
&&+3g_{bc}h_{ab}h_{cd}p_{bc}-g_{bd}h_{ab}h_{cd}p_{bc}+
3g_{bc}h_{ac}h_{cd}p_{bc}-g_{bd}h_{ac}h_{cd}p_{bc} -g_{cd}h_{ac}h_{cd}p_{bc}\nn \\
&&+
g_{bd}h_{ac}h_{cd}p_{cd}+g_{bd}h_{ad}h_{cd}p_{cd}+
g_{bd}h_{ad}h_{de}p_{cd})
 \\
\diagram{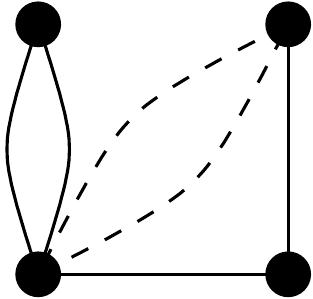}&=&
16(2g_{ab}h_{ab}h_{bc}p_{ab}-2g_{ac}h_{ab}h_{bc}p_{ab}-
2g_{bc}h_{ab}h_{bc}p_{ab}+g_{ac}h_{ab}h_{cd}p_{ac}-g_{ad}h_{ab}h_{cd}p_{ac}\nn \\
&&-
g_{cd}h_{ab}h_{cd}p_{ac}+3g_{bc}h_{ac}h_{cd}p_{ac}+g_{cd}h_{ab}h_{de}p_{ad}+
2g_{ac}{h_{bc}}^2p_{bc}+g_{bc}h_{ac}h_{cd}p_{bc}\nn \\
&&-g_{bd}h_{ac}h_{cd}p_{bc}
-g_{cd}h_{ac}h_{cd}p_{bc}\nn \\
&&+g_{bd}h_{ac}h_{cd}p_{cd}+g_{bd}h_{ad}h_{de}p_{cd})\ .
\end{eqnarray}
There are again the 2-sloop terms,
\begin{eqnarray}
\diagram{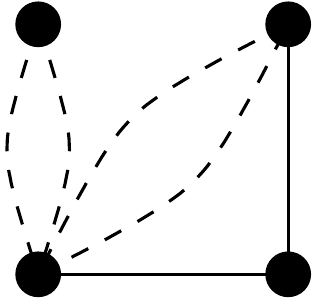} &=& \half \left[ \diagram{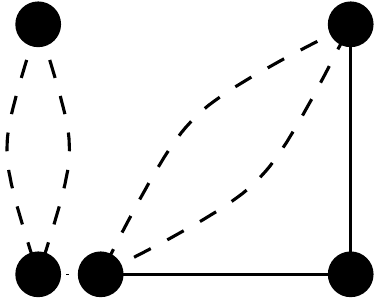} +
\diagram{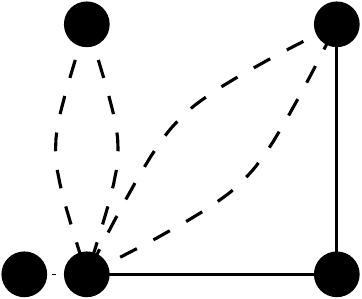} \right]   \\
\diagram{3loopp22sloopsA}&=&
16(2h_0g_{ab}h_{bc}p_{ab}-2h_0g_{ab}h_{bc}p_{ac}+2h_0g_{bc}h_{bc}p_{ac}+
2g_{ac}h_{ab}h_{cd}p_{ac}-2g_{ac}h_{ab}h_{cd}p_{ad}\nn \\
&&-2h_0g_{ab}h_{bc}p_{bc} +
2h_0g_{ac}h_{cd}p_{bc}+2g_{cd}h_{ad}h_{cd}p_{bd}-2g_{bc}h_{ac}h_{cd}p_{cd}+
2g_{bd}h_{ad}h_{de}p_{cd})\nn \\
&&
 \\
\diagram{3loopp22sloopsB}&=&
16(2h_0g_{ab}h_{bc}p_{ab}-2h_0g_{ab}h_{bc}p_{ac}+2h_0g_{bc}h_{bc}p_{ac}+
2h_0g_{bc}h_{cd}p_{ac}+2g_{cd}h_{ab}
h_{cd}p_{ad}\nn \\
&&+2g_{cd}h_{ab}h_{de}p_{ad} -2h_0g_{ab}h_{bc}p_{bc}+
2g_{bc}h_{ac}h_{cd}p_{bc}-2g_{bc}h_{ad}h_{cd}p_{bd}-2g_{bc}h_{ad}h_{cd}p_{cd})\nn \\
&&\ \\
 \diagram{3loopp22sloops}&=&
16(2h_0g_{ab}h_{bc}p_{ab}-2h_0g_{ab}h_{bc}p_{ac}+2h_0g_{bc}h_{bc}p_{ac}+
h_0g_{bc}h_{cd}p_{ac}+g_{ac}h_{ab}h_{cd}p_{ac}\nn \\
&&-g_{ac}h_{ab}h_{cd}p_{ad}+
g_{cd}h_{ab}h_{cd}p_{ad}+g_{cd}h_{ab}h_{de}p_{ad}-2h_0g_{ab}h_{bc}p_{bc}+
h_0g_{ac}h_{cd}p_{bc}\nn \\
&&+g_{bc}h_{ac}h_{cd}p_{bc}-g_{bc}h_{ad}h_{cd}p_{bd}
 +
g_{cd}h_{ad}h_{cd}p_{bd}-g_{bc}h_{ac}h_{cd}p_{cd}-g_{bc}h_{ad}h_{cd}p_{cd}
\nn \\
&&+g_{bd}h_{ad}h_{de}p_{cd})\ .
\end{eqnarray}
There are more sloops, but we find a simple expression with only the
above. It is
\begin{eqnarray}\label{Wickpfinal}
&&\diagram{3loopp}-\diagram{3looppl2sloop}-\diagram{3looppr2sloop}+
\diagram{3loopp22sloops}  \nn \\
&&\qquad =
-64h_0g_{ab}h_{ab}p_{ab}+64g_{ab}{h_{ab}}^2p_{ab}-
64h_0g_{ab}h_{bc}p_{ab}+64g_{ab}h_{ab}h_{bc}p_{ab}\ .
\end{eqnarray}
Trivially deslooping gives, as with the recursive-construction algorithm,
\begin{equation}
\diagram{3loopp} = 64\, R'''_{u}R^{(5)}_{u} (R''_{u}-R''_{0})^{2}\ .
\end{equation}

\subsubsection{Diagram (\textit{q})}\label{s:(q)}
Diagram ({\it q}) has  ${\rm Comb}=\frac1{2^3} \times \frac12 \times (\frac12)^2$.
We make contractions  (13)(13)(24)(24)(23)(34)
\begin{eqnarray}\label{lf11}{\parbox{0cm}{\rule{0mm}{7mm}}^1_3}
\diagram{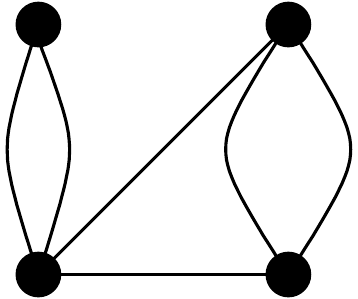}
{\parbox{0cm}{\rule{0mm}{7mm}}^2_4} &=&
16(4f_{ab}{g_{ab}}^2h_{ab}+4f_{ab}g_{ab}g_{ac}h_{ab}+6f_{ab}{g_{bc}}^2h_{ab}+
4f_{bc}{g_{bc}}^2h_{ab}+3f_{bd}{g_{bc}}^2h_{ab}\nn \\
&&+2f_{ab}g_{bc}g_{bd}h_{ab} +
2f_{bc}g_{bc}g_{bd}h_{ab}+f_{bd}g_{bc}g_{be}h_{ab}+3f_{ac}{g_{cd}}^2h_{ab}+
f_{ac}g_{cd}g_{ce}h_{ab}\nn \\
&&+2f_{bc}g_{bc}g_{bd}h_{ac}-2f_{cd}g_{bc}g_{bd}
h_{ac} - 2f_{bc}g_{ab}g_{ac}h_{bc})\ .
\end{eqnarray}
Sloop (13)(13), then (24)(24)(23)(34) gives
\begin{eqnarray}\label{lf12}
\diagram{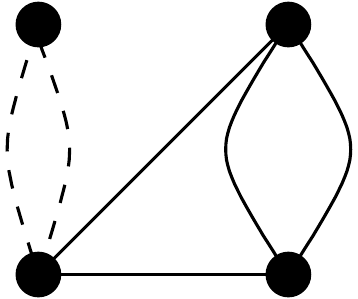}&=&
16(4h_0f_{ab}{g_{ab}}^2+4h_0f_{ab}g_{ab}g_{ac}-2h_0f_{bc}g_{ab}g_{ac}+
2h_0f_{ac}g_{ab}g_{ad}+6h_0f_{ab}{g_{bc}}^2\nn \\
&&+4f_{bc}{g_{bc}}^2h_{ab}+
3f_{bd}{g_{bc}}^2h_{ab}+2f_{bc}g_{bc}g_{bd}h_{ab}+f_{bd}g_{bc}g_{be}h_{ab}+
3f_{ac}{g_{cd}}^2h_{ab}\nn \\
&&+f_{ac}g_{cd}g_{ce}h_{ab}+2f_{bc}g_{bc}g_{bd}h_{ac} -
2f_{cd}g_{bc}g_{bd}h_{ac})\ .
\end{eqnarray}
Sloop (24)(24), then (13)(13)(34)(24) yields
\begin{eqnarray}\label{lf13}
\diagram{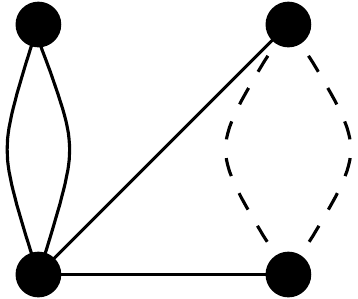}&=&
16(4f_{ab}g_{ab}g_{ac}h_{ab}+2f_{ab}g_{bc}g_{bd}h_{ab}+
2f_{bc}g_{bc}g_{bd}h_{ab}+f_{bd}g_{bc}g_{be}h_{ab}+ f_{ad}{g_{cd}}^2h_{ab}\nn\\&&+
f_{ac}g_{cd}g_{ce}h_{ab} +2f_{ac}{g_{bc}}^2h_{ac}+
2f_{bc}g_{bc}g_{bd}h_{ac}-2f_{cd}g_{bc}g_{bd}h_{ac}+
f_{cd}{g_{bd}}^2h_{ad}\nn\\ &&-2f_{bc}g_{ab}g_{ac}h_{bc})\ .
\end{eqnarray}
Sloops (13)(13)and (24)(24), then (23) (34) gives
\begin{eqnarray}
\diagram{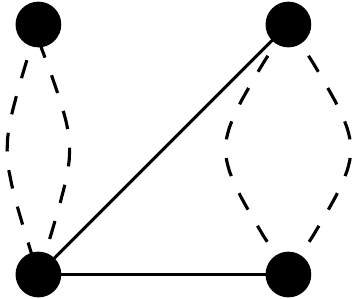}
&=&
16(4h_0f_{ab}g_{ab}g_{ac}-2h_0f_{bc}g_{ab}g_{ac}+2h_0f_{ac}g_{ab}g_{ad}
+2h_0f_{ac}{g_{bc}}^2+2f_{bc}g_{bc}g_{bd}h_{ab}\nn \\
&&+f_{bd}g_{bc}g_{be}h_{ab}
+f_{ad}{g_{cd}}^2h_{ab}+f_{ac}g_{cd}g_{ce}h_{ab}+
2f_{bc}g_{bc}g_{bd}h_{ac}-2f_{cd}g_{bc}g_{bd}h_{ac}\nn \\
&&+f_{cd}{g_{bd}}^2h_{ad})\ .
\end{eqnarray}
There are more sloops, but
our now acquired experience tells us that the simplest combination
should be
\begin{eqnarray}
&&\diagram{3loopq}-\diagram{3loopqB}-\diagram{3loopqC}+\diagram{3loopqD}\nn\\
&&\qquad =
64(f_{ab}{g_{ab}}^2h_{ab}+f_{ab}{g_{bc}}^2h_{ab}-h_0f_{ab}{g_{ab}}^2
-h_0f_{ab}{g_{bc}}^2)\ .
\end{eqnarray}
Trivial de-slooping yields in agreement with recursive construction
\begin{equation}
\diagram{3loopq} = 64\, [ {R'''_{u}}^{2}-  {R'''_{0}}^{2}] R''''_{u}
(R''_{u}-R''_{0}).
\end{equation}

\newcommand{\doi}[2]{\href{http://dx.doi.org/#1}{#2}}
\newcommand{\arxiv}[1]{\href{http://arxiv.org/abs/#1}{#1}}


\tableofcontents

\end{document}